\documentstyle[12pt]{article}
\input epsf

\def\percent{\%}
\def\Gbsg{\Gamma(b\to s\gamma)}
\def\msbar{\overline{\rm MS}}
\def\be{\begin{equation}}
\def\ee{\end{equation}}
\def\bea{\begin{eqnarray}}
\def\eea{\end{eqnarray}}
\def\lt{\lambda_t}
\def\lb{\lambda_b}
\def\ltau{\lambda_\tau}
\def\lN{\lambda_N}
\def\ltbtau{\lambda_{t,b,\tau}}
\def\lbtau{\lambda_{b,\tau}}
\def\lG{\lambda_G}
\def\vU{v_U}
\def\vD{v_D}
\def\db{\delta_b}

\def\GeV{{\rm GeV}}
\def\TeV{{\rm TeV}}
\def\Mgut{M_{\rm GUT}}
\def\mgl{m_{\tilde g}}

\def\msb{m_{\tilde b}}
\def\msq{m_{\widetilde Q}}
\def\mwi{m_{\widetilde W}}
\def\wi{\widetilde W}
\def\meff{m_{\rm eff}}
\def\mst{m_{\tilde t}}
\def\mstau{m_{\tilde \tau}}
\def\msl{m_{\widetilde L}}
\def\muc{m_{\tilde u,\tilde c}}
\def\mds{m_{\tilde d,\tilde s}}
\def\mqonetwo{m_{\widetilde Q_{1,2}}}
\def\memu{m_{\tilde e,\tilde \mu}}
\def\mlonetwo{m_{\widetilde L_{1,2}}}
\def\mcharj{m_{\widetilde{\chi}_j^-}}
\def\mcharone{m_{\widetilde{\chi}_1^-}}
\def\mchartwo{m_{\widetilde{\chi}_2^-}}
\def\mch{m_{H^-}}
\def\mtenh{M_{{\bf 10}_H}}
\def\msixth{M_{{\bf 16}_3}}
\def\mx{M_X}
\def\ms{m_S}
\def\mtenth{M_{{\bf 10}_3}}
\def\mfivth{M_{{\bf \overline{5}}_3}}
\def\mfivh{M_{{\bf 5}_H}}
\def\mfivbh{M_{{\bf \overline{5}}_H}}
\def\msufive{M_{\rm SU(5)}}
\def\epsz{\epsilon_Z}
\def\epsb{\epsilon_B}
\def\epsbsg{\epsilon_{b\to s\gamma}}
\def\epsc{\epsilon_c}
\def\achar{{\cal A}_{\widetilde\chi^-}}
\def\achartwo{{\cal A}'_{\widetilde\chi^-}}
\def\agluino{{\cal A}_{\widetilde g}}
\def\asm{{\cal A}_{\rm SM}}
\def\ach{{\cal A}_{H^-}}
\def\roughly#1{\,\,
\raise.3ex\hbox{$#1$\kern-.75em\lower1ex\hbox{$\sim$}}\,\,}
\def\half{{\textstyle{1\over2}}}
\def\fourth{{\textstyle{1\over4}}}

\def\thrfour{{\textstyle{3\over4}}}
\def\fivefour{{\textstyle{5\over4}}}
\def\ninefour{{\textstyle{9\over4}}}
\def\third{{\textstyle{1\over3}}}
\def\twothr{{\textstyle{2\over3}}}
\def\sixth{{\textstyle{1\over6}}}
\def\dtau{{d\over d\tau}}
\def\Xt{X_t}
\def\Xb{X_b}
\def\Xtau{X_{\tau}}
\def\At{A_t}
\def\Ab{A_b}
\def\Atau{A_{\tau}}
\def\case#1#2{{#1\over#2}}
\def\HH{{\cal H}}
\def\L#1{{\cal L}_{#1}}
\def\Ve{V_{\rm eff}}
\def\Vzero{V_0(\Lambda)}
\def\Vone{V_1(\Lambda)}
\def\CC{{\rm SU(3)_c\times U(1)_{em}}}
\def\bi{{\widetilde B}}
\def\spa{\phantom{3}}

\def\mph{\phantom{-}}

\def\zer{\phantom{.0}0}
\def\fone{{\,1\,}}
\def\fsev{{1\over7}\,}
\def\psev{\phantom{{1\over7}\,}}
\def\lhigh{\Lambda_{\rm HIGH}}
\def\llow{\Lambda_{\rm LOW}}
\def\vev#1{\langle #1 \rangle}

\begin{document}
\begin{titlepage}
\begin{center}
May 1995\hfill    SU-ITP-94-16 \\
               \hfill    RU-95-13 \\
               \hfill    hep-ph/9405428
\vskip .2in
{\large \bf
The Unified Minimal Supersymmetric Model with
Large Yukawa Couplings}
\vskip .3in
Riccardo Rattazzi\footnote{E-mail:
rattazzi@physics.rutgers.edu}\\[.03in]
{\em Department of Physics and Astronomy\\
     Rutgers University\\
     Piscataway, NJ  08855}
\vskip 10pt
Uri Sarid\footnote{E-mail:
sarid@squirrel.stanford.edu. Address after July 1995: Dept.~of
Physics,
University of Notre Dame, Notre Dame IN 46556.}\\[.03in]
{\em Physics Department\\
     Stanford University\\
     Stanford, California 94305}
\end{center}
\vskip .2in
\begin{abstract}
\medskip
The consequences of assuming the third-generation Yukawa couplings
are all large and comparable are studied in the context of the
minimal supersymmetric extension of the standard model. General
aspects of the RG evolution of the parameters, theoretical
constraints needed to ensure proper electroweak symmetry breaking,
and experimental and cosmological bounds on low-energy parameters
are presented. We also present complete and exact semi-analytic
solutions to the 1-loop RG equations. Focusing on SU(5) or SO(10)
unification, we analyze the relationship between the top and bottom
masses and the superspectrum, and the phenomenological implications
of the GUT conditions on scalar masses. Future experimental
measurements of the superspectrum and of the strong coupling will
distinguish between various GUT-scale scenarios. And if present
experimental knowledge is to be accounted for most naturally, a
particular set of predictions is singled out.
\end{abstract}
\end{titlepage}

\section{Introduction}
\label{sec:intro}
Unified theories are currently the most promising candidates for
physics beyond the standard model. The marriage of force
unification, namely grand unified theories (GUTs) or perhaps string
theory, and spin unification, by which we mean a supersymmetry
relating fermions and bosons, has been a fruitful and prolific area
of research in the last few decades. Grand unification allows the
understanding of the electroweak and strong forces as low-energy
manifestations of a single microscopic force, in particular
explaining the quantization and the assignments of electromagnetic
charges for all known particles. The simplest GUTs \cite{gg,gn},
based on SU(5)
or SO(10) gauge groups, unify some [SU(5)] or all [SO(10)] of the
quarks and leptons in each generation. The unified matter multiplets
neatly encompass the known standard-model matter particles --- no
new particles are needed, and no known particles are left out. The
one exception is the right-handed neutrino, which must be included
in SO(10) unification. If neutrinos have masses, this potential
embarrassment becomes a
boon. In fact, SO(10) not only  favors typical ranges for
their masses via the seesaw mechanism, but in specific models
can also lead to detailed predictions about the flavor structure
of the mass matrix. Moreover, SO(10) beautifully incorporates both
the Pati-Salam idea \cite{ref:pati}  of leptons as the fourth color
and
an underlying symmetry between the left- and right-handed quarks and
leptons \cite{ref:moha}. String theory aims to go beyond GUTs and
unify all forces including gravity, perhaps without any adjustable
parameters. Though still in
their infancy, string models would presumably reproduce the
successes of GUTs by either implying a grand unified theory as a
``low-energy'' consequence or by furnishing the appropriate boundary
conditions to mimic grand unification predictions.
GUTs, and even more so string theory, ambitiously span energy scales
some 13 to 15 orders of magnitude above the highest scales at which
the standard model has been directly tested.  For compelling
esthetic reasons such a span requires these theories to be
supersymmetric: fermions and bosons, present in equal numbers, with
mirror (and therefore highly restricted) interactions.
Supersymmetric theories are theoretically attractive on their own,
but when wedded to unified theories they can yield quantitative
phenomenological successes.  Supersymmetry (SUSY) cannot be an exact
symmetry of nature, but it must be approximately valid down to
roughly the electroweak scale if such a low scale is to have a hope
of being naturally embedded in a GUT- or string-scale theory
\cite{ref:witten}.
The minimal candidate model for unification is then given at low
energies by the supersymmetrized standard model, which we will call
the MSSM\footnote{Here we use the term MSSM to refer only to the
particle content and interactions of the minimal supersymmetric
extension of the (essentially) experimentally extablished standard
model, without any assumptions about the boundary conditions on its
parameters at the GUT or string scales.}, having a squark for each
quark, a slepton for each lepton, a gaugino for each gauge boson,
and Higgsinos for the requisite two Higgs doublets $H_u$ and $H_d$
which give masses to the up- and down-type
quarks and leptons (respectively). For a given effective theory
below the unification scale, such as the MSSM, the renormalization
group (RG) evolution determines the relationship between the physics
of the unified theory and the physics observed at the electroweak
scale. If one chooses as the effective theory the particular
particle content of the MSSM, and embeds this minimal model in an
SU(5) GUT, one arrives at a remarkably successful predicted
relationship between the three low-energy gauge couplings
\cite{ssq,costa,unification}. There are also various predictions for
the quarks and leptons, some more robust than others. In particular,
the bottom quark and tau lepton Yukawa couplings are in most cases
predicted to be equal at the GUT scale \cite{ceg}; a successful
prediction for the bottom mass at low energies, especially given a
heavy top quark, can then be easily obtained with precisely the two
Higgs doublets needed for the MSSM
\cite{twoHiggs}. The large Yukawa coupling of the heavy top quark
can also
trigger a correct breakdown of electroweak symmetry at low energies
\cite{ref:radbreak,ref:apw}. Thus the various pieces of the
unification
puzzle interlock tightly to produce a framework which we find both
theoretically and phenomenologically compelling.

Numerous authors have explored in detail the issues associated with
gauge coupling unification, in GUTs and lately also in string
theories. The present paper takes a different path and seeks the
consequences of Yukawa coupling unification
\cite{alsalt,ref:carbeta}. We will only be concerned with the Yukawa
couplings of the third generation, namely
the top and bottom quarks and the tau lepton, for two reasons:
first, because they are larger than the Yukawas of
the lighter generations, so it is natural to expect that they arise
directly and from renormalizable operators and so are robustly
predicted by the unified model, whereas the smaller couplings
presumably arise from more complicated mechanisms and are thus more
model-dependent; and
second, because the third generation Yukawa couplings are the only
ones big enough to appreciably influence the rest of the MSSM via
the RG
evolution. The focus of our research parallels these two
motivations: we have seen in previous work \cite{ref:hrsI}, as
summarised below, that the top mass can be predicted from
approximate or exact unification at the GUT scale; and we expand on
our previous observations \cite{ref:hrsII} that the large bottom and
tau Yukawa
couplings which result from such unification qualitatively change
the expected features of the MSSM at low energies.

The assumption underlying this work is that, at the unification
scale, either (I) $\ltau = \lb = \lt$  or at least  (II) $\ltau =
\lb \sim \lt$, where $\ltbtau$ are the Yukawa couplings to the
appropriate Higgs doublet, and ``$\sim $'' means that the couplings
differ by a factor of order one. When is this assumption valid? In
the simplest SO(10) scenario, in which both light Higgs doublets
arise from a single SO(10) multiplet, the tree-level Yukawa
couplings are exactly equal at the GUT scale, as in assumption (I)
\cite{ref:als,ref:hrsI}. Threshold corrections will typically lift
this equality somewhat, and thereby can facilitate proper
electroweak symmetry breaking, as we show below. In more involved
SO(10) models, the light Higgs doublets may come from mixtures of
several SO(10) multiplets.  Nevertheless, we expect assumption (II)
to often hold. In the simplest GUT scenarios based upon SU(5) the
bottom and tau
couplings are equal, but are unrelated to the top coupling. Most of
the work on the MSSM has usually assumed that the top coupling was
much larger than the other two, resulting in the observed hierarchy
between the top and bottom quarks. From a
GUT-scale model-building perspective it seems to us at least as
natural {\it a priori} to assume that all three Yukawas are
comparable, as in case (II); then the observed lightness of the
bottom bottom and tau must
result from the small vacuum expectation value (VEV) of the Higgs
doublet to which they couple. Finally, the jury is still out on the
predictions string theory makes for the Yukawa couplings. If the
effective field theory which describes physics below the string
scale is a
GUT, then one of the above scenarios may hold
\cite{ref:lykken}. Otherwise, there are still reasons to believe
that the Yukawa couplings for the third generation are approximately
equal, at least in some string-inspired models.

In addition to the above more theoretical motivations, there is also
a phenomenological astrophysical advantage to large $\tan\beta$, at
least in SO(10) models \cite{ref:vissani,ref:bmr}. Various
astrophysical and
cosmological data, such as the neutrino solar flux deficit and the
density
fluctuations at large scale observed by COBE, can be explained if
the left-handed neutrinos acquire a mass via the seesaw mechanism
from right-handed neutrino Majorana masses in the
$10^{10}-10^{13}\,\GeV$ range. A third-generation right-handed
neutrino with such an intermediate mass can significantly affect
bottom-tau unification
(through the RG evolution of Yukawa couplings). If SO(10)-type
boundary conditions $\lb^G=\ltau^G$ and $\lt^G=\lN^G$ are assumed,
it is difficult to reproduce the experimental value of $m_b/m_\tau$
for small $\tan\beta$ $(<10-20)$. Interestingly, though,
for larger values of $\tan\beta$, the intermediate-scale right
handed neutrino does not significantly alter the successful
prediction of $m_b/m_\tau$.

Motivated by the phenomenological successes of the
unified MSSM and by the wide-ranging contexts displaying approximate
or exact Yukawa unification, we analyze in this work various
implications of a Yukawa-unified MSSM. In Sec.~\ref{sec:topmass} we
review the
prediction of the top mass as a sensitive function not only of the
GUT-scale boundary conditions and the bottom and tau masses, but
also (perhaps surprisingly) of the
superpartner masses. Reversing the argument, we find bounds on
the superspectrum as functions of the top mass. Sec.~\ref{sec:bsg}
treats the related predictions for the radiative bottom quark decay
$b\to
s\gamma$ (and comments on $\tau\to\mu\gamma$). We point out the
importance of seemingly subdominant diagrams which are also enhanced
by large $\tan\beta$. Satisfying the recent experimental bounds on
this process places certain constraints on the superspectrum if a
delicate fine-tuning is to be avoided. In Sec.~\ref{sec:electro} we
outline the basic implications of $\lb \sim \lt$ for electroweak
symmetry breaking. Not only must the symmetry be broken radiatively
without losing the $\CC$ gauge symmetries, but
also a large hierarchy must be generated in the Higgs VEVs to
account for the top-bottom mass hierarchy. Sec.~\ref{sec:hier} deals
with the various options for generating this hierarchy. Two
symmetries (PQ
and R) can make this hierarchy more natural---and lead to a favored
superspectrum---but there is always a necessary fine-tuning of at
least one part in $\sim 50$ ($\sim m_t/m_b$) somewhere in the
Lagrangian \cite{ref:nelran,ref:hrsII}. We return in
Sec.~\ref{sec:correct} to the problem of properly breaking the
electroweak symmetry in the presence of the PQ and R symmetries. The
various conditions which must be satisfied at appropriate scales to
guarantee the proper spontaneous symmetry breaking are discussed in
some detail in subsection \ref{sec:general}. In addition to the
various individual mass-squared parameters, we examine the two flat
directions in the scalar potential (and the scales at which they may
destabilize the vacuum) and the trilinear scalar couplings from $A$
and $\mu$ terms. These can be important, even for the third
generation, when there is some hierarchy between $m_Z$ and the SUSY
scale. We then turn to the general behavior of the soft scalar
masses as they evolve down from the GUT scale, focusing in
subsection \ref{sec:homog} on the homogeneous part of the RG
equations which dominates when the PQ and R symmetries are
approximately valid. The favored scenarios with and without these
symmetries are briefly summarized in subsection \ref{sec:darwin},
without any assumptions about the soft scalar masses at the GUT
scale. These assumptions are introduced in Sec.~\ref{sec:gut}. The
GUT-scale constraints on scalar masses in minimal SO(10) theories,
and in SU(5) or nonminimal SO(10) models, are presented in
subsection \ref{sec:bc} using a common notation. The ramifications
of these SO(10)- or SU(5)-type GUT relations are explored in detail
in subsections \ref{sec:soten} and \ref{sec:sufive}, respectively.
These include the ease of obtaining proper electroweak symmetry
breaking for different values of the parameters, the possible and
probable superspectra, and the implications of lifting the PQ or R
symmetries. We also study the effects of a right-handed neutrino
mass below the GUT scale, briefly examine the consequences of
universal scalar masses, and reconsider SU(5)-type boundary
conditions as a perturbation on the SO(10)-type conditions. We then
turn in Sec.~\ref{sec:cosmo} to the astrophysical and
cosmological constraints on the Yukawa-unified MSSM. We address
both the electric neutrality of the lightest superpartner and its
relic abundance. To estimate this abundance, we adapt previous
analyses to the large $\tan\beta$ scenario, and in the case of a
bino-like LSP classify the dominant operators contributing to LSP
annihilation in order to clarify its suppression. We present our
conclusions in Sec.~9. In particular, we summarize the
phenomenological expectations from Yukawa unification, comment on
the most natural and therefore favored scenarios, and outline some
directions for future
investigation. In Appendix A we present the exact and complete
semi-analytic solutions to the 1-loop RG equations for the MSSM with
large $\ltbtau$. They are semi-analytic in that they are given in
terms of integrals over the dimensionless (gauge and Yukawa)
couplings, which must be evaluated numerically or approximated
analytically, as we show for several examples. Appendix B is devoted
to
a study of one of the potentially flat directions in the scalar
potential of the MSSM, and to the scales at which it can impose a
constraint on the scalar mass parameters. Finally, Appendix C
justifies the approximation we have made in using the RG-improved
tree-level scalar potential.

\section{Top mass: Prediction and Constraint}
\label{sec:topmass}
At tree level, the observed masses of the third family fermions are
related to their Yukawa couplings and to the VEVs of the up- and
down-type Higgs doublets via:
\bea
m_t &=& \lt \vU \equiv \lt v \sin\beta \nonumber\\
m_b &=& \lb \vD \equiv \lb v \cos\beta \nonumber\\
m_\tau &=& \ltau \vD \equiv \ltau v \cos\beta
\label{eq:treeyuk}
\eea
where $v = 174\GeV$ and $\tan\beta \equiv v_U/v_D \equiv
\vev{H_U^0}/\vev{H_D^0}$. The Yukawa couplings are in
turn determined through the renormalization group evolution by the
Yukawa couplings $\ltbtau^G$ at the GUT scale $\Mgut$. And finally,
in
the grand unified theory these couplings are related to each other
according to:
\be
\ltau^G = \lb^G = \lt^G \equiv \lG\,,
\label{eq:guttreeone}
\ee
or
\be
\ltau^G = \lb^G \sim \lt^G \equiv \lG\,.
\label{eq:guttreetwo}
\ee
In the minimal SO(10) scenario $\lt^G/\lb^G = 1$, while Higgs mixing
or an SU(5) model could suggest that $\lt^G/\lb^G$ is of order
1. The RG evolution requires as additional inputs the scale of
unification $M_G$ and the unified value $g_G$ of the gauge
couplings, both of which are already fixed by gauge unification (but
see Ref.~\cite{ref:hrsI} for the treatment of $\alpha_s$). Thus the
four low-energy
observables $m_t$, $m_b$, $m_\tau$ and $\tan\beta$ and the two
GUT-scale parameters $\lG$ and $\lt^G/\lb^G$ are related by three
(RG) equations; fixing $m_b$ and $m_\tau$ from experiment leaves
three
equations in the four remaining variables $m_t$, $\tan\beta$, $\lG$
and $\lt^G/\lb^G$, yielding a single prediction for the top mass as
a function of the angle $\beta$. (In principle, of course, the Higgs
doublets VEVs $\vU$ and $\vD$ and hence also
$\beta$ are predicted in terms of the GUT-scale parameters of the
Higgs sector, but at this stage those
parameters are completely unknown; we will return to them below.)
If, as in minimal SO(10), the initial ratio $\lt^G/\lb^G$ is fixed,
then the top mass and $\tan\beta$ are predicted individually. We
have previously presented a detailed study of this prediction
\cite{ref:hrsI} when $\lt^G$ and $\lb^G$ are split at most by
threshold corrections; we will return to the more general case
$\lt^G\sim\lb^G$ below. A crucial finding of that work (valid more
generally for any $\lt^G\sim\lb^G$) was related to the fact that,
since the mass of the  bottom quark
results from a large coupling to a small VEV rather than a small
coupling to a large VEV, any chiral symmetries protecting
this mass should act on the VEV and not on the Yukawa coupling as in
the usual case. Such  approximate symmetries, discussed in detail
below, are not respected by the parameters of the {\it
generic} MSSM, which therefore exhibits large corrections to $m_b$
from 1-loop couplings to the other Higgs VEV (\cite{ref:hrsI}; see
also \cite{ref:hempone}). In
particular, we found that the usual suppression factor of
$1/16\pi^2$ in the leading 1-loop corrections is {\it a priori
completely undone} by the enhancement $\sim\tan\beta$ from the
larger VEV of the up-type Higgs doublet.
The two dominant contributions are given by the diagrams shown in
Fig.~1. Keeping only these corrections (and therefore
dropping similar but smaller corrections to $m_\tau$), the third
generation mass relations now read
\bea
m_t &=& \lt\left[\lG,\lt^G/\lb^G\right]\,v \sin\beta \nonumber\\
m_b &=& \lb\left[\lG,\lt^G/\lb^G\right]\, v \cos\beta\,\left(1 +
{\delta m_b\over m_b}\right)
\label{eq:oneloopyuk}\\
m_\tau &=& \ltau\left[\lG,\lt^G/\lb^G\right]\, v \cos\beta
\nonumber
\eea
where we have explicitly shown the dependence of the three Yukawa
couplings at low energies on the two GUT-scale parameters. The exact
form of the corrections to the bottom mass was given in our previous
work. A useful approximation is given by $\delta m_b/m_b =
\left(\tan\beta/50\right)\, \db$, where
\be
\db \simeq {50\over16\pi^2} {\alpha_G\over\alpha_W}
{\mu\mwi\over\meff^2}
\,\left[{8\over3} {\alpha_s\over\alpha_G} g_s^2
\,f\left(\mgl^2\over\meff^2\right) -
        2 \lt^2 \,f\left(\mu^2\over\meff^2\right)
\right],
\label{eq:dmb}
\ee
$f(x) = (1-x + x \ln x) /(1-x)^2$, $\mgl$ and $\mu$ are the gluino
mass and the $\mu$ term evaluated at the electroweak scale,
$\alpha_W = g_2^2/4\pi$ and $\alpha_s = g_s^2/4\pi$ are the SU(2)
and $\rm SU(3)_c$ coupling strengths respectively  and $\meff^2
\equiv {1\over2} (\msb^2+\msq^2)$ is the average of the squared
masses of the  SU(2)-singlet bottom squark and the
SU(2)-doublet third generation squarks. [In the second (subdominant)
term we used $A_t \simeq 2 M_{1/2} = 2 {\alpha_G\over\alpha_W} \mwi$
and approximated $\msb^2+\msq^2 \simeq \mst^2+\msq^2$. Also, the
expressions are considerably modified when one of the stop or
sbottom eigenvalues becomes very small, but this will not be
relevant for the cases we study.] We see that even if $\lt^G/\lb^G$
is fixed, for example in minimal SO(10), we now have the additional
unknown $\db$, which precludes a separate determination of $m_t$ and
$\tan\beta$. In other words, if we don't know enough about the
superspectrum to pin down $\db$, we cannot fix $\lG$ by comparing
the prediction of $m_b$ with experiment, and hence we lose the
independent prediction of the top mass.

\begin{figure}[tb]
\centering
\leavevmode
\epsfxsize=18cm \epsfbox[0 450 690 650]{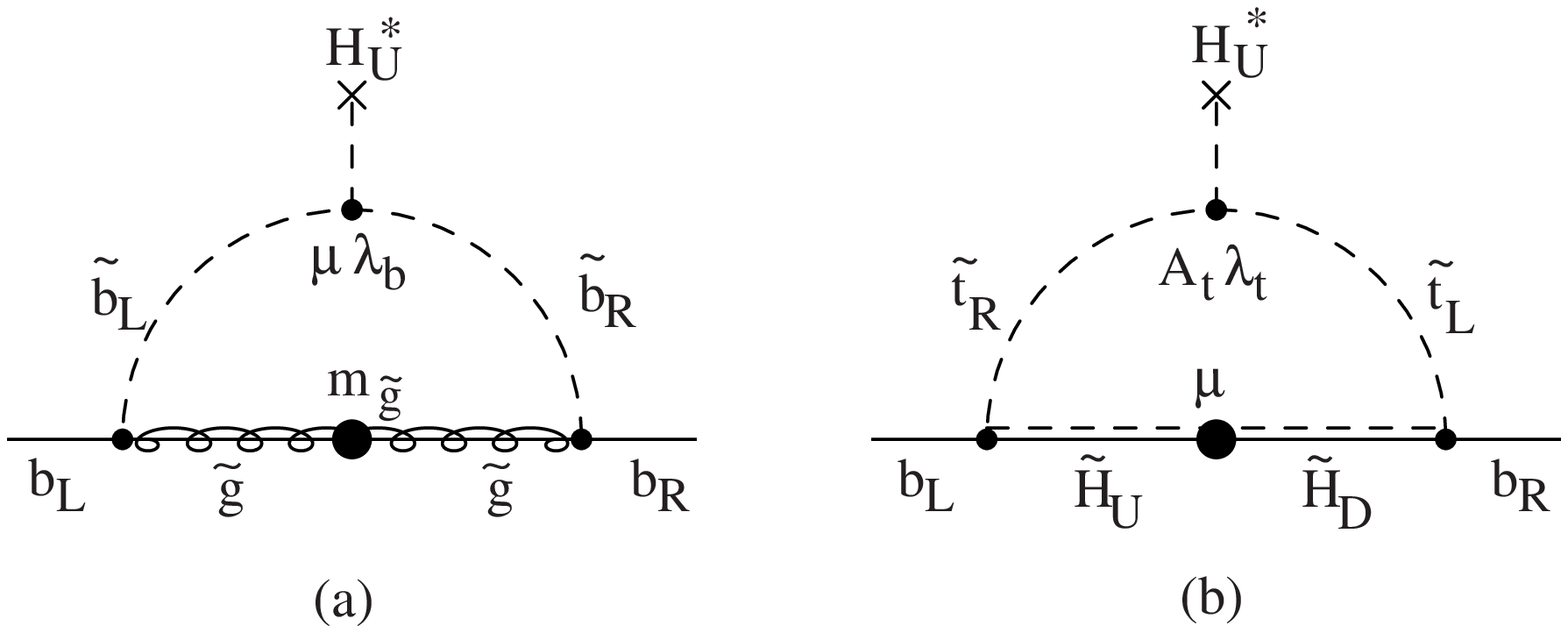}
\begin{quote}
{\small
Fig.~1. The leading (finite) 1-loop MSSM contributions to the $b$
quark mass.}
\end{quote}
\end{figure}

It will prove very helpful to understand the magnitude of the
$\delta m_b/m_b$ corrections, and in fact of all large $\tan\beta$
enhancements, from the point of view of symmetries. To this end we
recall the Peccei-Quinn (PQ) and R symmetries introduced in
Ref.~\cite{ref:hrsI}. The PQ symmetry amounts to setting $\mu = 0$ ,
while the R symmetry requires the vanishing of gaugino masses
$M_{1/2}$, of the SUSY-breaking trilinear scalar couplings $A_i$,
and of the bilinear SUSY-breaking Higgs coupling $B$. If either
symmetry were exact, then when the up-type Higgs acquired a VEV the
down-type VEV would remain zero, so $\tan\beta$ would be infinite.
(Of course we have in mind the usual scenario in which the Higgs
mass
matrix has a negative eigenvalue in the $H_u$ direction only.)
Also, down-type quarks and leptons such as the bottom and tau would
be exactly massless to all orders. We will
see that these symmetries are the key to making large $\tan\beta$ as
technically natural as possible: just as the bottom mass can be made
as light as needed by imposing the usual chiral symmetry, so
the PQ and R symmetries can be imposed on the Lagrangian to varying
degrees. Unfortunately, current LEP bounds set strong bounds on how
natural the large $\tan\beta$ scenario can be
\cite{ref:hrsII,ref:nelran}, as we discuss below. But PQ and R are
still the key to alleviating as much as possible the need for
fine-tuning, and are also useful for classifying the various
superspectra and discussing their phenomenological consequences.

If the symmetries are only approximately valid, one must specify at
what scale this approximation holds: we will see that in very
fine-tuned scenarios,
the squark masses $m_0$ evaluated at the electroweak scale are much
smaller than their values $M_0$ at the GUT scale, so an
approximately symmetric GUT Lagrangian having $\mu \sim M_{1/2} \ll
M_0$
could yield a spectrum at observable energies $\mu \sim M_{1/2} \sim
m_0$ having no observable symmetries. Thus for those cases we will
distinguish between having PQ and R symmetries at all scales, and
having them only at high scales.

\begin{figure}[htb]
\centering
\leavevmode
\epsfysize=10cm \epsfbox[10 370 680 730]{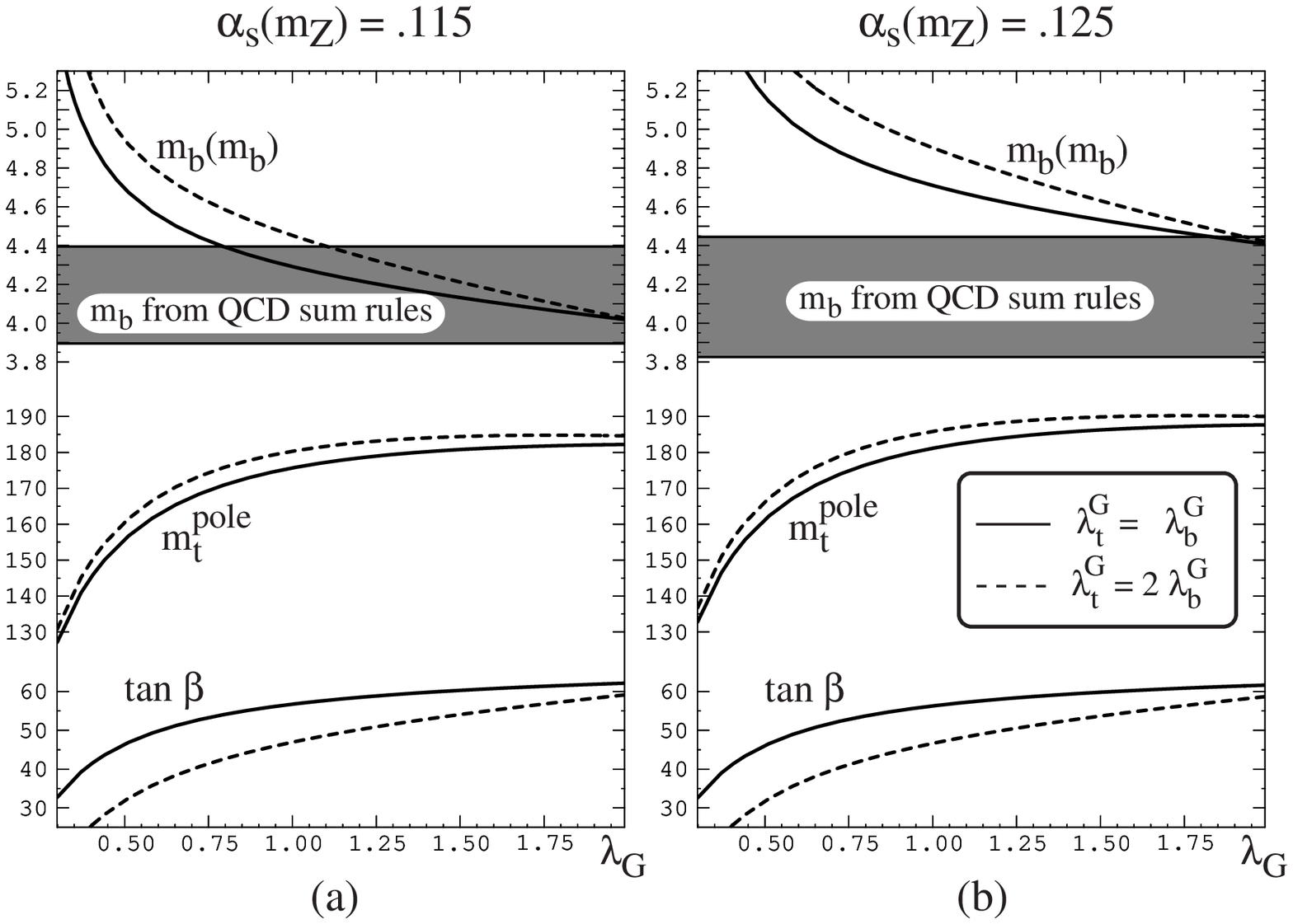}
\begin{quote}
{\small
Fig.~2. The predictions of 2-loop RG evolution (plus 1-loop
threshold corrections) for the running bottom quark mass (in GeV),
the pole top quark mass (in GeV), and $\tan\beta$, as functions of
the GUT-scale Yukawa coupling $\lG$. To compute threshold
corrections, we assumed the preferred superspectrum of
Eq.~(\ref{eq:symspec}). The solid lines correspond to exact Yukawa
unification, while the dashed lines indicate $\lG \equiv \lt^G = 2
\lb^G$ as an example of approximate Yukawa unification. The values
of $m_b$, after the finite corrections $\delta m_b$ are added,
should fall within the shaded bands in order to agree with the
bottom mass as extracted from QCD sum rules \cite{ref:hrsI}.}
\end{quote}
\end{figure}

In Fig.~2, we present the results of a detailed 2-loop analysis,
following Ref.~\cite{ref:hrsI}, of the running $\msbar$ bottom quark
mass
$m_b(m_b)$, the top quark pole mass, and the approximate ratio
$\tan\beta$ as functions of $\lG$ within the perturbative regime. We
chose two sample values of the QCD coupling $\alpha_s(m_Z)$, and
considered either exact or approximate Yukawa unification, where the
former is defined to be $\lt^G = \lb^G$ while the latter is
exemplified by $\lt^G/\lb^G = 2$. (There is in addition a slight
logarithmic dependence on the superspectrum, as calculated in
Ref.~\cite{ref:hrsI}; for definiteness we have assumed the spectrum
singled out below, namely all squarks and sleptons and the
pseudoscalar Higgs at $\sim 600\,\GeV$, while the SU(2) gaugino and
the Higgsinos are at $\sim m_Z$.) Also shown are the corresponding
allowed ranges for $m_b(m_b)$ extracted from the data on $e^+ e^-\to
b\bar b$ using QCD sum rules \cite{ref:mbI}. We use the values
obtained in the analysis of Ref.~\cite{ref:hrsI}, which update those
in Ref.~\cite{ref:gl}. The uncertainty on $m_b(m_b)$ is essentially
theoretical, being dominated by our ignorance of ${\cal
O}(\alpha_s^2)$
corrections to QCD sum rules. In the absence of  $\db$, we could
read off an
allowed range of $\lG$ by requiring agreement between the
theoretical and experimental values of $m_b$; then both $m_t$ and
$\tan\beta$
could be predicted within some range. Instead, we can only determine
the top mass (and $\tan\beta$) as functions of $\db$.

\begin{table}[tb]
\centering
\renewcommand\arraystretch{1.2}
\noindent
\begin{tabular}{||c||r|c|c||r|c|c||}
\hline
\multicolumn{7}{||c||}{$\lt^G = \lb^G$} \\
\hline
$m_t^{\rm pole}$ & \multicolumn{3}{c||}{$\alpha_s(m_Z) = .115$}
& \multicolumn{3}{c||}{$\alpha_s(m_Z) = .125$} \\
\cline{2-7}
GeV & $\lG$ & $\delta m_b\over m_b$ (max,min) & $\db$ (max,min) &
$\lG$ & $\delta m_b\over m_b$ (max,min) &
$\db$ (max,min)\\
\hline
150 & $.44$ & $(-.09 , -.19)$ & $(-.10 , -.22)$ & $.39$
& $(-.19 , -.30)$ & $(-.23 , -.37)$ \\
160 & $.56$ & $(-.05 , -.15)$ & $(-.05 , -.16)$ & $.48$
& $(-.14 , -.26)$ & $(-.16 , -.29)$ \\
170 & $.77$ & $(\mph\zer , -.12)$ & $(\mph\zer , -.11)$ &
$.63$ & $(-.11 , -.23)$ & $(-.10 , -.23)$ \\
180 & $1.4$ & $(\mph.05 , -.07)$ &  $(\mph.05 , -.05)$ & $.94$ &
$(-.06 , -.19)$ &
$(-.06 , -.17)$ \\
\hline
\hline
\multicolumn{7}{||c||}{$\lt^G = 2 \lb^G$} \\
\hline
$m_t^{\rm pole}$ & \multicolumn{3}{c||}{$\alpha_s(m_Z) = .115$}
& \multicolumn{3}{c||}{$\alpha_s(m_Z) = .125$} \\
\cline{2-7}
GeV & $\lG$ & $\delta m_b\over m_b$ (max,min) & $\db$ (max,min)&
$\lG$ & $\delta m_b\over m_b$ (max,min) &
$\db$ (max,min)\\
\hline
150 & $.40$ & $(-.16 , -.26)$ & $(-.32 , -.50)$ & $.36$
& $(-.27 , -.37)$ & $(-.60 , -.84)$ \\
160 & $.49$ & $(-.11 , -.21)$ & $(-.18 , -.34)$ & $.44$
& $(-.22 , -.32)$ & $(-.39 , -.58)$ \\
170 & $.64$ & $(-.07 , -.18)$ & $(-.09 , -.23)$ & $.55$
& $(-.17 , -.29)$ & $(-.25 , -.42)$ \\
180 & $.98$ & $(-.02 , -.13)$ & $(-.02 , -.14)$ & $.75$
& $(-.13 , -.25)$ & $(-.16 , -.30)$ \\
\hline
\end{tabular}
\caption[Table 1]{The consequences of Yukawa unfication using 2-loop
RG evolution and 1-loop logarithmic threshold corrections from the
preferred spectrum of Eq.~(\ref{eq:symspec}). For every value of the
top quark pole mass we list the unified Yukawa coupling at the GUT
scale, the minimim and maximum values of $\delta m_b/m_b$ needed to
bring the bottom quark prediction into agreement with experimental
data, and the corresponding values of the $\tan\beta$-independent
quantity $\db$.}
\end{table}

\begin{figure}[tb]
\centering
\leavevmode
\epsfysize=12cm \epsfbox[65 360 715 720]{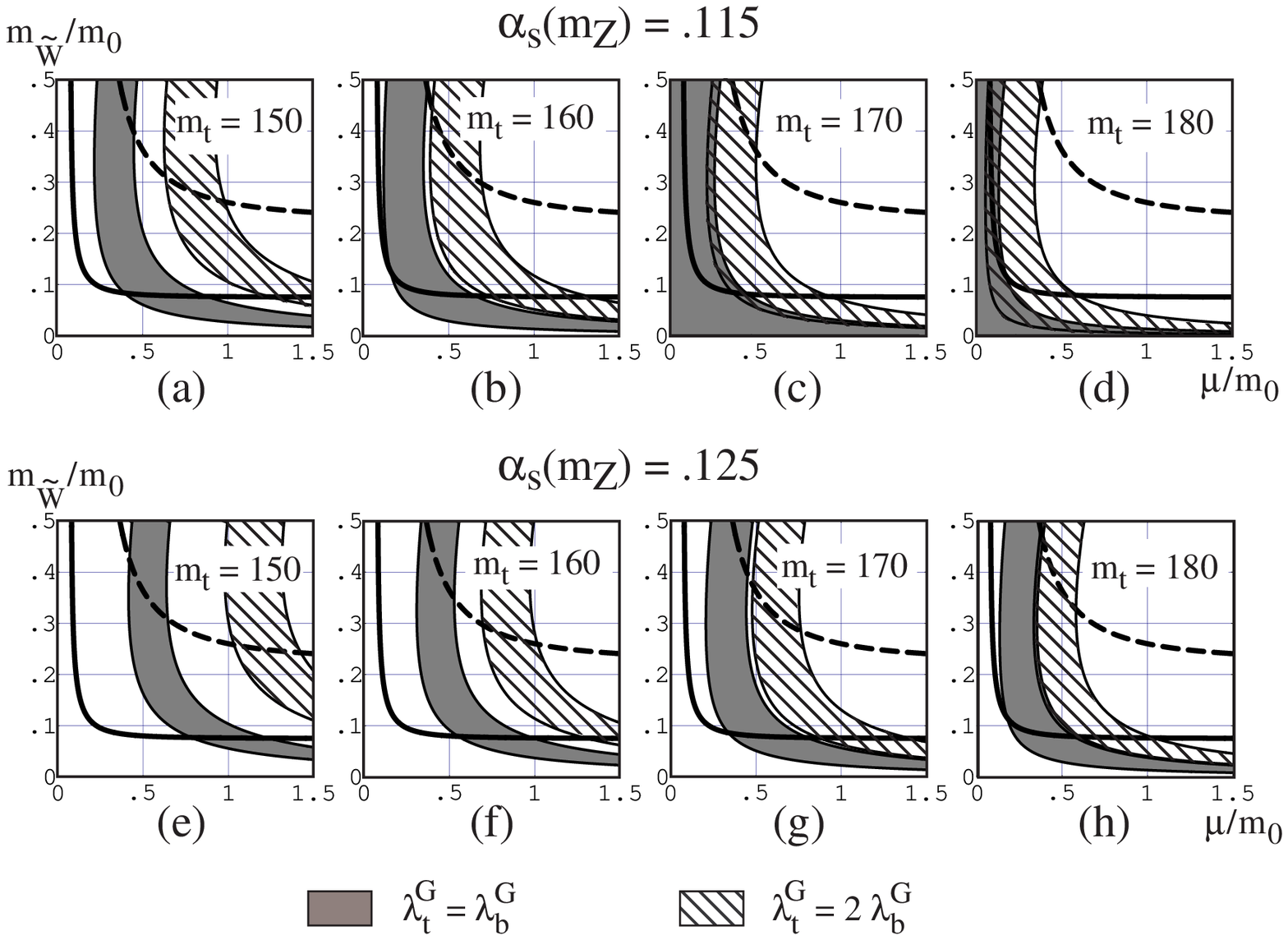}
\begin{quote}
{\small
Fig.~3. The values of $\mu$ and the wino mass, normalized to a
typical squark mass (specifically $m_0^2 = \half (\msb^2 +
\msq^2)$), which allow proper bottom-tau unification for large
$\tan\beta$. The shaded and hatched regions correspond to exact and
approximate Yukawa-unified scenarios, respectively, while the solid
and dashed lines represent the lower bounds imposed by LEP chargino
limits assuming $m_0$ is 600 (the preferred value) or 200 GeV,
respectively. }
\end{quote}
\end{figure}

Turning the argument around, for a given top mass we can calculate
the amount of correction $\delta m_b\over m_b$ needed to bring the
bottom mass
prediction into agreement with experiment. We can then remove the
$\tan\beta$ dependence, leaving only the spectrum-dependent quantity
$\db$. Table~1 displays these minimal and maximal allowed values of
$\delta m_b\over m_b$ and $\db$. (Actually, these bounds on $\delta
m_b\over m_b$ and $\db$
themselves depend on the spectrum due to threshold corrections, but
this dependence is a weak logarithmic one; typically, the
logarithmic variation in $m_t$ is at most a few GeV for the more
interesting higher values of $m_t$. To obtain precise predictions,
though, all thresholds corrections should be included using a
definite superspectrum.)  We learn that a positive $\db$ must be
quite small, while a negative $\db$ may be large enough in magnitude
to bring high predictions of $m_b$ back into agreement with
experiment. For example, when $\alpha_s(m_Z) = .115$ and $\lt^G =
\lb^G$,  superpartner spectra
for which $|\db| \roughly{>} 5\percent$ {\it allow} a light top
quark, whereas spectra for which $|\db| \roughly{>} 16\percent$ {\it
mandate} a light
top, where by light we mean $m_t \roughly{<} 160\,\GeV$. Conversely,
when $\lt^G = 2 \lb^G$, a superspectrum for which $|\db| \roughly{<}
15\percent$ mandates a heavy top and favors a small $\alpha_s(m_Z)$.
Fig.~3
translates this information into constraints on $\mwi$ (the mass of
the gaugino superpartner of the W) and $\mu$ at the electroweak
scale, normalized to a typical squark mass $m_0$ (taken specifically
to be $\meff$), for various values of the top quark mass,
$\alpha_s(m_Z)$ and $\lt^G/\lb^G$. Also shown are the lower bounds
on $\mwi/m_0$ and $\mu/m_0$ imposed by LEP chargino limits, for $m_0
= 600\,\GeV$ (solid lines) or $m_0 = 200\,\GeV$ (dashed lines).
Evidently the large, nonlogarithmic threshold correction
$\db$ is actually of great use: since, unlike the typical
logarithmic corrections, it is very sensitive to the superspectrum,
we can use experimental measurements of the top mass to learn about
the
hierarchies in the superspectrum. And from a model-building
viewpoint, we can exclude those theories in which the gaugino mass
and $\mu$ parameter do not fall into the allowed ranges shown in
Fig.~3.

To reemphasize the importance of the corrections to the bottom mass
(see also \cite{ref:mttb}, we have studied the consequences of
bottom-tau unification for arbitrary $\lt^G/\lb^G$, imposing only
that $\lG \equiv \lt^G < 2$ to ensure the validity of a perturbative
expansion. (A larger $\lt^G$ would result in a weak-scale top mass
very close to the fixed-point value, regardless of bottom-tau
unification. This fixed-point value depends on $\lb^G$ and
$\ltau^G$, as well as on $\alpha_s$.) In Fig.~4 we show the
predictions for $\tan\beta$ and $m_t^{\rm pole}$, fixing the values
of $\alpha_s(m_Z)$ and $m_b(m_b)$ and integrating over all possible
superpartner and pseudoscalar Higgs masses between $m_Z$ and 1 TeV.
The black regions correspond to $\db = 0$, while in the gray regions
$-25\percent < \db < 20\percent$. The effects of finite $\db$ are
striking.

\begin{figure}[htb]
\centering
\leavevmode
\epsfysize=12cm \epsfbox[70 180 550 625]{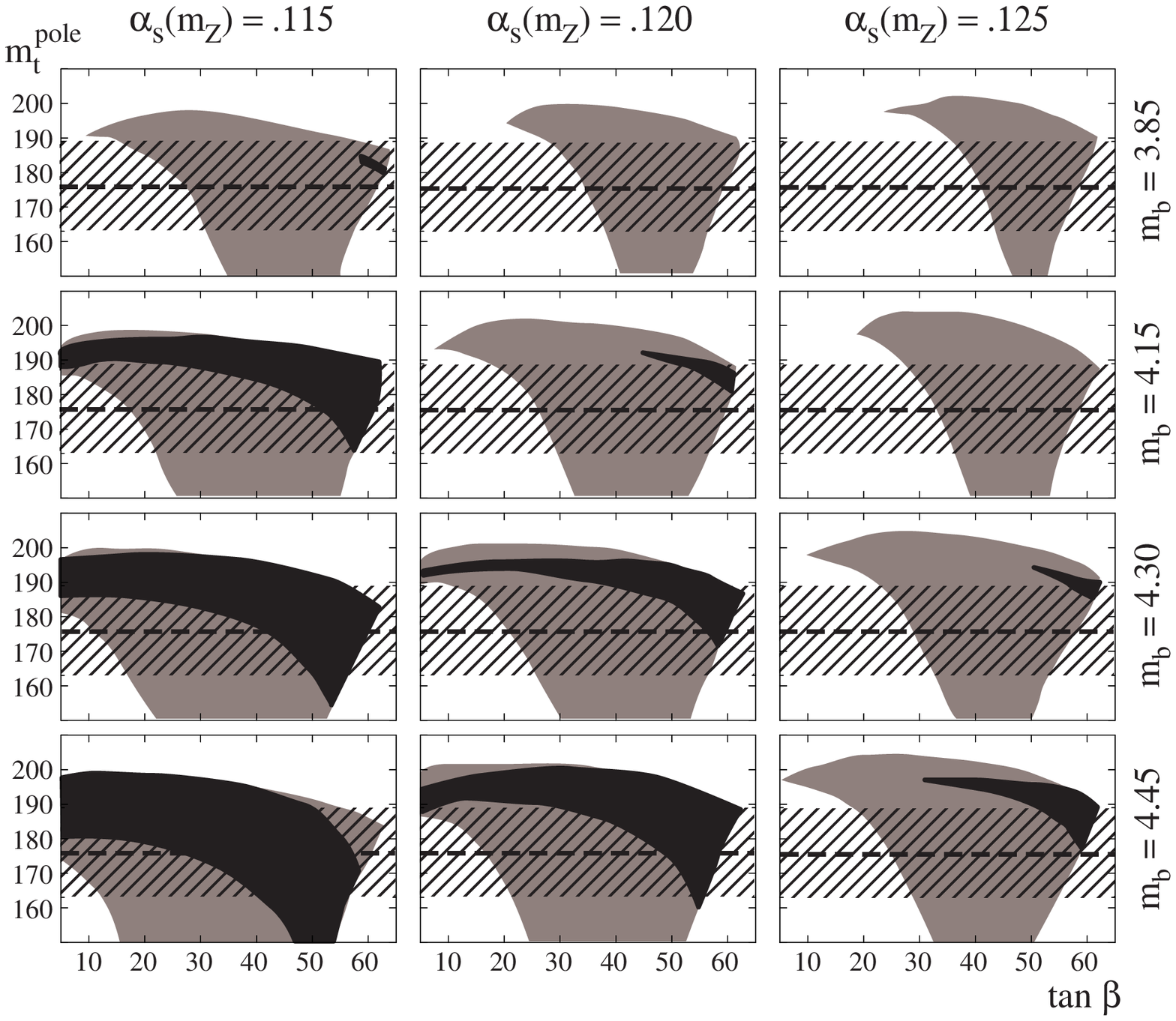}
\begin{quote}
{\small
Fig.~4. The ranges of top quark pole mass and $\tan\beta$ allowed by
bottom-tau unification at $\Mgut$, for different values of the
strong coupling and of the running bottom quark mass. We scan over
all possible superspectra between $m_Z$ and 1 TeV, allow arbitrary
$\lt^G/\lb^G$, and restrict $\lG \equiv \lt^G < 2$ to ensure
perturbativity. The effects of the 1-loop corrections $\db$ are
evident by comparing the black regions, for which $\db = 0$, with
the gray ones, in which $-25\percent < \db < 20\percent$. These
predictions are expected to be accurate to within a few GeV. The
dashed horizontal line and hatched band are the top mass central
value and an estimate of its uncertainty, respectively, as recently
announced by the CDF collaboration \cite{ref:cdf}.}
\end{quote}
\end{figure}

\section{$b\to s\gamma$}
\label{sec:bsg}
There is another immediate phenomenological implication of the large
1-loop corrections in the $\tan\beta\gg1$
framework \cite{ref:hrsI,ref:bsg}: the same diagrams, but with
a photon attached in all possible ways and with a
flavor-changing vertex as shown in Fig.~5, contribute to the bottom
quark radiative decay $b\to s\gamma$. These contributions, which
for small $\tan\beta$ are typically somewhat smaller than or
comparable to the 2-Higgs standard model contribution, are
parametrically enhanced by a factor of $\tan\beta\sim50$ in the
amplitude. But the CLEO bound \cite{ref:cleo} on the inclusive
branching ratio ${\rm BR}(b\to s \gamma) < 4.2\times 10^{-4}$ (at
$95\percent$ CL) is already
roughly saturated by the 2-Higgs standard model amplitude even if
the charged Higgs is rather heavy (and is in fact oversaturated with
a light charged Higgs). Hence the large $\tan\beta$ contribution
must not be too large.

\begin{figure}[tb]
\centering
\leavevmode
\epsfxsize=18cm \epsfbox[20 450 660 650]{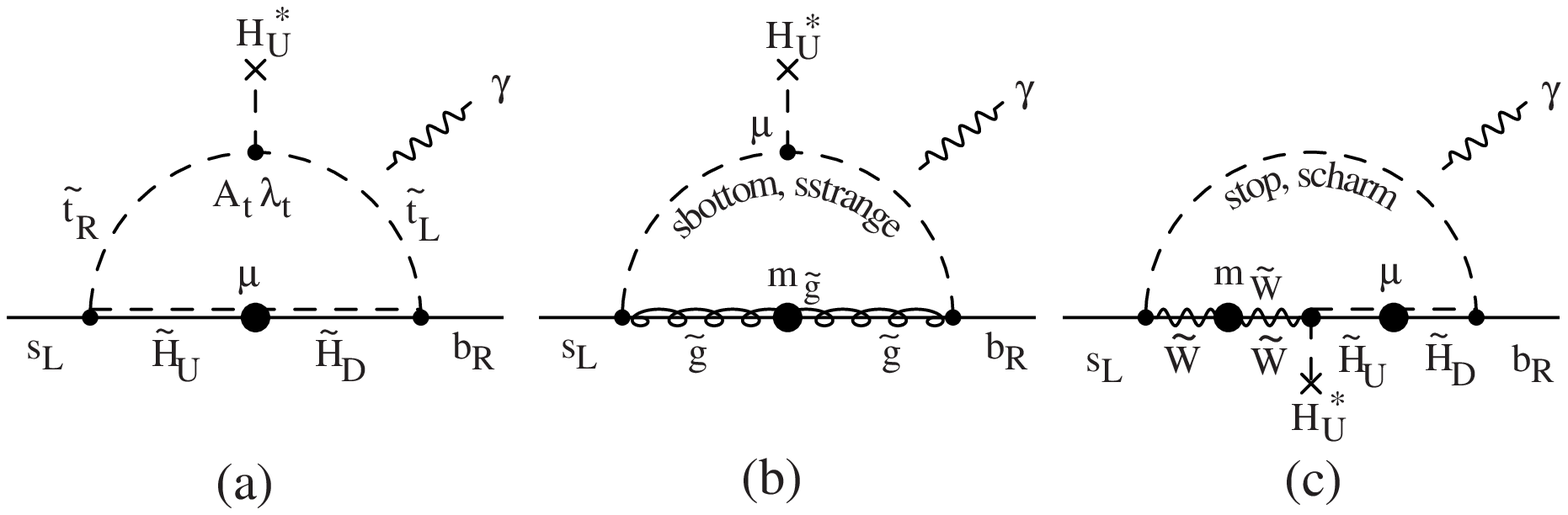}
\begin{quote}
{\small
Fig.~5. The primary chargino, gluino, and secondary chargino
diagrams which contribute to radiative bottom quark decay and are
enhanced by large $\tan\beta$. Note that while the amplitude in (a)
is completely predicted by third-generation parameters and $V_{cb}$,
the other two depend sensitively on intragenerational squark mixing
parameters and may be equally important.}
\end{quote}
\end{figure}

For two reasons \cite{ourbsg}, we will focus our attention on the
Yukawa-coupled (``primary'') chargino-exchange diagram of Fig.~5(a)
rather than the gluino-exchange diagram of Fig.~5(b) or the
gauge-coupled (``secondary'') chargino-exchange diagram of
Fig.~5(c).
First, throughout the relevant regions of parameter space, the
primary chargino-exchange amplitude exceeds or approximately equals
the other amplitudes in magnitude. Second, it is predominantly
determined by the third generation, namely by stop exchange, and
hence its magnitude is fixed by the Kobayashi-Maskawa quark-mixing
matrix element $V_{ts} \simeq V_{cb}$; in contrast, the other two
diagrams arise from squark mixing between the second and third
generation, and therefore depend on an independent mixing angle,
which we shall call $V_{23}$. The primary chargino diagram also
depends on the $A$ parameter which mixes the \mbox{SU(2)-doublet}
and
\mbox{-singlet} stop squarks, but as we noted above, the RG
equations typically fix $A$ at low energies almost entirely in terms
of the gaugino mass (independent of the GUT-scale value of $A$). So
the sign and magnitude of this diagram is completely calculable
\cite{ref:carbeta,ref:hrsII} in terms of the same parameters which
enter
$\db$; we have found that, when $\db < 0$, as must be the case for
any sizeable $|\db|$, the chargino exchange amplitude $\achar$
interferes {\it constructively} with the standard model and
charged-Higgs amplitudes ($\asm$ and $\ach$, respectively). Hence
there can be
no cancellations between these, and the constraint on $\achar$ is
more severe. On the other hand, the new angle $V_{23}$, which
determines the gluino and secondary chargino amplitudes $\agluino$
and $\achartwo$, gets not only a contribution $\sim V_{cb}$ through
the RG evolution, but also one from the flavor structure at the GUT
scale. Since we would like to remain as model-independent as
possible, we will not make any assumptions about this structure, and
thus $V_{23}$ will not be determined. By tuning the flavor
parameters and thereby $V_{23}$ one could cancel the various large
$\tan\beta$ contributions to $b\to s\gamma$ against each other and
avoid any bounds from this process. However, one should also be
careful about other phenomenological implications of this new
source of flavor violation. Since we are dealing with a grand
unified theory, there is also a leptonic counterpart of the new
mixing angle
$V_{23}$, and this gives relevant contributions to the rate for
$\tau\to \mu\gamma$. A more detailed analysis of the potential for
cancellations and of $\Gamma(\tau\to \mu\gamma)$ is presented
elsewhere \cite{ourbsg}. Our approach
here will be to take into account only the sizeable and calculable
primary chargino contribution: in those parameter ranges for which
it is small, the other amplitudes are typically also small and there
is no conflict with experiment, while in those ranges for which
$\asm + \ach + \achar$ exceeds the experimental bounds, the other
diagrams must be tuned to cancel against these amplitudes. Our goal
is to study when and to what degree such a cancellation is needed.

We will use the following expressions \cite{ref:bgbsg} to calculate
the limits on the MSSM parameters for large $\tan\beta$:
\be
{\rm BR}(b\to s \gamma) \simeq {\rm BR}(b\to c e
\bar{\nu})  {(6\alpha_{\rm em}/\pi) \left[\eta^{16/23} A_\gamma +
C\right]^2 \over I(m_c/m_b) \left[1-(2/3\pi) \alpha_s(m_b)
f_{\rm QCD}(m_c/m_b)\right]}
\ee
where ${\rm BR}(b\to c e \bar{\nu}) = 0.107$, $\eta =
\alpha_s(m_Z)/\alpha_s(m_b) = 0.520$ [using $\alpha_s(m_Z)=0.12$ as
a reference value], $C = -0.191$, $I(m_c/m_b) = 0.45$, $f_{\rm
QCD}(m_c/m_b) =
2.41$. The dimensionless amplitude is given by $A_\gamma =
\asm+\ach+\achar$ where
\bea
\asm &=& {3\over2} {m_t^2\over m_W^2}
f_\gamma^{(1)}\left({m_t^2\over
m_W^2}\right) \label{eq:bsgsm}\\
\ach &=& {1\over2} {m_t^2\over m_{H^-}^2}
f_\gamma^{(2)}\left({m_t^2\over m_{H^-}^2}\right)
\label{eq:bsgtwoh}\\
\achar &=& \tan\beta\, {m_t^2 A_t \mu\over \mst^4}\, \sum_{j=1}^2
U_{j2} V_{j2}^* {\mcharj\over\mu} F\left({\mst^2\over
\mcharj^2}\right)
\label{eq:bsgmssm}
\eea
and the various functions are
\bea
f_\gamma^{(1)}(x) &=& {7-5x-8x^2\over 36(x-1)^3} + {x(3x-2)\over
6(x-1)^4}\ln x \\
f_\gamma^{(2)}(x) &=& {3-5x\over 6(x-1)^2} + {3x-2\over 3
(x-1)^3}\ln
x \\
F(x) &=& {x\over 6(1-x)^3} \left[5-12x+7x^2+2x(2-3x)\ln x\right]\,.
\label{eq:bsgfns}
\eea
In Eq.~(\ref{eq:bsgmssm}) we assume that the two stop mass
eigenvalues are roughly degenerate. (This is in particular
a good approximation for the interesting situation in which the stop
is heavier then the top and the diagonal soft masses are almost
degenerate.) Notice that the crucial off-diagonal $\tilde t_R\tilde
t_L$ mixing has been factored out in $\achar$.
We have kept the exact dependence on the chargino mass matrix:
\be
U^* \left(\begin{array}{cc}
\mwi & m_W \sqrt{2} \sin\beta \\
m_W \sqrt{2} \cos\beta & \mu \end{array}\right) V^{-1}
= \left(\begin{array}{cc}
\mcharone & 0 \\
0 & \mchartwo \end{array}\right)\,.
\label{eq:charmat}
\ee

We can now determine how large $\achar$ can be without requiring
some cancellation to avoid conflict with experiment. We find that if
$\achar = 0$ then the 2-Higgs standard model exactly saturates the
experimental bound for $\mch = 1\,\TeV$ and $m_t = 170\,\GeV$; for
those values $\ach
\simeq .15 \asm$. Assuming some theoretical uncertainty allows the
charged Higgs to be significantly lighter: a $30\percent$
uncertainty in
the branching ratio would allow a charged Higgs of $350\,\GeV$
($\ach \simeq .5 \asm$), while with a $50\percent$ uncertainty the
charged
Higgs can be as low as $250\,\GeV$ ($\ach \simeq .75 \asm$). For the
ranges of top quark mass we are considering, the $m_t$ dependence is
much smaller than such theoretical uncertainties. Now, if we add
constructively a supersymmetric contribution equal to $50\percent$
of the
standard-model amplitude, the branching ratio is $30\percent$ above
the
experimental bound without including any charged-Higgs amplitude, or
$50\percent$ above the experimental bound if we include a
$25\percent$
charged-Higgs contribution. (These values correspond to taking the
soft SUSY-breaking scalar masses to be roughly degenerate at $\sim
700\,\GeV$ while the $\mu$ and wino mass are at $\sim m_Z$, which we
argue is the favored scenario.) The charged Higgs mass cannot be
made too large without fine-tuning the Z mass, so a $25\percent$
charged-Higgs contribution is a reasonable lower bound. Thus, to be
conservative, we conclude that {\it there must be some cancellations
whenever} $\achar > .5 \asm$, either from the other diagrams
[Figs.~5(b) and 5(c)] due to tuning of the flavor physics or from
cancellations in the $A$ parameter due to
tuning of the GUT values of $A$ and the gaugino mass. We define a
measure of
the need for cancellations as\footnote{Since $\achar = .5 \asm$
requires no more tuning than $\achar < .5 \asm$, we define $\epsbsg$
so it saturates when $\achar = .5 \asm$.  For similar reasons we
will define $\epsb \equiv \min(B/\mwi,1)$ and $\epsb \equiv
\min(m_Z^2/m_A^2,1)$.}:
\be
\epsbsg \equiv
\min\left(\left|{.5 \asm\over\achar}\right|,1\right)\,.
\label{eq:epsbsg}
\ee
For example, if the chargino amplitude is 10 times greater than the
saturating amplitude $.5 \asm$, then some other contribution must be
adjusted with a precision of $0.1$ to cancel against it. In fact, we
will never need the exact definition of $\epsbsg$;
all that we will require of it is that it be small
whenever the chargino amplitude is considerably too big.

\section{Electroweak Symmetry Breaking}
\label{sec:electro}
Thus far, the existence of the electroweak-breaking Higgs VEVs $\vU$
and $\vD$ has been simply assumed. We must study the generation of
those VEVs in order to understand how compatible is the idea of
Yukawa unification with radiative electroweak breaking, how the
SO(10) or SU(5) symmetry constraints on the soft SUSY-breaking mass
parameters affect this breaking, how the top-bottom mass hierarchy
is obtained, and how natural or unnatural is such a scenario. This
study will reveal favored ranges for the Yukawa couplings and the
soft SUSY-breaking masses, and highlight the central role of the PQ
and R symmetries in discussing electroweak breaking for
Yukawa-unified models.

Electroweak symmetry breaking is governed by the scalar potential
for the neutral components of the two Higgs fields $H_U$ and $H_D$,
which at tree level is of the form
\bea
V_0 &=& m_U^2 |H_U|^2 + m_D^2 |H_D|^2 + \mu B \left(H_U H_D + {\rm
h.c.}\right) + \nonumber\\
&{}&
{{g_1^2+g_2^2}\over 8}
\left (|H_U|^2- |H_D|^2\right )^2
\label{eq:v0}
\eea
where $g_2$ and $g_1$ are respectively the $SU(2)$ and hypercharge
gauge couplings. The parameters $m_{U,D}^2 = \mu_{U,D}^2 + \mu^2$
contain the soft-breaking Higgs masses as well as the $\mu$
parameter from the superpotential, while $B$ is the soft-breaking
bilinear mass parameter. The conditions for proper electroweak
breaking are well-known:
\be
m_U^2 + m_D^2 \geq 2 |\mu B|
\label{eq:bndbel}
\ee
ensures that the potential is bounded from below, and
\be
m_U^2 m_D^2 < \mu^2 B^2
\label{eq:ewsb}
\ee
guarantees the existence of a minimum away from the origin and so
breaks the symmetry. In practice, since $|\mu B|$ will always be
much less than or at most comparable to $|m_U^2|$ and $|m_D^2|$, we
can
reduce these requirements to $m_A^2 = m_U^2 + m_D^2 > 0$ (using the
expression for the pseudoscalar Higgs mass) and $m_U^2 < 0$ (noting
that large $\tan\beta$ means that the up-type Higgs gets the large
VEV).

It is useful to interpret the two above conditions for proper
symmetry breaking in terms of the usual custodial symmetry
exchanging up- and down-type
particles. In practice we need $m_U^2<0<m_D^2$, which represents
a substantial violation of this symmetry.
In the usual scenario, with the initial condition $\lt \gg \lbtau$,
this custodial violation is provided by the Yukawa couplings
themselves.  The large top Yukawa coupling drives negative the
soft-breaking mass parameter of the ``up-type'' Higgs $H_U$ to which
it couples,  while the other Higgs is largely unaffected. Hence,
even with custodially symmetric soft masses at $M_G$, it is very
natural to obtain \cite{ref:radbreak,ref:apw}, at the electroweak
scale, $m_U^2 < 0$ while the other Higgs mass is sufficiently
positive to ensure $m_U^2 + m_D^2 > 0$. On the other hand, with
unified or approximately-unified Yukawa couplings ($\lt^G/\lb^G \sim
1$), it is hard to see why the two Higgs mass parameters should run
differently, so the natural explanation for $m_U^2 < 0 < m_U^2 +
m_D^2$ is lost. In fact, with the boundary condition $\lt^G=\lb^G$,
the only sources of custodial breaking in the couplings are the
hypercharges and the
absence of a right-handed neutrino (but presence of a right-handed
tau). These furnish only a tiny splitting even when integrating from
the
GUT to the electroweak scales. If such a splitting between $m_U^2$
and $m_D^2$ is to be $\sim m_Z^2$ then the soft-breaking masses
themselves must be considerably bigger than $m_Z^2$.
To break the symmetry more naturally, custodial breaking must be
enhanced.
To this end, either $m_U^2$ can be split by various means from
$m_D^2$ already at the GUT scale, or one may relax the requirement
$\lt^G = \lb^G$, which in any case is bound to be modified---either
by a little, due to GUT-scale threshold corrections, or by a lot, in
the case of mixed-SO(10)
or SU(5) models.

Furthermore, in the usual scenario, if all mass parameters in the
scalar
potential are comparable then so are the VEVs of the two Higgs
doublets; but comparable Higgs VEVs are perfectly compatible in the
usual scenario with the experimental hierarchy $m_t\gg m_b$ since
that is furnished by the assumed hierarchy of Yukawa couplings
$\lt\gg\lb$. By contrast, in a unified scenario it is the ratio of
VEVs which must be large. Upon minimizing the
scalar potential $V_0$, one easily obtains (for large $\tan\beta$)
\be
{1\over\tan\beta} \equiv {v_D\over v_U} = -{\mu B\over m_u^2+m_D^2}
= -{\mu B\over
m_A^2}\,,
\label{eq:tanbet}
\ee
as well as $m_Z^2 = -2 m_U^2$ (which sets the scale). A large VEV
hierarchy requires a small coupling between the two Higgs doublets,
namely $\mu B \ll m_U^2+m_D^2 = m_A^2$, so that an expectation value
in one is only weakly fed into the other. But this small Higgs
coupling, as we now discuss,
implies \cite{ref:nelran,ref:hrsII} a necessary degree
of cancellation between some parameters at the GUT scale.

\section{Generating a Hierarchy}
\label{sec:hier}
{}From Eq.~(\ref{eq:tanbet}) it is clear that to generate the
hierarchy of VEVs ($\tan\beta\sim 50$) necessary for the top-bottom
(and top-tau) mass ratio, we need a small $\mu B$ or a large
$m_A^2$.
This is difficult: on one hand, $m_A^2 = m_U^2 + m_D^2$ and
$m_Z^2 = -2 m_U^2$ run quite similarly and are tightly coupled
through the RG evolution, so it is difficult to make
$m_A^2$ much larger than the electroweak scale; on the other hand,
$\mu$ cannot be much below the Z mass since otherwise a light
chargino (or neutralino) would have been detected at LEP, and the RG
equations imply that $B$ is naturally at least as large as
$\mwi$---which again cannot be much below the Z mass without
producing a light chargino or neutralino. To make these arguments
concrete, we can rewrite Eq.~(\ref{eq:tanbet}) as
\be
{1\over50} \sim -{\mu\,B\over m_A^2} =
-\underbrace{\mu\,\mwi\over m_Z^2}_{\displaystyle \roughly{>} 0.9}
\cdot
\underbrace{B\over \mwi}_{\displaystyle \equiv \epsb}
\cdot
\underbrace{m_Z^2\over m_A^2}_{\displaystyle \equiv \epsz}\,.
\label{eq:tanbetII}
\ee
The most natural scenario would display approximate PQ or R
symmetries, with either $\mu\ll m_Z$ or $\mwi\sim B\sim A \ll m_Z$
or both, while all other masses would be around $m_Z$.
When $\tan\beta\gg1$, such small $\mu$ or $\mwi$ result in a
chargino much lighter than $m_Z$. Alas, this scenario is now
experimentally ruled out: specifically, LEP \cite{ref:lep} places a
lower bound of roughly $\half
\,m_Z$ on the lightest chargino mass. A lower bound of $\half \,m_Z$
on the
lightest eigenvalue of the chargino mass matrix in
Eq.~(\ref{eq:charmat}) translates into a bound on the Lagrangian
parameters of
\be
\left[\mwi^2-\left(\half \,m_Z\right)^2\right]
\left[\mu^2-\left(\half \,m_Z\right)^2\right] \roughly{>}
2 \,m_W^2 \left(\half \,m_Z\right)^2\,.
\label{eq:leplimit}
\ee
Subject to this bound, the prefactor $\mu \mwi/m_Z^2$ in
Eq.~(\ref{eq:tanbetII}) is minimized when $\mu = \mwi$, yielding
$\mu \mwi/m_Z^2 \roughly{>}1/4 + 2^{-1/2} m_W/m_Z \simeq .88$.
Therefore, the parameters at the GUT scale must be adjusted so that,
at the electroweak scale, either $B$ is much lighter than its
natural
minimum value $\sim\mwi$, or $m_Z^2$ is much less than its natural
value $\sim m_A^2$. (Note that in the first case $\tan\beta$ is the
quantity that is tuned to be small, while in the second case $m_Z^2$
is tuned to be small.) We quantify these tunings by $\epsb$ and
$\epsz$, respectively [see the footnote for Eq.~(\ref{eq:epsbsg})].
Thus the bounds from LEP imply that some GUT parameters must be
adjusted to cancel with an accuracy of at least
\be
\epsb \epsz\,\roughly{<}\,{1\over\tan\beta} \sim {1\over50}\,;
\quad\mbox{saturated when}\, \mu\sim\mwi\sim m_Z
\label{eq:mintune}
\ee
and the least amount of tuning is required when $\mu$ and $\mwi$ are
roughly as light as they can be, namely both near the Z mass.

\begin{table}
\centering
\renewcommand\arraystretch{1.1}
\begin{tabular}{|r|rrr|rrr|}
\hline
&$\mu\sim\fsev m_0$ & & & $\mu\sim \psev m_0$ & & \\
&$\mwi\sim \psev m_0$ & & & $\mwi\sim \psev m_0$ & & \\
&$m_Z\sim \fsev m_0$ & $\Longrightarrow$ & $\epsz\sim {1\over50}$ &
{\bf either} \phantom{$\sim m_0$} & & \\
&$m_i\sim \psev m_0$ & $\Longrightarrow$ & $\epsbsg\sim \fone$ &
  $m_Z \sim \fsev m_0$ & $\Longrightarrow$ & $\epsz\sim {1\over50}$
\\
no R & & & $\epsb \sim \fsev$ &
  $m_i \sim \psev m_0$ & $\Longrightarrow$ & $\epsbsg\sim \fone$ \\
&& & & & & $\epsb \sim {1\over50}$ \\
&& & & {\bf or}\phantom{$m_i\sim {1\over7}m_0$} & & \\
& &  &  &
  $m_Z \sim  \psev m_0$ & $\Longrightarrow$ & $\epsz\sim \fone$ \\
& &  &  &
  $\phantom{X}m_i \sim \psev m_0$ & $\Longrightarrow$ & $\epsbsg\sim
{1\over50}$ \\
&& &  &  & & $\epsb \sim {1\over50}$ \\
\hline
&$\mu\sim\fsev m_0$ & & & $\mu\sim \psev m_0$ & & \\
&$\mwi\sim \fsev m_0$ & & & $\mwi\sim \fsev m_0$ & & \\
R&$m_Z\sim \fsev m_0$ & $\Longrightarrow$ & $\epsz\sim {1\over50}$ &
  $m_Z\sim \fsev m_0$ & $\Longrightarrow$ & $\epsz\sim {1\over50}$
\\
&$m_i\sim \psev m_0$ & $\Longrightarrow$ & $\epsbsg\sim \fone$ &
  $m_i\sim \psev m_0$ & $\Longrightarrow$ & $\epsbsg\sim \fone$ \\
&& &$\epsb \sim \fone$ &  & & $\epsb \sim \fsev$ \\
\hline
&& PQ & & &no PQ & \\
\hline
\end{tabular}
\caption[Table 2]{Typical scenarios, and consequent fine-tunings,
for generating the Higgs VEV hierarchy $v_D \sim 1/50 v_U$ with or
without the PQ and R symmetries.}
\end{table}

Eq.~(\ref{eq:tanbetII}) by itself does not distinguish whether $B$
or $m_Z^2$ should be tuned small; that is decided by $\epsbsg$,
which requires considerable tuning of the flavor sector or of the
$A$
parameter at the GUT scale if the superpartners and the pseudoscalar
Higgs are near $m_Z$. Table~2 sketches four possible scenarios,
distinguished by whether PQ and R are good symmetries (at low
energies---$m_i$ are the
sfermion masses at the electroweak scale). Note that $\epsb\epsz$
favor the PQ- and R-symmetric case {\it or} the asymmetric case
having all masses near $m_Z$, but that $\epsbsg$ favors the former
over the latter. Thus the most natural scenario as measured by these
three criteria is the maximally symmetric one: a Lagrangian
(\cite{ref:hrsII}; see also the first reference in
\cite{ref:pocolsp}) which
is maximally PQ- and R-symmetric while obeying the LEP bounds and
keeping the superpartners as light as possible, that is,
\be
\left\{\begin{array}{l}
\vphantom{1\over2}m_A \sim m_0 \sim \sqrt{\tan\beta}\,m_Z
\,(\sim 600\,\GeV) \\
\vphantom{1\over2}\mu\sim A\sim B\sim\mwi\sim\third\mgl\sim m_Z
\end{array}\right.
\label{eq:symspec}
\ee
where by $m_0$ we mean the typical mass of the squarks and sleptons
evaluated at the electroweak scale. (To reiterate: we chose
$\mu\sim\mwi\sim{1\over7}m_0$ because this is the most natural case
allowed by LEP---if $\mu$ or $\mwi$ were smaller, $m_Z$ would
require further tuning to make it sufficiently light, while if they
were much larger we would lose the advantages of the PQ and R
symmetries and the tuning would again be exacerbated.) This spectrum
implies a small correction to the bottom mass, $\db \sim 5\percent$,
and
hence (from Table~1) a heavy top and preferably a somewhat low value
of $\alpha_s(m_Z)$.

\section{Correct Symmetry Breaking}
\label{sec:correct}
\subsection{The General Problem}
\label{sec:general}
We return now to the question of how the electroweak symmetry may be
correctly broken, while preserving the $\CC$
gauge symmetries. In principle, what needs to be done is to
study the effective potential $\Ve$ for field values
$\ms\leq\phi\leq M_G$, where $\ms$ denotes collectively the soft
SUSY-breaking masses and $\phi$ is the set of scalar fields in
the MSSM. In practice, since we are dealing with a perturbative
theory, we need only consider the RG-improved
tree level potential $\Vzero$ renormalized at scales $\Lambda$
between
$\ms$ and $\Mgut$. First, we must make sure that the potential is
stable for $\Lambda\gg \ms$, i.e. that $\Vzero$ is bounded from
below
at high scales. If this were not the case, the scale of gauge
symmetry breaking would be $\sim\Lambda\gg
\ms$, which is phenomenologically unacceptable, and possibly even
$\CC$ would be broken. Second, we need to guarantee that at
scales $\Lambda\sim \ms$ the minimum of $\Vzero$
is such that $H_u$ and $H_d$, and no other fields, acquire
nonvanishing expectation values. This amounts to imposing some
positivity constraints on the sfermion mass-squared parameters,
which we discuss below. Finally, since we are interested in large
rather than small $\tan\beta$, the
instability should arise in the $H_u$ direction, while the small
$H_d$ VEV is generated through the mixing mass parameter $B\mu$.

An essentially technical comment is in order here. We will be mainly
concerned with the parameters related to the third family
and to the Higgs sector.  This is because, in the limit in which
flavor mixings and the Yukawa couplings of the two light families
are neglected, the SUSY-breaking masses of these families have a
numerically small impact on the RG evolution of the parameters of
the third generation and the Higgs sector. The only effect
\cite{ref:lleyda} is via the hypercharge D-term $S$ (see Appendix
A), which is small because of the small hypercharge gauge coupling.
Moreover the effects of $S$ are completely determined by its GUT
scale value $S_G$, since it renormalizes multiplicatively. Indeed
with SU(5) or SO(10) boundary conditions on scalar masses, $S_G$
itself is completely specified by the soft masses of $H_u$ and $H_d$
(since these are the only light incomplete matter SU(5) multiplets).

The relevant parameters then consist of
the seven soft-breaking scalar masses $\mu_U^2 = m_U^2 -
\mu^2$, $\mu_D^2 = m_D^2 - \mu^2$, $\mst^2$, $\msb^2$, $\msq^2$,
$\mstau^2$ and $\msl^2$, the three trilinear soft-breaking
parameters $A_{t,b,\tau}$, the single bilinear soft-breaking
parameter $B$, the $\mu$ term in the superpotential, and the three
gaugino masses. Their 1-loop RG equations are given for reference in
Appendix A.

Let us now discuss in more detail the constraints which these
parameters need to satisfy. We begin with those which must be
satisfied at scales $\Lambda\gg \ms$. It is well known that the
MSSM,
like any generic SUSY model, possesses a host of ``approximately
flat'' directions in the space of scalar fields $\phi$. By
``approximately
flat'' we mean that the potential, at the renormalizable level,
is only quadratic in those directions. In general, though,
irrelevant operators suppressed by inverse powers of a large mass
$M$ such as
$\Mgut$ or $M_{\rm Planck}$ can give an additional stabilizing
contribution \cite{ref:cfgz}. To be conservative, we will always
assume the superpotential contains an appropriate operator of the
form $\phi^4/M$. Then along any such direction parametrized by a
field $\phi_\alpha$, the potential is essentially (see below) given
by $m_\alpha^2|\phi_\alpha|^2 +|\phi_\alpha|^6/M^2$,  where
$m_\alpha^2$ is equal to a sum of squared masses. Regardless of the
sign of $m_\alpha^2$, there is no minimum for $\phi_\alpha
\roughly{>} \sqrt{m_\alpha M}\equiv \lhigh$ ($\sim 10^9\,\GeV$ for
$M = \Mgut$), so parameters normalized at scales $\Lambda > \lhigh$
can never yield an unwanted minimum. For $\phi_\alpha < \lhigh$, the
potential in dominated by the quadratic piece, though there may be a
scale $\llow$ below which a linear term may again stabilize the
potential. In the absence of a linear term, the lowest scale of
interest is $\llow \sim \ms$, at which the superpartners are
integrated out. If $m_\alpha^2 >0$ were to become negative at a
critical scale $\llow \ll\Lambda_c \ll  \lhigh$, dimensional
transmutation \cite{ref:CW} would take place: the VEV of
$\phi_\alpha$  would be fixed by the one loop correction to the
effective potential $\Vone$ to be of order $\Lambda_c$ (times a
coupling constant). (Notice that, in the absence of the irrelevant
operator, if $m_\alpha^2$ were to be negative already at the GUT
scale, then we would clearly be expanding around the wrong vacuum in
the GUT theory.) In order to get acceptable low-energy physics we
have then to impose $m_\alpha^2(\Lambda)\geq 0$, for all
approximately-flat directions $\alpha$ and for all scales $\Lambda$
between the $\lhigh$ and $\llow$ relevant to that $\phi_\alpha$.

When we restrict our attention to the fields of the third family and
the Higgs sector, there are only two such flat directions:
\begin{enumerate}
\item $\vev{H_u}=\vev{H_d}=\phi_1$, with all other fields at zero;
and
\item $\vev{H_u}=\phi_2$, $\vev{\widetilde{L}}= (\phi_2^2 +\phi_2
\mu/\lb)^{1/2}$, $\vev{\widetilde{Q}} = \vev{\widetilde{b_c}} =
(\phi_2\mu/\lambda_b)^{1/2}$, with all other fields at zero
\cite{ref:komatsu}.
\end{enumerate}
The color and isospin orientations are determined by imposing
vanishing D and F terms. Along $\phi_1$ the potential is purely
quadratic, $V_0(\phi_1) = m_1|\phi_1|^2$ with $m_1^2=
m_U^2+m_D^2-2|B\mu|$. The stability constraint has already been
given in Eq.~(\ref{eq:bndbel}). This constraint should be satisfied
between $\lhigh \sim 10^9\,\GeV$ (to be conservative) and $\llow\sim
\ms$. Along direction $\phi_2$ there is also a linear term:
\bea
V_0(\phi_2)&=&(\mu_U^2+\msl^2)|\phi_2|^2+(\msl^2+\msq^2+\msb^2)
\left|{\mu\phi_2\over \lambda_b}\right| \nonumber\\
&\equiv& m_2^2 |\phi_2|^2 + m_3^2 \left|{\mu\phi_2\over
\lambda_b}\right|\,.
\label{eq:flat}
\eea
For this flat direction, the dominant stabilizing term at high
scales is the left-handed neutrino mass operator $(H_u L)^2/M_N$,
where $M_N$ is the right-handed neutrino mass. Indeed the effect of
this operator can be important down to $\lhigh\sim 10^7\,\GeV$,
since $M_N$ could be as low as $10^{12}-10^{13}\,\GeV$. At low
scales, the linear term will
stabilize the potential (provided $m_3^2>0$, which will always be
the case). Therefore, as we show in Appendix B (see also
Ref.~\cite{ref:grz}), the $\phi_2$ flat direction can only pose a
problem at scales above $\llow \sim (2\pi/\alpha) \mu/\lb \sim
10^4-10^5\,\GeV$. So we need to impose $m_2^2(\Lambda)>0$ at least
for all $\Lambda$ between $\lhigh\sim 10^7\,\GeV$ and $\llow\sim
10^5\,\GeV$.

A general scan of the values of $m_{1,2}^2(\Lambda)$ at all
intermediate scales would be numerically arduous.
Fortunately, with only minor assumptions on
the initial parameters,  $m_1^2(\Lambda)$ and
$m_2^2(\Lambda)$ decrease essentially monotonically with $\Lambda$.
Imposing positivity just at low $\Lambda$ then guarantees the
absence of unwanted vacua at all scales.
Consider for instance the PQ- and R-symmetric limit of the RG
equations in Appendix A. Monotonicity of $m_{1,2}^2$
is clearly satisfied when $X_{t,b,\tau}$ are positive throughout the
running. In turn this condition is satisfied when the $X_i$ start
out
positive and of comparable magnitudes (check for instance the
entries in the matrix $\cal H$ in Appendix A, whose behavior in
monotonic in $\Lambda$). In most interesting cases, the necessary
positivity of
masses at low energy will imply positive $X_i$ at the GUT scale [for
instance in minimal SO(10)]. Introducing a finite $\mu$ does not
alter the conclusions, as long as R symmetry is preserved.
For small $\mu$ and large gaugino masses, the situation is also
unchanged: in the first stage of the running, their
contribution to $m_{1,2}^2$ is positive, but very small; however it
soon becomes negative and its absolute value increases monotonically
when $\Lambda$ is lowered, so again checking positivity of
$m_{1,2}^2$ at low scales suffices. Finally, when both PQ and R are
broken, the above discussion applies straightforwardly to $m_2^2$,
but not to $m_1^2$ due to the additional inhomogeneous piece $B\mu$.
For this situation we have explicitly verified monotonicity for a
wide range of initial parameters. We thus conclude that, quite
generally, the imposition of the constraints at a low scale is
sufficient to ensure their validity throughout the RG evolution. Our
analysis is thereby considerably simplified: we need only impose
$m_1^2(\Lambda = \ms\sim m_Z) \simeq m_A^2 > 0$ and
\be
m_2^2(\Lambda \sim 10^5\,\GeV)>0
\label{eq:mtwoconstr}
\ee
to avoid an instability in the $\phi_1$ and $\phi_2$ directions.

We note in passing that the constraints from flat directions
involving also the fields in the first two families are not a
problem for us. This is because their $m_\alpha^2$
always involve the soft masses for these fields, which are for us
essentially arbitrary and can thus be taken large enough to
stabilize the flat directions.

We next turn to the constraints on the potential at the
electroweak scale. In what follows we will use just the tree
level potential $\Vzero$. This approximation, discussed in detail in
Appendix C, is motivated by the fact that we are not  concerned with
precise predictions for the various masses, but rather with the
characteristic
hierarchies in the spectrum, with the rough bounds on the various
parameters and with comparing the naturalness of various parameter
choices.

First, the scalar configuration with $\vev{H_{u,d}}\not = 0$, and
all the other fields at zero, should be a local minimum. This will
be the case if we impose that the MSSM parameters, evaluated at the
electroweak scale, satisfy:
\be
m_i^2(\Lambda = m_Z) > 0,\quad i = Z,A,
\tilde{t},\widetilde{b},\widetilde{Q},
\widetilde{\tau},\widetilde{L}
\label{eq:constrs}
\ee
where as before $m_Z^2 = -2 m_U^2$ and $m_A^2 = m_U^2 + m_D^2$.
(Notice that in the above equation we have neglected any
contribution to the sfermion masses coming from the Higgs VEV. We
have also ignored the phenomenological bounds on these masses, which
yield somewhat stronger constraints: $m_i^2 \roughly {>} m_Z^2$.
However we stress once more that, for the purpose of studying the
spectrum hierarchies and the naturalness of different scenarios, the
above contraints are sufficient. Indeed, in most situations we will
end up with sfermion masses well above $m_Z$.)

A second class of constraints is needed to avoid having other minima
with electroweak- or color-breaking VEVs of order $\ms$. Such minima
can arise, even for positive sfermion masses, from the destabilizing
effect of the trilinear terms in the scalar potential. These are
given by the soft
$A$-terms and also by the trilinear terms
proportional to $\mu$ in the supersymmetric part of the scalar
potential. In what follows we will mainly be concerned with
necessary
constraints, and will not enter into a comprehensive discussion of
the
sufficient ones. Let us consider the effect of $A$ terms first.
These
were discussed in Ref. \cite{ref:deren} where a
necessary condition to avoid unwanted minima was given:
$m_a^2+m_b^2+m_c^2\geq |A|^2/3$ (where $a,b,c$ represent
any three fields having a Yukawa coupling $\lambda$, and $A$ is the
corresponding soft-SUSY-breaking trilinear coupling). When this
condition is not satisfied, there is a color- and charge-breaking
minimum with
energy density $\sim -\ms^4/\lambda^2$. In the case of a light
fermion this vacuum is considerably deeper than the usual Higgs one.
For the top quark,
$\lambda$ is sufficiently large, and $\ms$ is often assumed to be
sufficiently small, that this extraneous vacuum is not deeper
(and typically shallower) than the ordinary vacuum. This is why the
$A$-term requirement is usually not applied to the stop. However,
when there is a hierarchy $\ms\gg m_Z$, the extraneous minimum, when
it exists, is indeed
parametrically  deeper than the ${\cal O}(g^2 v^4)$ Higgs minimum,
so
the necessary condition given above must be applied also to the soft
parameters of the third generation. Similar arguments can be made
for the trilinear $\mu$ terms, though we are not aware of previous
discussions
in the literature. Now the triplets of fields in danger of
developing expectation values are those entering the various $\mu$
couplings: $(H_u,\widetilde L, \widetilde \tau)$, $(H_u,\widetilde
Q,\widetilde b)$ and $(H_d,\widetilde Q,\widetilde t)$ (where the
last member of each triplet is the SU(2)-singlet scalar field). For
instance, along the direction $\vev{H_u}=\vev{\widetilde
L}=\vev{\widetilde \tau}=\phi$ the potential is given by
\be
V=(m_U^2+\msl^2+\mstau^2)|\phi|^2 -2\mu\ltau
|\phi|^3+\hat\ltau^2|\phi|^4
\label{eq:trimu}
\ee
where $\hat\ltau^2=\ltau^2+(g_1^2+g_2^2)/2$.  To avoid a minimum
away from the origin in the above potential, $\mu$ must not be too
big\footnote{Indeed one can find more general constraints by
considering an arbitrary direction in the $(H_u,\tilde L,\tilde
\tau_c)$ space. We are not interested here with such a general
study---all we want to point out is that $\mu$ cannot be much larger
than the sfermion masses. Similar considerations apply to the $A$
terms. }:
\be
\ltau^2\mu^2 < \hat\ltau^2 \left(m_U^2+\msl^2+\mstau^2\right)\,.
\label{eq:trimubnd}
\ee
Notice that, because of the D-term contribution to $\hat\ltau^2$,
the bound (\ref{eq:trimubnd}) is irrelevant when the Yukawa
couplings are small (namely for sfermions of the first two families,
or even $\tau$ and $b$ at small $\tan\beta$.) When $\mu$ is somewhat
above this bound, an unwanted
minimum with $V\sim -\mu^4\ltau^4/\hat\ltau^6$ is present. Again,
for scenarios with a hierarchy $\mu\gg m_Z$, this new vacuum is much
deeper than the correct one.

In the course of our study we have verified that the positivity
constraints of Eqs.~(\ref{eq:mtwoconstr}) and (\ref{eq:constrs}) are
always stronger than those coming from the trilinear $A$ and $\mu$
terms, at least {\it for the parameter ranges of interest to us}.
Thus, while important in principle, the instabilities arising from
trilinear terms in the scalar potential do not impose any
constraints in practice.

Next, we examine the evolution of the Lagrangian parameters down to
the electroweak scale. The form of the RG equations dictates that
the soft-breaking scalar mass-squared parameters at the electroweak
scale are
linearly related to their initial values, to the square of the
GUT-scale gaugino mass $M_{1/2}$, and to the square of the $\mu$
parameter. (In fact there are also terms proportional to the
GUT-scale values of the $A$ parameters, namely $\propto A_G^2$ and
$\propto A_G M_{1/2}$. As shown in Appendix A they can be neglected
unless $A_G$ is at least an order of magnitude bigger than the other
GUT-scale
parameters.) Thus the constraints of Eqs.~(\ref{eq:constrs}) and
Eq.~(\ref{eq:mtwoconstr}), when saturated, constitute a set of eight
hyperplanes in the space of initial scalar mass-squared parameters.
When $M_{1/2} = \mu = 0$, the low-energy masses $m_i^2$ are just
homogeneous linear combinations of the GUT-scale parameters $M_i^2$,
so the various constraint planes determine a cone---or rather,
technically, a pyramidal surface---within which those constraints
are satisfied. Such a cone, drawn in the 3-dimensional space of
initial parameters for the minimal SO(10) theory as discussed below,
is shown in Fig.~6. For finite gaugino mass, $A$ or $\mu$, the
constraint planes are shifted by finite amounts. If there was any
allowed solid angle for $M_{1/2}=A=\mu=0$, the new allowed region
will be a truncated cone shifted from the origin. If there was no
allowed solid angle for $M_{1/2}=A=\mu=0$, turning these parameters
on can allow a finite (hyper-)polyhedron. In the absence
of running, that is, if the constraints of Eqs.~(\ref{eq:constrs})
are evaluated at the GUT scale, the cone (or polyhedron) they
determine spans a solid angle of order unity. As the parameters in
Eqs.~(\ref{eq:constrs}) are evolved to lower energies, the planes
turn about the origin and the cone changes; it may even close
completely, in which case proper electroweak breaking becomes
impossible. We will of course be interested in the cone evolved all
the way to the electroweak scale; it is useful to remember that it
is drawn in the space of parameters in the effective GUT-scale
Lagrangian, and its boundaries correspond
to those GUT-scale parameter values which lead to the vanishing of
particular scalar masses at the electroweak scale or of $m_2^2$ at
$10^5\,\GeV$---in other words, it encompasses the GUT-scale
parameters which would lead to proper breaking at the electroweak
scale. A narrow cone means that it is difficult to find GUT-scale
parameters which will lead to a
low-energy world similar to ours.

\begin{figure}[tb]
\centering
\leavevmode
\epsfysize=10.1cm \epsfbox[0 400 720 785]{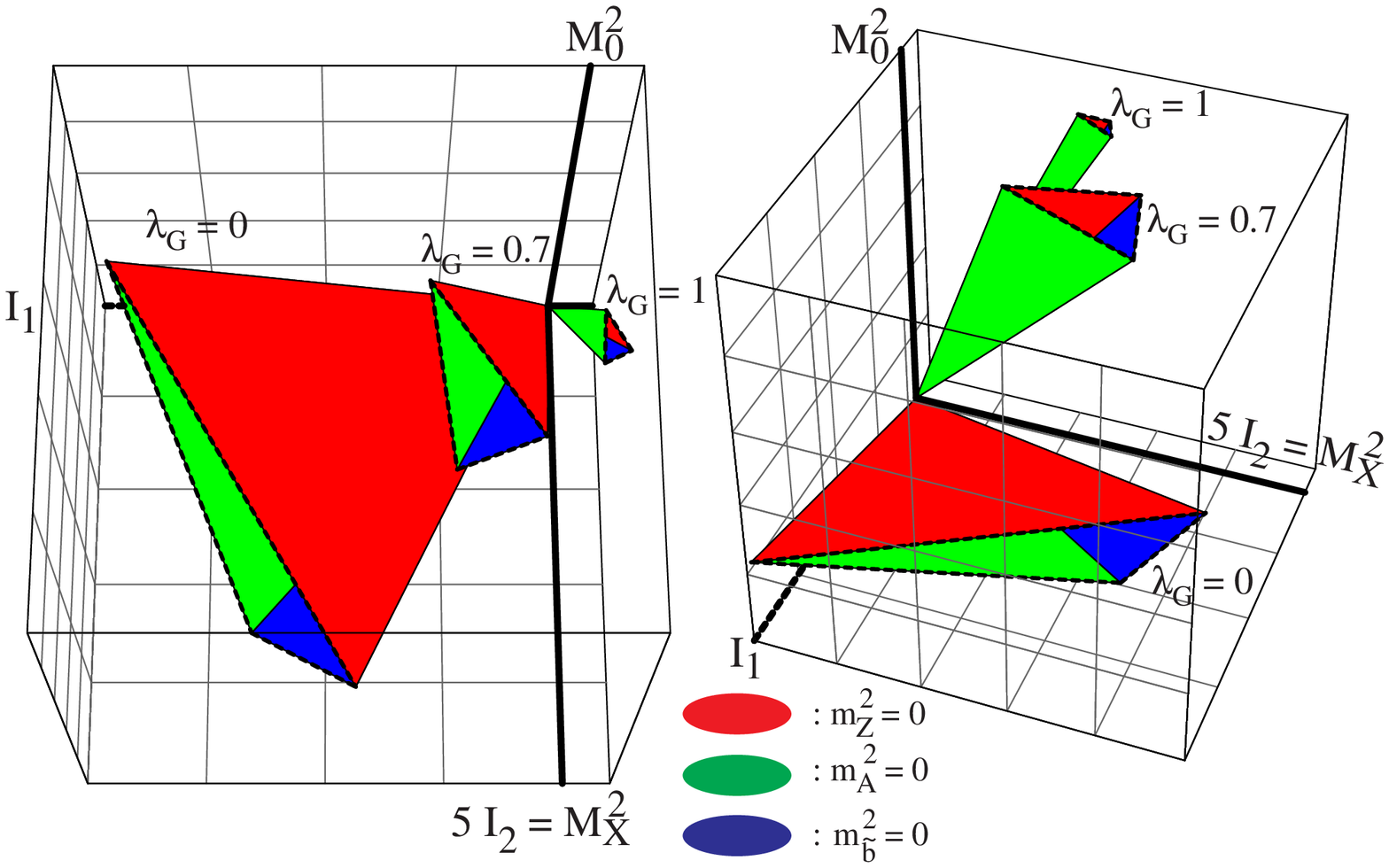}
\begin{quote}
{\small
Fig.~6. The allowed ``cones'' in the space of scalar GUT mass
parameters in the minimal SO(10) scenario with exact PQ and R
symmetries, within which electroweak symmetry is correctly broken.
Only the dominant (planar) constraints are shown: $m_Z^2 > 0$,
$m_A^2 > 0$ and $\msb^2>0$. Note the focusing of the solid angle for
increasing $\lG$, a consequence of the exponential homogeneous
evolution.}
\end{quote}
\end{figure}

\subsection{The Homogeneous Evolution}
\label{sec:homog}
To understand the evolution of the cone, for the time being {\it we
restrict our attention to the homogeneous parts of the differential
RG equations for the soft parameters}, neglecting the gaugino, $A$
and $\mu$ contributions. This not only illuminates the functional
behavior of the solutions, but is also directly relevant for the
case we have so far espoused, the PQ- and R-symmetric one in which
the $\mu$ and $A$ parameters and the gauginos are much lighter than
all
the other masses (except $m_Z$). In this maximally-symmetric case
the
evolution is driven by the Yukawa couplings, which in turn depend on
$\lG$ (and
$\rho_G$):
\begin{itemize}
\item For small $\lG$ there is little evolution, and the cone
remains wide. Here there are no generic difficulties in satisfying
the constraints for a wide range of initial parameters. Whether such
a range is available in particular GUT models is a question that
will be answered in the next section.\\
\item To understand the behavior for large $\lG$, it is useful to
change basis, considering certain fixed linear combinations ${\cal
M}
\vec m_i^2$ (where ${\cal M}$ is a constant matrix) of the seven
soft-breaking parameters. Appendix A contains the solutions, in this
new basis, of the RG equations for the seven soft scalar masses, as
well as
for the three $A$ terms and the $B$ parameter, in terms of integrals
over only the gauge and Yukawa couplings.  The matrix ${\cal M}$ is
chosen to
separate the homogeneous part of the seven scalar mass RG equations
into two
classes: three of the linear combinations, denoted $X_t$, $X_b$ and
$X_\tau$, evolve essentially multiplicatively, contracting
exponentially as they evolve down to the electroweak scale. The
other four, $I_{1,2,3,4}$, are approximately invariant. (Actually
one linear combination of the $I_i$ evolves slightly due to the
hypercharge D-term $S$ discussed in Appendix A, while  three
other independent combinations are truly invariant; but for the
present
purposes we can neglect $S$ and consider all four $I_i$ to be
invariant.) Thus we find that, for large $\lG$, the three $X_i$ are
exponentially suppressed at the electroweak scale relative to their
GUT-scale
values, and hence, generically, also relative to the four invariants
$I_i$. In the limit of very large $\lG$, at which the Yukawa
couplings tend to
their ``fixed points'' at low energies, the constraint equations
constitute seven constraints in only four variables $I_i$:
\be
0 < \vec m_i^2 =
{\cal M}^{-1}\left(\begin{array}{c} \vec X \\ \vec I
\end{array}\right) \sim
{\cal M}^{-1}\left(\begin{array}{c} e^{-{1\over\pi^2}\int \lambda^2
dt} \vec X_G \\ \vec I \end{array}\right) \sim
{\cal M}^{-1}\left(\begin{array}{c} \vec 0 \\ \vec I
\end{array}\right)\,.
\label{eq:mgenconstr}
\ee
Given the value of ${\cal M}$ in Appendix A, it is simple to check
that, in the limit $\lambda\to \infty$, there is only a trivial
solution: $I_i \equiv 0$. This solution is highly nontrivial,
however, in the sense that it
requires adjusting the soft masses at the GUT scale such that they
just cancel when combined into the $I_i$. At finite $\lambda$ the
degree of cancellation needed is just the degree to which the $X_i$
are suppressed; schematically, then, we must adjust the GUT-scale
parameters to satisfy
\be
\left|I_i\right| \roughly{<} e^{-{1\over\pi^2}\int \lambda^2
dt}\,X_{G}\,,\quad i=1,2,3,4\,.
\label{eq:genconstr}
\ee
How much tuning is {\it actually} required in order to satisfy
Eqs.~(\ref{eq:constrs})? We will determine numerically the allowed
region of parameter space under various GUT-scale assumptions.
Generically, one would expect a factor of $e^{-{1\over\pi^2}\int
\lambda^2 dt}$ for each invariant which needs a cancellation, hence
$\left(e^{-{1\over\pi^2}\int \lambda^2 dt}\right)^4$. However, since
the leptonic Yukawa coupling is considerably smaller than the other
two over much of the RG evolution, and since one of the invariants
and one of the $X$'s are essentially determined by leptonic masses,
a better estimate is
$\left(e^{-{1\over\pi^2}\int \lambda^2 dt}\right)^3$. To alleviate
this tuning problem, we should stay away from the fixed-point regime
of large Yukawa couplings, or relax the PQ or R symmetries (but then
$\epsb\epsz$ is made worse). Away from the fixed point, a splitting
$\lt^G > \lb^G$ also helps.
In the end, of course, the question itself only makes sense in a
particular
context: so far we have assumed there is no theoretical bias towards
any relationship between the initial parameters, except for the
approximate PQ and R symmetries. In a GUT context there will be some
definite biases.
\end{itemize}

The large-$\lG$ fine tuning for the symmetric case is illustrated
graphically in Fig.~6, assuming the SO(10) scenario discussed below
in which only one contracted direction $X$ and two invariants
$I_{1,2}$ play a role. The RG equations for the seven soft masses
are solved for three
cases: $\lG = 0$ (that is, without running), an intermediate value
of $\lG$, and a large $\lG$. The figure shows the cone in which
the electroweak-breaking constraints of Eqs.~(\ref{eq:constrs}) are
satisfied, drawn in the space spanned by the three linear
combinations of GUT parameters $X_G$, $I_1$ and $I_2$. Note that
only a small number---three or four---of the constraints are the
decisive ones, and once they are satisfied all others are as well;
in this case, they are $m_Z^2 > 0$, $m_A^2 > 0$ and $\msb^2 > 0$. We
see that near the fixed point, the cone closes up\footnote{Actually,
as evident from the figure, the cones close up before they reach the
ray $I_1 = I_2 = 0$. This ``premature focusing'' is a property of
the specific GUT boundary conditions and will be discussed below.}
around the ray
$I_1 = I_2 = 0$, meaning that if we don't tune the GUT parameters to
lie in
this ray, then the soft-breaking parameters at the electroweak scale
will not satisfy the constraints. Then either electroweak symmetry
will not break ($m_Z^2 < 0$), or the potential will not be bounded
from below ($m_A^2 < 0$, leading to dimensional transmutation at
a scale much larger than the SUSY scale, and also to
$\tan\beta\simeq 1$), or an electromagnetically-charged scalar will
acquire a VEV.

The contractions may also be seen analytically in terms of sum
rules,
which are particular linear combinations of the electroweak-scale
soft-breaking masses having only positive coefficients and chosen to
be independent of all the invariants. Linear combinations
of the $X_i$ with positive coefficients can give such  sum rules,
for
example $X_t+X_b$. These sum
rules have the property that on one hand they are phenomenologically
constrained to be positive (since $\vec m_i^2 > 0$), but on the
other hand they are driven to zero by the RG equations as $\lG$
increases. Schematically, we have for our example
\be
0 < 2\msq^2 + \mst^2 + \msb^2 + m_A^2 = X_t + X_b \sim
e^{-{1\over\pi^2}\int \lambda^2 dt}\,(X_t+X_b)_{G}\,.
\label{eq:gensumrule}
\ee
Since each of the mass terms in this equation should be positive,
for large $\lG$ they must each be made to evolve towards zero at low
energies. Now, each mass term can itself be expressed as a sum of
the $X_i$ and the $I_i$; since the $X_i$ are exponentially reduced
while the $I_i$
remain invariant, the soft masses can only evolve towards zero if
the various $I_i$ are tuned to be small already at the GUT scale. In
fact, the sum rules embody the same information as the planes in the
constraint equations (\ref{eq:constrs}) and in Fig.~6. In
particular, there is a dominant sum rule corresponding to the
innermost set of planes which define
the constraining cone; for the case of Fig.~6, this dominant sum
rule is $m_Z^2 + {4\over3} (m_A^2 + \msb^2)$. It indicates which are
the masses closest to saturating the constraints, and hence which
are
typically the lightest. Notice also that all the soft scalar masses
except the sleptons appear in the sum rule of
Eq.~(\ref{eq:gensumrule}), indicating that essentially all these
parameters contract for large $\lG$. The slepton masses also
contract, according to another sum rule, but to a lesser degree.

A very useful graphical way to describe the allowed domain in the
space of initial parameters is to project the constraint planes onto
the (hyper-)plane spanned by the various invariants. For the SO(10)
scenario illustrated in Fig.~6 and described in detail below, the
result is a set of lines in the plane of $(I_1/X_G,I_2/X_G)$ . To
normalize the
axes, we define a ``typical soft scalar mass'' $M_0^2 \equiv \third
X_G = \third \mtenh^2 + \twothr \msixth^2$, and use the more direct
Lagrangian parameter $\mx^2$ (see below) instead of $I_2$. The
horizontal and vertical axes are thus shown in units of
$\mx^2/M_0^2$ and $I_1/M_0^2$, respectively. Each line forms the
boundary of the half=plane where one of the $m_i^2$ is positive.
Fig.~7
shows all eight lines, and emphasizes the region allowed by
the constraint equations (\ref{eq:mtwoconstr}) and
(\ref{eq:constrs}), for three values of $\lG$, and for various
values of $\lt^G/\lb^G$. In general the allowed region is a polygon;
for the PQ- and R-symmetric SO(10) case, it is usually a triangle
bounded by the lines corresponding to the three masses which appear
in the dominant sum rule $m_Z^2 + {4\over3} (m_A^2 + \msb^2)$. Near
an edge of the triangle, the corresponding mass parameter is much
smaller than all the others. The hatched region in Fig.~7(a) is
where $m_Z^2$ is $\sim \tan\beta
\sim 50$ times less than $m_A^2$, which is the favored scenario (as
discussed in more detail below). Similar regions are indicated for
the other triangles.

\begin{figure}[tb]
\centering
\leavevmode
\epsfysize=8cm \epsfbox[25 495 740 760]{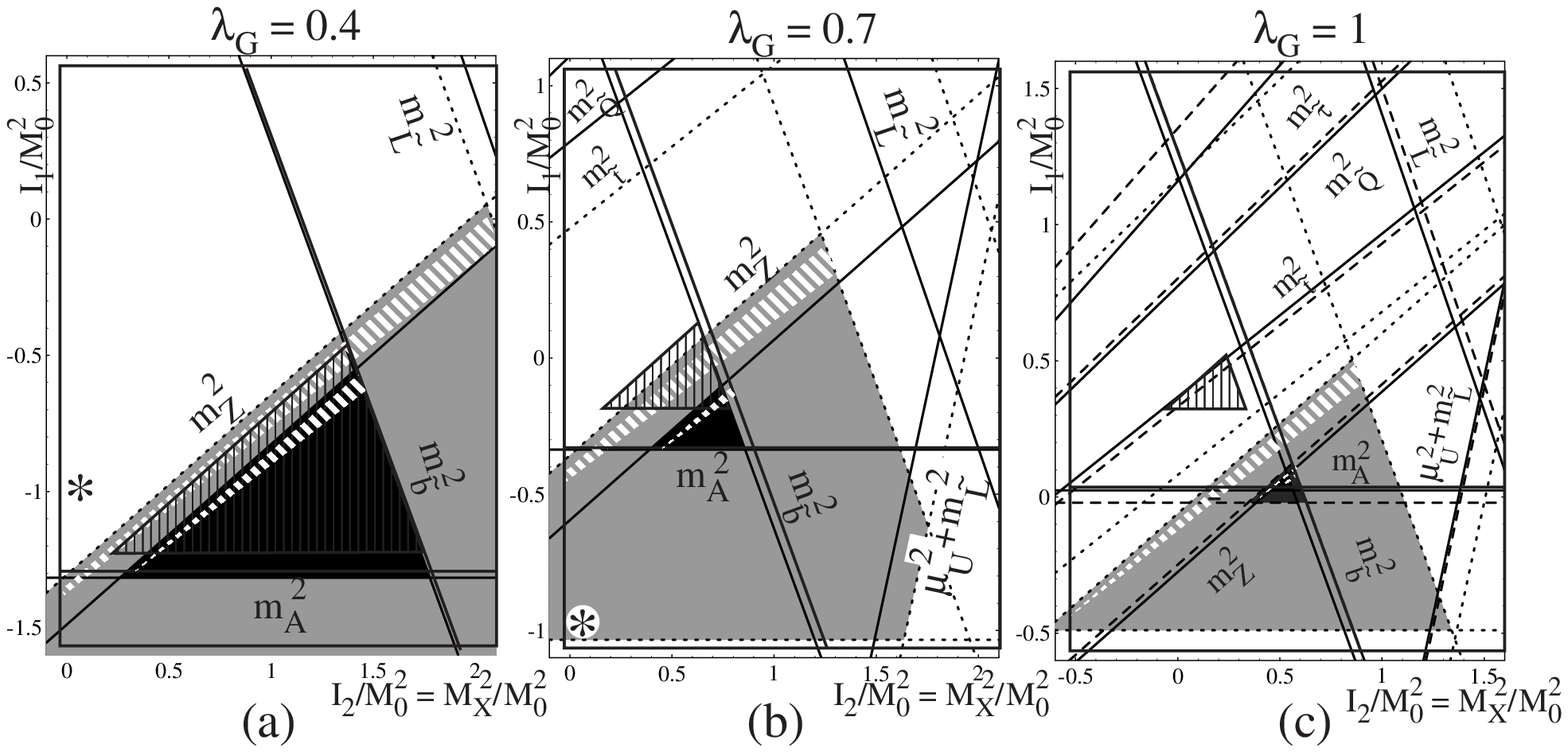}
\begin{quote}
{\small
Fig.~7. The allowed regions for the same scenario as Fig.~6, but
projected into two dimensions by dividing by the typical GUT squark
mass $M_0$. The GUT-scale Yukawa coupling is increased from left to
right, and both exact Yukawa unification (black areas, solid lines)
and approximate unification (light gray areas, dotted lines) are
considered. For $\lG = 1$ we also show the perturbed case
$\lb^G/\lt^G = 1.1$ as the dark gray area and dashed lines. The
lines delineate the half plane in which the corresponding
mass-squared is positive and hence acceptable. In the
diagonally-hatched regions $m_Z^2 \ll m_A^2$, as discussed in the
text. Finally, the vertically-hatched triangles are the allowed
areas assuming exact unification and a right-handed neutrino at $M_N
= 10^{12}\,\GeV$.}
\end{quote}
\end{figure}

\subsection{Evolution and Natural Selection}
\label{sec:darwin}
{}From the previous subsection, we conclude that if $\mu$ and $\mwi$
are
chosen much smaller than the typical soft scalar masses so PQ and R
are approximately valid, then we expect the allowed triangular area
in the space of GUT-scale scalar masses to be small if
$\lG\roughly{>} 0.7$ and $\lt^G = \lb^G$ (the focused case), and
large
otherwise. In fact, as we show below, both SO(10)- and SU(5)-type
boundary conditions on the scalar masses result in {\it premature}
focusing: the triangles close up for finite $\lG$ values,
not far above unity. Within the allowed triangle, therefore,
a few particles---namely those which bound the triangle itself---are
very much lighter than the rest. (In contrast, for $\lG\to\infty$
focusing all masses must become very light.)

Now, if there {\it is not} much focusing, then all the scalar masses
are comparable throughout most of the triangle, while in a narrow
($\sim 1/\tan\beta$) strip within that triangle $m_Z^2$ is $\sim
1/\tan\beta$ times lighter than $m_A^2$. Tuning the GUT scalar
parameters to lie within this strip suppresses the large $b\to s
\gamma$ amplitudes, and allows us to generate large $\tan\beta$ with
no further tuning of $B$ and without violating LEP bounds by taking
$\mu^2\sim M_{1/2}^2 \sim m_A^2/\tan\beta$. If there {\it
is} significant focusing, then within the small allowed triangle the
particles which bound the triangle have masses---indicated
collectively
with $m_{0,L}^2$---which are much lighter than the others,
so a hierarchy $m_{0,L}^2 \ll m_{0,H}^2$ is invariably present. As
we will see in Tables 3 and 4, $m_Z^2$ and $m_A^2$ are always among
the light masses, while $\msq^2$ is always one of the heavy masses.
Hence throughout the triangle the dangerous $b\to s\gamma$
amplitudes are somewhat suppressed, depending on the degree of
focusing. There is now no naturalness criterion to
distinguish between the following situations: either the GUT scalar
parameters are tuned to lie within a narrow strip in this triangle,
resulting in $m_Z^2 \sim m_A^2/\tan\beta \sim m_{0,L}^2$ and
requiring no further tuning of $B$ since we can select $\mu^2 \sim
M_{1/2}^2 \sim m_Z^2 \sim m_A^2/\tan\beta$ and meet the LEP
constraints; or the GUT scalar parameters are not adjusted to be in
the strip, so $m_Z^2 \sim m_A^2 \sim m_{0,L}^2$, but $B$ must be
tuned to $\sim1/\tan\beta$ of its natural value because LEP
requires us to select $\mu^2 \sim M_{1/2}^2 \sim m_Z^2 \sim m_A^2$.
To summarize: when the PQ and R symmetries hold at the GUT scale, if
there is no focusing then they hold at the electroweak scale and the
overall tuning need only be $\sim 1/\tan\beta$; but if there is
strong focusing then they can either hold or not hold at low scales,
and in either case the need for tuning is greater than $1/\tan\beta$
by the degree of focusing.

Strictly speaking, the exactly PQ- or R-symmetric scenarios are
never allowed by LEP limits, so we should in principle always take
$\mu$ and $M_{1/2}$ into account in evolving the cone to low
energies. But this would not qualitatively change the discussion.
Clearly, approximating $\mu$ or $M_{1/2}$ by zero is valid when
there is no focusing and the symmetries are approximately valid at
all scales. But even when there is focusing and some of the
particles end up with small masses $m_{0,L}$, introducing finite
$\mu$ or $M_{1/2}$ comparable to those masses will only change the
allowed triangle area and the light masses by ${\cal O}(1)$; in
fact, if $\mu$ or $M_{1/2}$ are then increased beyond $m_{0,L}$, the
focusing is alleviated such that the new value of $m_{0,L}$ is again
comparable to $\mu$ or $M_{1/2}$.

If the PQ or R symmetries are not approximately valid at the GUT
scale, the planes $m_i^2 = 0$ which delimit the allowed volume are
shifted, and the focusing is alleviated. (The exception is when
$\mu$ or $M_{1/2}$ are too big in certain GUT scenarios---then the
triangles once again close for finite values of $\lG$.) Irrespective
of
focusing, the tuning is worse than $\sim1/\tan\beta$, as
demonstrated schematically in Table 2. However, we will reserve
judgment on the breaking of PQ and R symmetries until after we have
studied the SO(10) and SU(5) symmetry constraints in the scalar mass
sector.

\section{Grand Unified Soft Masses}
\label{sec:gut}
\subsection{SO(10) and SU(5) Boundary Conditions}
\label{sec:bc}
We now address the question of what values of initial parameters
arise from theories at the GUT scale.
Since the idea of Yukawa unification is largely based on symmetry
principles, it behooves us to consider the implications of those
same symmetries for the various soft SUSY-breaking parameters. We
have already employed, in our solutions to the RG equations, the
assumption that gauge coupling unification is accompanied by gaugino
mass unification, in order to reduce the number of independent
gaugino mass parameters to one. Let us now examine the implications
of SO(10) or SU(5) symmetries for the soft-breaking masses. As a
special case, we comment briefly on the universal scenario. We also
examine the threshold corrections due to a light right-handed
neutrino.

Consider first the simplest SO(10) scenario, in which both light
Higgs doublets originate from a single ${\bf 10}_H$ multiplet, or
more generally any SO(10) model in which all the GUT fields from
which the light
doublets arise have degenerate soft masses. When SO(10) breaks, in
general both D and F terms could split the scalar masses in a single
SO(10) multiplet. D-term \cite{ref:dterms} splittings are
generically present because the rank of the gauge group is reduced,
but F-term splittings are more model dependent and need not arise.
For example, when SUSY breaking is communicated from a hidden sector
only via gravitational interactions, the soft terms are very
constrained \cite{ref:hito}.  This property leads, in the minimal
missing-VEV models \cite{ref:dimwil}, to exclusively D-term
splittings. Therefore when we refer to SO(10)-type boundary
conditions we will only include D-term splittings, whereas more
general F-term splittings will be encompassed by the discussion of
SU(5) boundary conditions, or when necessary by the completely
general discussion. Accordingly, in SO(10)-type models the seven
soft-breaking masses are determined at the GUT scale by only three
soft-breaking parameters: the soft Higgs mass $\mtenh$, the
third-generation soft squark and slepton mass $\msixth$, and a soft
mass parameter $\mx$ from a D-term that is left over when the $\rm
U(1)_X$ symmetry in SO(10) is spontaneously broken. Recall that the
rank of SO(10) is
higher by one than that of SU(5) and of $\rm SU(3)\otimes
SU(2)\otimes U(1)_Y$, thus $\rm SO(10) \supset SU(5) \otimes
U(1)_X$, where the generator of $\rm U(1)_X$ is proportional to $\rm
3(B-L)+4T_{3R}$. One common way to break this $\rm U(1)_X$ and
reduce the rank to that of the standard model is to introduce a pair
of 16-dimensional Higgs representations having GUT-scale masses,
${\bf 16}_H$ and $\overline{{\bf 16}}_H$, which acquire
VEVs in their ``$\nu_R$'' components, thus preserving the SU(5)
symmetry. When $\rm U(1)_X$ breaks this way, its D-term acquires a
VEV proportional to the difference of the soft masses of the
${\bf 16}_H$ and $\overline{\bf 16}_H$. This D-term
then contributes to the soft masses of the fields which couple to
$\rm U(1)_X$ in proportion to their $\rm U(1)_X$ charges. To
quantify this contribution, we define a mass parameter
\be
\mx^2 = M_{{\bf 16}_H}^2 - M_{\overline{{\bf 16}}_H}^2\,.
\label{eq:mx}
\ee
(Note that this definition differs by a factor of 10 from the
definition in our previous work \cite{ref:hrsII}. We chose the
present
definition because it is closer to the fundamental parameters of the
grand-unified theory, and so is on the same footing as $\mtenh$ and
$\msixth$.) It is the presence of $\mx$ which allows the up- and
down-type Higgs masses to be split at the GUT scale in almost any
SO(10)-unified scenario, and thus greatly facilitates proper
electroweak breaking.

In SU(5), of course, the two Higgs doublet superfields originate
from different representations, ${\bf 5}_H$ and
$\overline{{\bf 5}}_H$, so their soft SUSY-breaking masses are
generically split. This is also the case in some SO(10) models, for
example when the light doublets are mixtures of different GUT
multiplets having different soft masses, or even when they both lie
in the same SO(10) multiplet
but the soft terms are general enough to induce F-type splittings.
Note that
in this second case the Yukawa couplings are indeed exactly unified.
For
brevity, however, we will call any such boundary conditions on
scalar masses SU(5)-type. The seven soft masses in SU(5)-type models
are determined by four parameters: the two Higgs masses $\mfivh$ and
$\mfivbh$, and the soft masses for the two representations which
contain the
third-generation MSSM squarks and sleptons, $\mfivth$ and $\mtenth$.
For comparison with the SO(10) case, we will recombine these four
parameters into the same three combinations which occur in SO(10)
plus an extra
degree of freedom $\msufive$, as follows: $\mtenh^2 \equiv
\half(\mfivh^2 + \mfivbh^2)$, $\msixth^2 \equiv \thrfour\mtenth^2 +
\fourth\mfivth^2$, $\mx^2 \equiv \fivefour(\mfivbh^2-\mfivh^2) +
\fivefour(\mtenth^2-\mfivth^2)$, and $\msufive^2 \equiv
\half(\mfivh^2-\mfivbh^2) + \half(\mtenth^2-\mfivth^2)$.  (To
reiterate: in an SU(5) context, the four quantities $\mtenh$,
$\msixth$, $\mx$ and $\msufive$ should be regarded just as
convenient linear combinations of the underlying soft masses.)

With these redefinitions we may write the seven soft masses
at the GUT scale in either SO(10) or SU(5) as follows:
\bea
M_{U\phantom{,t,b}}^2 &=& \mtenh^2 \phantom{ .+ \msixth^2}
-{2\over10}\mx^2  + {1\over2}\msufive^2\label{eq:MU}\\
M_{D\phantom{,t,b}}^2 &=& \mtenh^2 \phantom{ .+ \msixth^2}
+{2\over10}\mx^2  - {1\over2}\msufive^2\label{eq:MD}\\
M_{{\widetilde Q},{\tilde t},{\tilde \tau}}^2 &=& \phantom{ \mtenh^2
+.} \msixth^2
+{1\over10}\mx^2  + {1\over4}\msufive^2\label{eq:MQttau}\\
M_{{\tilde b},{\widetilde L}\phantom{,t}}^2 &=& \phantom{ \mtenh^2
+.} \msixth^2
-{3\over10}\mx^2  - {3\over4}\msufive^2\label{eq:MbL}\,,
\eea
(Capital letters denote the parameter values at the GUT scale.)
Their linear combinations $X_i$ and
$I_i$ then take the form:
\bea
X_t^G &=& \phantom{2} \mtenh^2 + 2 \msixth^2
\phantom{.+{1\over10}\mx^2}  +
\phantom{1\over3}\msufive^2\label{eq:xt}\\
X_b^G &=& \phantom{2} \mtenh^2 + 2 \msixth^2
\phantom{.+{1\over10}\mx^2}  -
\phantom{1\over3}\msufive^2\label{eq:xb}\\
X_\tau^G &=& \phantom{2} \mtenh^2 + 2 \msixth^2
\phantom{.+{1\over10}\mx^2}  -
\phantom{1\over3}\msufive^2\label{eq:xtau}\\
I_1 &=& \vphantom{1\over3}2\,\mtenh^2 - 3\,\msixth^2
\phantom{.+{1\over2}\mx^2}+ 2\,\,\msufive^2\label{eq:ione}\\
I_2 &=& \phantom{2 \mtenh^2 - 3 \msixth^2 + \half} {1\over5}\mx^2 +
{1\over2}\msufive^2\label{eq:itwo}\\
I_3 &=& \phantom{2 \mtenh^2 - .} \,\,\msixth^2 - {3\over10}\mx^2 +
{5\over4}\msufive^2\label{eq:ithree}\\
I_4 &=& \phantom{2\,\mtenh^2 - {1\over3} \msixth^2 - \half\mx^2 \,}
\phantom{.}-\phantom{1\over3}\msufive^2\,.
\label{eq:ifour}
\eea

A common assumption in much of the previous work on unified
supersymmetric theories, and in particular in large $\tan\beta$
models \cite{alsalt}, is a ``universality'' of soft SUSY-breaking
scalar masses at the GUT scale. Under the universality assumption,
$\mtenh = \msixth$ while $\mx = \msufive = 0$. We have not made this
assumption because we do not expect it to hold at the GUT scale
(whether or not it is a good approximation at the Planck or string
scales) and because it requires \cite{ref:hrsII} tremendous
fine-tuning of parameters to achieve proper electroweak
symmetry-breaking with large $\tan\beta$. We will have more to say
about this  case in the discussion in Sec.~\ref{sec:soten}.

A much better motivated assumption is that the third-generation
right handed neutrino, that is, the electroweak-singlet
superfield which couples to $\nu_L$ through a Yukawa coupling with
$H_U$, has a Majorana mass $M_N$ smaller than $\Mgut$.
Theoretically, such a mass may arise\footnote{This happens, for
instance, in the absence of ${\bf 126}+\overline{\bf 126}$ SO(10)
Higgs multiplets
and of R-odd gauge singlets. Note that in order to
preserve the validity of perturbation theory it is better to
avoid the ${\bf 126}$ or bigger representations,
since they contribute a large positive term to the gauge
$\beta$-function \cite{ref:bmr}.} from a high-dimensional
operator induced at some scale $M' > \Mgut$ (such as the string or
Planck scale) and therefore be suppressed by a power of $1/M'$.
Phenomenologically, a lower $M_N$ leads through the see-saw
mechanism to a higher mass for the observed neutrinos, which may
then explain various cosmological and astrophysical puzzles. In any
case, as long as $M_N$ is not too far below $\Mgut$, its effects can
be approximated by threshold corrections to the Yukawa couplings and
to the soft SUSY-breaking scalar masses. If the SO(10)-type boundary
conditions $\lt^G = \lN^G \equiv \lG$ are valid, we can calculate
the size of the corrections:
\bea
\Delta \ltau = \Delta \lt &=& - {1\over 2} \lG^3 \,{\ln
\left({\Mgut/ M_N}\right)\over 8\pi^2}\,; \nonumber\\
\Delta M_U^2 &=& -\lG^2 X_t^G \,{\ln \left({\Mgut/ M_N}\right)\over
8\pi^2}\,; \label{eq:rhn}\\
\Delta M_{\widetilde L}^2 &=& -\lG^2 X_\tau^G \,{\ln \left({\Mgut/
M_N}\right)\over 8\pi^2}\,.
\nonumber
\eea
The consequences will be discussed in the following section.

\subsection{SO(10)-type GUT Masses}
\label{sec:soten}
We first consider the symmetric, minimal SO(10) scenario: (PQ- and
R-) symmetric in that $\mu$, $A$, $B$ and the gaugino masses are
negligible relative to the various squark, slepton and pseudoscalar
masses; minimal in that the two light Higgs doublets are contained
in a ${\bf 10}_H$ of SO(10), so $\lt^G = \lb^G$ (up to small
threshold corrections); and SO(10) in that the soft-breaking masses
are those that arise in a minimal SO(10) theory, hence $\msufive
\equiv 0$.
Three independent dimensionful GUT-scale initial parameters specify
the electroweak-scale consequences, and we
choose them (as noted above) to be $M_0^2$ ($ = \third \mtenh^2 +
\twothr \msixth^2 = \third X_t^G = \third X_b^G = \third X_\tau^G
\equiv \third X^G$), $\mx^2$ ($ = 5 I_2$) and $2 \mtenh^2 - 3
\msixth^2 = I_1$. The third invariant is a linear combination of
$M_0$ and the first two invariants; it is not very constrained,
since it is mostly associated with $X_\tau$,
which does not contract much. The fourth and last
invariant vanishes identically by the SO(10) symmetry. The evolution
from the GUT scale to the electroweak scale, and the contraction of
the allowed triangle towards small $\mx$ and $I_1$ as $\lG$
increases, were illustrated in Fig.~6, or in projected form (in the
space of $\mx^2/M_0^2$ and $I_1/M_0^2$) in Fig.~7. Three features
are
worth noting:
\begin{enumerate}
\item In the general case of seven independent initial soft mass
parameters, we expect that the allowed region closes asymptotically
as $\lG \to \infty$, when all four invariants must be set to zero to
allow the three $X_i$ to contract. Graphically, this means that the
eight planes (or lines) corresponding to $m_i^2(\Lambda = m_Z) = 0$
and $m_2^2(\Lambda = 10^5\,\GeV) = 0$ all cross at
one hyper-ray $I_{1,2,3,4} = 0$ (namely the origin in the projected
space). Instead, we see that in the symmetric minimal SO(10) case
the allowed region of parameter space is bounded by the planes (or
lines) $m_Z^2 = 0$, $m_A^2 = 0$ and $\msb^2 = 0$ which cross
prematurely, closing the allowed triangle for a
finite value of $\lG$. The reason is that SO(10) boundary conditions
are compatible with $I_{1,2,3,4} = 0$ only if {\it all} initial
masses vanish, a trivial and uninteresting scenario. For
nonvanishing $M_0$, the restrictive SO(10) boundary conditions can
only be satisfied for sufficiently small $\lG$. Just how small?
Consider the sum-rule-like combination $m_Z^2 + {4\over3} (m_A^2 +
\msb^2)$, which was chosen so that the invariant part is a
negative definite quantity: $(-2 I_1 -7 I_2 -14 I_3 +14 I_4)/183 =
-4/61 M_0^2$. The contracting part $(-97 X_t +216 X_b +7
X_\tau)/183$
starts out positive, with a value $126/61 M_0$, and contracts
monotonically to zero as $\lG$ increases. Hence the sum $m_Z^2 +
{4\over3} (m_A^2 + \msb^2) = (-97 X_t +216 X_b +7 X_\tau)/183 - 4/61
M_0^2$ must vanish for a finite $\lG$. The critical value turns out
to be $\lG \simeq 1.2$. Therefore in this minimal symmetric
SO(10) case it is important that $\lG$ is not only small, but in
particular is well below $\sim 1.2$, for there to be a significant
allowed region of parameter space. [If we wish to include the small
contribution to
this ``sum rule'' of the hypercharge D-term, we can again find the
proper
combination of masses in which the invariant part is negative and
proportional to $M_0^2$. The result is:
\be
m_Z^2 + {4\over3} (m_A^2 + \msb^2) + \left[
{\sixth \alpha_G \ln(\Mgut/m_Z)\over 1+
3\alpha_G \ln(\Mgut/m_Z)}\right]\,
(m_Z^2 + m_A^2)\,,
\label{zab}
\ee
but there is hardly any change in the conclusions.]
\item When the parameter space contracts, it does so around a
nonzero
value of $\mx^2/M_0^2$. This highlights the important role of the
$\rm U(1)_X$ D-term mass parameter in allowing both a negative
$m_U^2$ and a sufficiently positive $m_D^2$, even though the RG
equations drive $m_D^2$ down more than $m_U^2$.
\item Notice that the allowed cone in the PQ- and R-symmetric SO(10)
scenario of Fig.~6 is confined to the $M_0^2 > 0$ half plane. This
result was used when dividing by $M_0^2$ to determine the allowed
area in the projected space of Fig.~7. We have found that, in order
to satisfy the low-energy constraints of Eqs.~(\ref{eq:constrs}) and
Eq.~(\ref{eq:mtwoconstr}), $M_0^2$ must always be positive for all
values of $\lG$ and $\lt^G/\lb^G$ under consideration, and for SU(5)
as well as SO(10) boundary conditions on the soft scalar masses, as
long as the PQ and R symmetries hold. To meet the low-energy
constraints with negative $M_0^2$ inevitably requires very large
values of $|\mu^2/M_0^2|$ and sometimes of $|M_{1/2}^2/M_0^2|$;
since these would require very delicate fine-tuning of the GUT-scale
parameters, we will assume $M_0^2 > 0$ for the remainder of this
paper.
\end{enumerate}

{}From Fig.~7 we can infer some properties of the universality
assumption when PQ and R are valid. The asterisk ($\ast$) indicates
the coordinate $(\mx^2,I_1) = (0,-M_0^2)$ corresponding to universal
scalar mass boundary conditions at the GUT scale. Proper electroweak
symmetry-breaking
occurs only if $\lt^G$ and $\lb^G$ are widely split, and then only
for intermediate values of $\lG$. Furthermore, to meet LEP
constraints in the approximately PQ- and R-symmetric scenario, the
value of $\lG$ must be tuned to achieve $m_Z^2 \sim \mu \mwi \ll
m_0^2$. If on particular we set $\mu^2\sim\mwi^2\sim
m_A^2/\tan\beta$ and tune $\lG$ with a precision $\sim 1/\tan\beta$
to get $m_Z^2$ light enough, we end up with the minimally fine-tuned
scenario at the lower-left corner of Table 2. Thus the PQ- and
R-symmetric universal case is allowed and only minimally tuned (via
$\lG$) {\it if} the Yukawas are widely split at the GUT scale.

Fig.~7 also illustrates the effects of threshold corrections due to
a right-handed neutrino with a mass $M_N \ll \Mgut$. The
vertically-hatched triangles in Fig.~7 shows the area allowed when
$M_N \sim 10^{12}\,\GeV$. The $\sim 5\percent$ correction to the
Yukawa
couplings lowers $\lt^G$ and so would reduce the allowed area, but
the $\sim 40\percent$ negative correction to $M_U^2$ is far more
significant and increases the area. The result is that the area
remains $\sim 0.1$ even at $\lambda_G\simeq 1.0$, comparable
to that of the dark-shaded triangle [shown only in Fig.~7(c)] which
would arise from a typical $\sim 10\percent$ threshhold correction
to $\lt^G/\lb^G$.
Notice, however, that the neutrino effect, in addition to being well
motivated, also favors a small $\mx^2\sim 0$; in other
words, the $\Delta M_U^2$ shift can substitute for the shift
produced by the $\rm U(1)_X$ D-term which was needed for proper
radiative symmetry breaking. A vanishing $\mx^2$ could conceivably
be achieved naturally by means of a symmetry. Finally, note that a
light right-handed neutrino threshold is not sufficient
to allow for the PQ- and R-symmetric universal case when
$\lt^G=\lb^G$.

In Table 3 we display some characteristics of the SO(10) scenario
for various values of $\lG$ and of $\lt^G/\lb^G$. The heavy-dashed
boxes correspond to the PQ- and R-symmetric case: the three on the
left are for the minimal Higgs choice ($\lt^G = \lb^G$), while the
three on the right allow for large Higgs mixings at the GUT scale
($\lt^G = 2\lb^G$). The top entry in each box gives the area of the
allowed triangle using the coordinates of Fig.~7, namely
$\mx^2/M_0^2$ and $I_1/M_0^2$. Also shown in some interesting cases
are the larger areas that would result from a slight Yukawa
splitting ($\lt^G = 1.1\lb^G$) due to some slight mixing or
threshold effect. Note that the area decreases rapidly as $\lG$
increases, indicating the aggravated need for fine-tuning of the
GUT-scale parameters. Of course, the value of the area depends on
the choice of coordinates and the metric, which are to some extent a
matter of taste. We use these particular coordinates because we
expect them to be a priori of order unity, and so if the triangle
area is much smaller than 1 then some tuning is apparently
necessary. A crossed-out box indicates that the corresponding
parameter choice leads to a value for $\db$ incompatible with
bottom-tau unification and the low-energy values for $m_b$ and
$m_\tau$ (though some of those boxes are nevertheless filled in for
reference).

\begin{figure}[tb]
\centering
\leavevmode
\epsfysize=11.5cm \epsfbox[30 350 645 730]{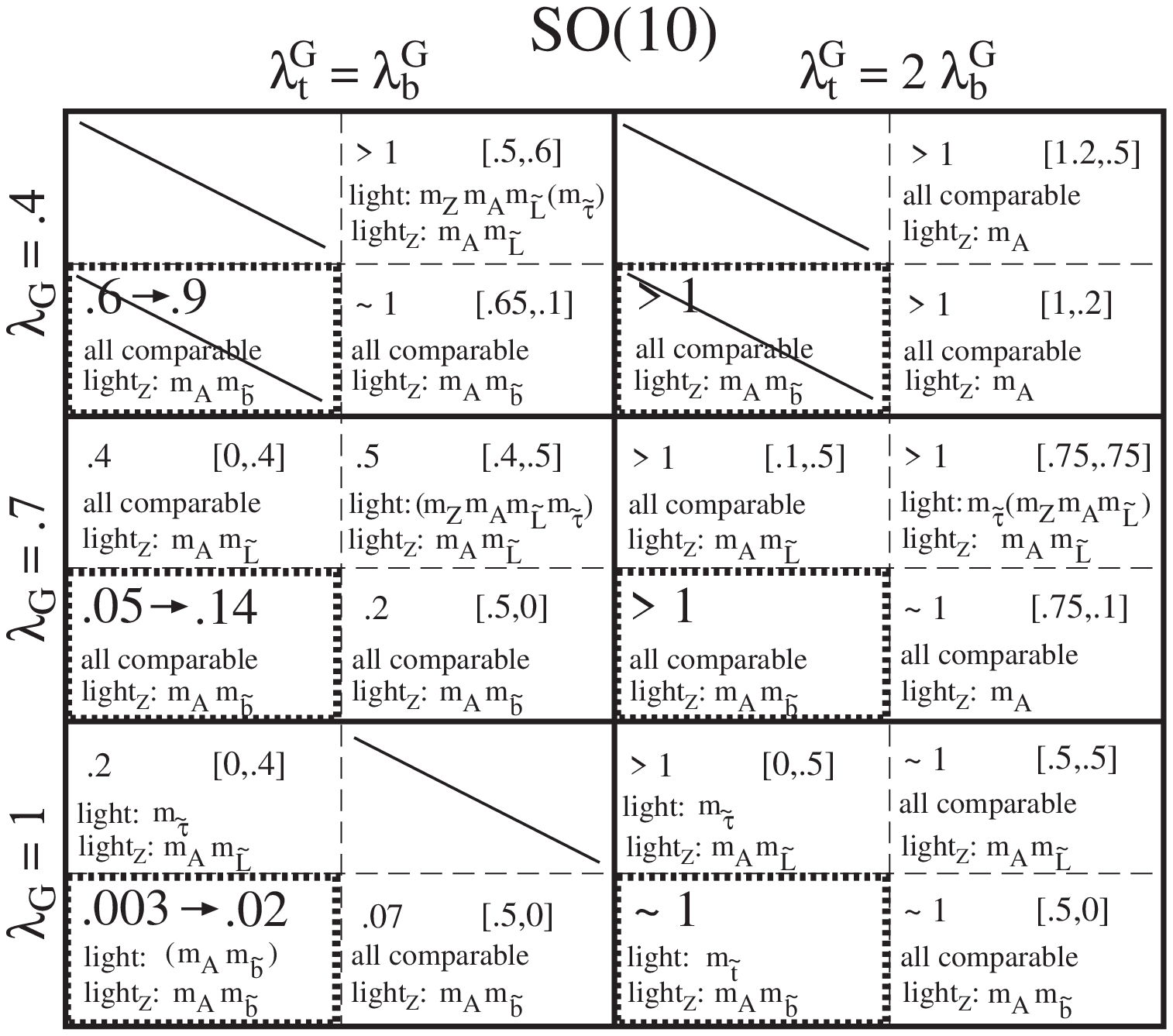}
\begin{quote}
{\small
Table 3. The characteristics of exact and approximate Yukawa
unification with SO(10)-type boundary conditions on the scalar
masses. For each of the six choices of $\lt^G/\lb^G$ and $\lG$, four
boxes are shown, corresponding to the presence of absence of
approximate PQ and R symmetries (with $\mu/M_0$ and $M_{1/2}/M_0$
shown respectively in square brackets). The first entry in each box
is the allowed area, the second is a typical spectrum, and the third
lists the masses than can be decreased simultaneously with $m_Z$.
See the text for further details.}
\end{quote}
\end{figure}

The full superspectrum is completely determined, up to an overall
scale, by choosing a specific point within the allowed triangle.
For the PQ- and R-symmetric scenarios we are now considering, the
middle of each box indicates the superspectrum that is typical near
the light-$m_Z^2$ portion of the triangle, shown as the hatched
region in Fig.~7. [Recall that, while this region is $\sim 50$ times
smaller than the triangle and hence by definition requires that much
fine-tuning, it leads to a hierarchical spectrum which is both
phenomenologically allowed and requires no further tuning to achieve
acceptable $\tan\beta$ and $\Gamma(b\to s\gamma)$. Furthermore, far
away from this region, $m_Z$ is greater than $\mu$ and $\mwi$ and
hence is in conflict with LEP.] In most cases all the superpartners
and the pseudoscalar Higgs have similar masses. However, when $\lG$
is large and $\lt^G \simeq \lb^G$, the allowed triangle is small and
therefore its bounding particles, the pseudoscalar Higgs and the
SU(2)-singlet bottom squark, are somewhat lighter than the other
particles. And when $\lG$ is large but there is a large top-bottom
Yukawa splitting, the SU(2)-singlet stop becomes relatively light.
The last item in each box is the list of masses which can vanish
simultaneously with $m_Z$: for the symmetric SO(10) case, they are
always $m_Z$, $m_A$ and $\msb$, as discussed above. In other words,
we may choose parameters at the corner of the allowed triangle such
that $m_A$ and $m_Z$ are much lower than all the other superpartner
masses, or $\msb$ and $m_Z$ are much lower than the others.

Fig.~8(a) shows contours of fixed allowed area as functions of the
size
of the GUT-scale Yukawa coupling $\lG$ and the amount of top-bottom
splitting splitting $\lt^G/\lb^G$. The sharp bends occur when the
$\mst^2 > 0$ constraint becomes more restrictive that the $\msb^2 >
0$ constraint, so the rate at which the triangle closes is
determined by the evolution of $m_Z^2$, $m_A^2$ and $\mst^2$ rather
than of $m_Z^2$, $m_A^2$ and $\msb^2$. Note the dramatic decrease in
area as the maximal value of $\lG$ is reached for fixed
$\lt^G/\lb^G$---this is the premature focusing implied by SO(10)
[and SU(5), as we shall see] boundary conditions on the scalar
masses. The large-$\lG$, small-splitting region clearly requires
very precise adjustment of the GUT-scale parameters.

\begin{figure}[tb]
\centering
\leavevmode
\epsfysize=8.8cm \epsfbox[70 250 550 560]{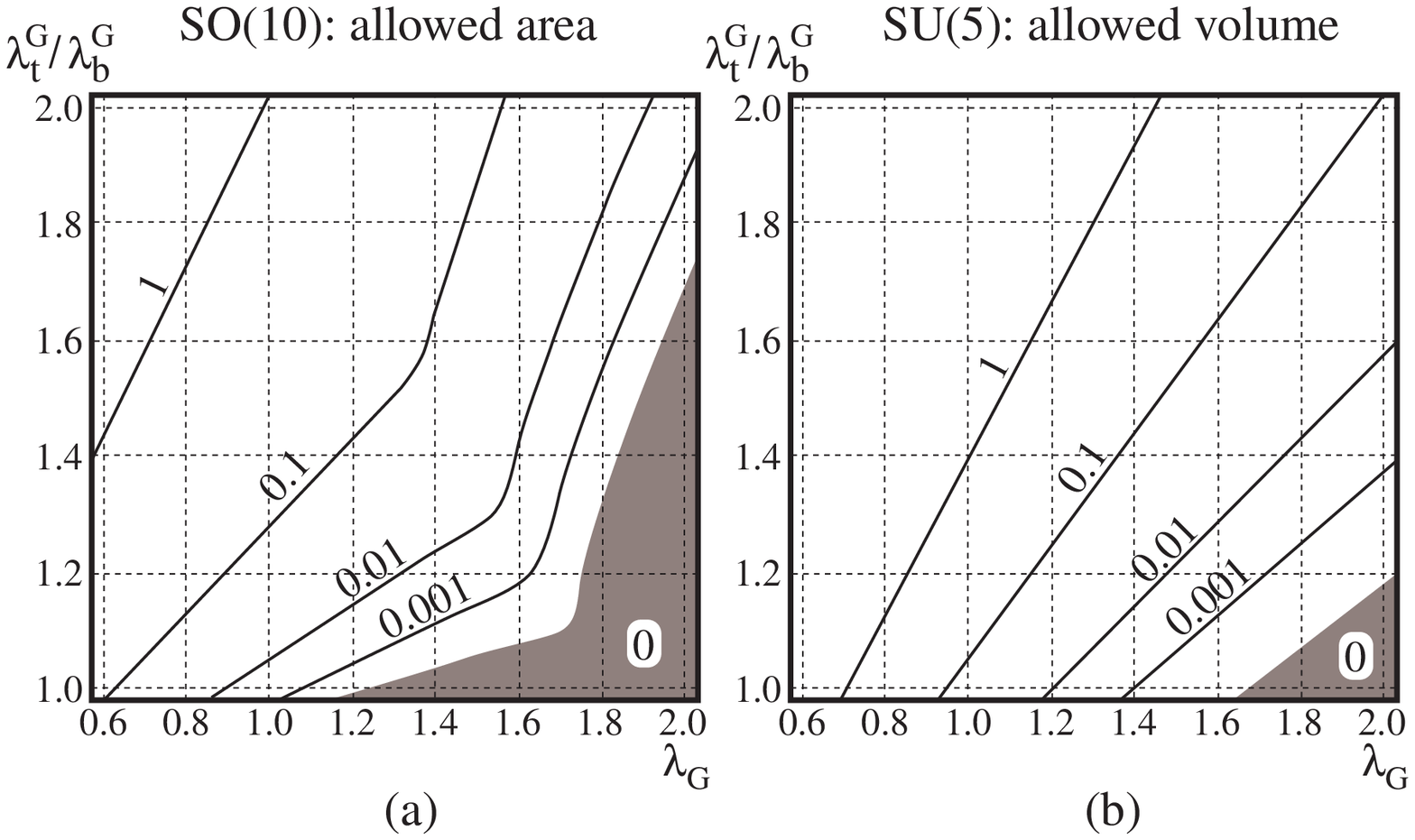}
\begin{quote}
{\small
Fig.~8. Contours of constant allowed areas or volumes for SO(10)- or
SU(5)-type boundary conditions, respectively, assuming exact PQ and
R symmetries. Notice that small $\lG$ and large $\lt^G/\lb^G$ values
are favored, since they alleviate the premature focusing of the
homogenous RG evolution.}
\end{quote}
\end{figure}

We next relax either the PQ or the R symmetry, or both, and once
again ask for the allowed region in the parameter space of soft
scalar masses. For fixed $\mu$ and $M_{1/2}$, the planes which
delimit the allowed region are now shifted by fixed amounts $\sim
\mu^2$ and $\sim M_{1/2}^2$, so that they no longer intersect at the
origin. We are actually most interested in the relative quantities
$\mu/M_0$ and $M_{1/2}/M_0$, that is, the amount of PQ and R
breaking relative to the other (soft scalar) mass parameters. So for
fixed $\mu$ and $M_{1/2}$, we should consider various slices of
constant $M_0$ in the scalar mass parameter space. (Such slices are
in fact the projections shown in Fig.~7.) Small PQ and R breaking
corresponds to looking at large-$M_0$ slices: at such large
distances from the origin, the small displacements of the planes are
insignificant, and the allowed portion of the slice is essentially
the same as in the symmetric case. Larger PQ and R breaking
correspond to slices of smaller $M_0$, nearer the origin: the
allowed portion of the slice may be large relative to the distance
from the origin, in which case there is no fine-tuning, or the slice
may not even intersect the allowed region, in which case those
values of $\mu/M_0$ and $M_{1/2}/M_0$ are not allowed.

We must also ensure that, within the allowed region, $\db$ is within
the range
allowed by bottom-tau unification. [To calculate $\db$ for a
particular choice of $\mu/M_0$ and $M_{1/2}/M_0$ and a given value
of $\alpha_s(m_Z)$, we first determine the range of values of
$\msb^2 + \msq^2$ within the allowed region of parameter space, and
then use Eq.~(\ref{eq:dmb}); if any resulting $\db$ is acceptable,
we allow that choice of $\mu/M_0$ and $M_{1/2}/M_0$.] In Fig.~9 we
show the values of $\mu/M_0$ and $M_{1/2}/M_0$ which lead to proper
electroweak symmetry breaking (i.e. there is a finite allowed
triangle) and acceptable bottom-tau unification, for three
representative
values of $\lG$, two values of $\alpha_s(m_Z)$, and either $\lt^G =
\lb^G$ (shaded) or $\lt^G = 2 \lb^G$ (hatched). Recall that
$M_{1/2}$ is the gaugino mass at the GUT scale (which happens to
roughly equal the wino mass: $\mwi = g_2^2/g_G^2 M_{1/2} \simeq 0.85
M_{1/2}$) and $M_0$ is the typical scalar mass also at the GUT
scale, while $\mu$ is evaluated at the electroweak scale. The
allowed regions are all roughly ``L''-shaped. At their top and on
their far right (when applicable), they are cut off by the
requirement of proper electroweak symmetry breaking, while on their
lower-left and upper-right sides they are bounded by the limits on
$\db$.

\begin{figure}[tb]
\centering
\leavevmode
\epsfysize=12cm \epsfbox[0 270 635 730]{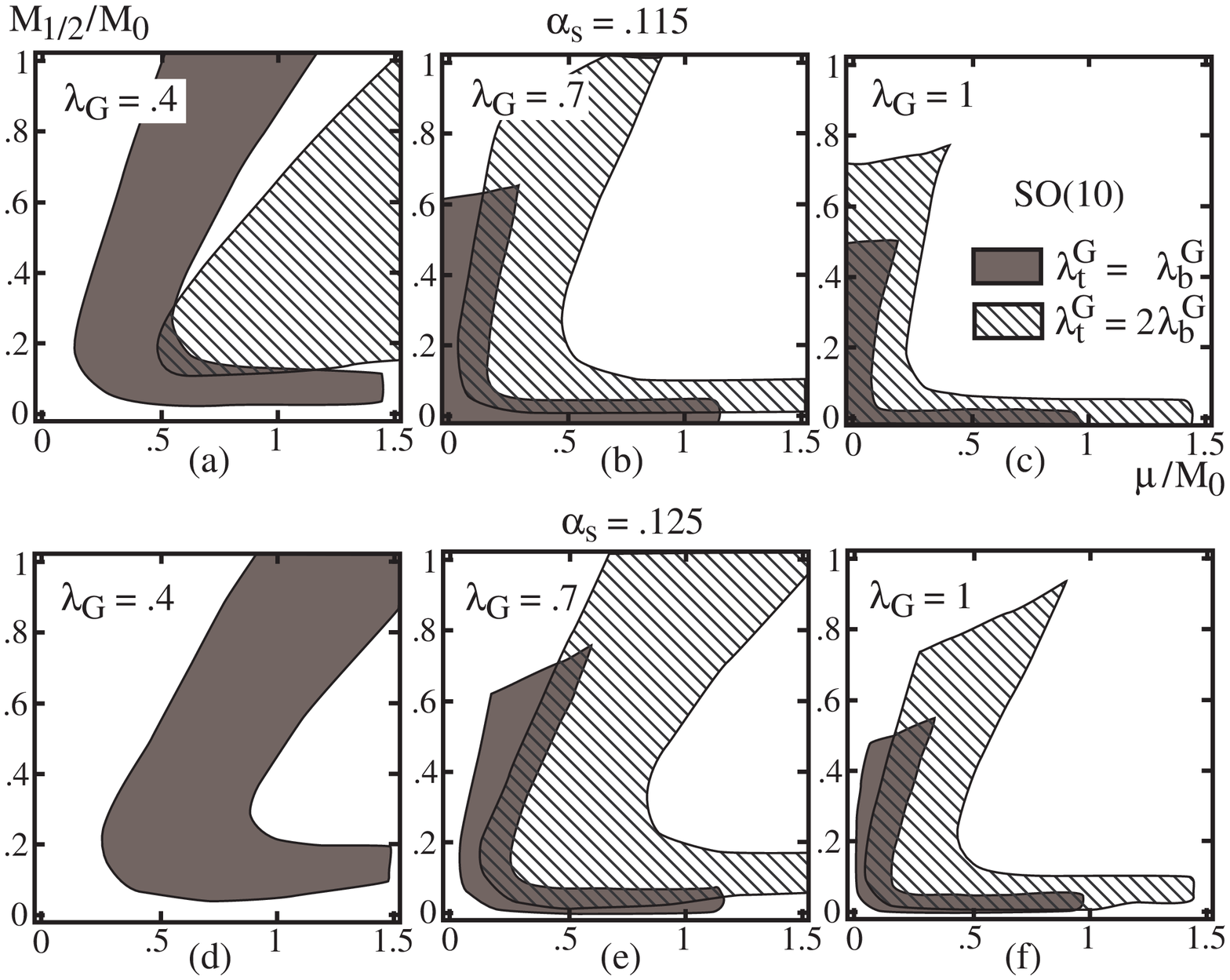}
\begin{quote}
{\small
Fig.~9. The regions of GUT-scale PQ and R breaking allowed by the
constraints of proper electroweak breaking and bottom-tau
unification. The gray areas give the ranges of $\mu/M_0$ (where
$\mu$ is actually evaluated at a scale $m_Z$) and $M_{1/2}/M_0$
assuming exact Yukawa unification, while the hatched regions assume
$\lb^G/\lt^G = 2$. All assume SO(10)-type scalar mass boundary
conditions.}
\end{quote}
\end{figure}

Particular values of $\mu/M_0$ and $M_{1/2}/M_0$ are examined in
more detail in Table 3. Once again we consider three different
values of $\lG$ and either $\lt^G = \lb^G$ or $\lt^G = 2 \lb^G$,
which lead to the six major boxes of the table. Each is divided into
four sub-boxes: the lower-left one is the symmetric case described
above; in the lower-right one we break PQ, in the upper-left one we
break R, and in the upper-right sub-box we break both the PQ and the
R symmetries. (Note that these correspond spatially to the four
corners of Fig.~9, as well as to Table 2.) For each sub-box we have
chosen an appropriate pair of $[\mu/M_0,M_{1/2}/M_0]$ values from
the allowed region of Fig.~9, and have indicated the allowed area of
the triangle for those values. When choosing these values we avoided
the boundaries of the allowed regions, because there the area is
typically very small and hence (in some sense) unlikely. Clearly
these choices are somewhat arbitrary, and there can correspondingly
be some variation in the spectrum. Notice that when the allowed area
is small in the symmetric case, raising $\mu$ or $M_{1/2}$ has the
expected effect of increasing the area (but recall the price one
pays in $\epsb\epsz$), since the focusing described above is
alleviated. When either the PQ or R symmetries are approximately
valid, it is
still necessary to focus on the light-$m_Z$ part of the triangle
(see Table 2); when both symmetries are broken, much of the triangle
is allowed by LEP and no part is selected by naturalness criteria,
so we have arbitrarily chosen to look at its center. The middle
entry in each sub-box indicates as before the typical superspectrum
for the appropriate region of the triangle. If the various squark,
slepton and pseudoscalar masses for a given sub-box are all within
roughly a factor of two, we characterize them as ``all comparable'',
and otherwise we indicate which ones are significantly lighter;
masses in parenthese are only marginally lighter (i.e. somewhat less
than half the heavier masses). The bottom entry of each box shows
once again the particles which can become light simultaneously with
the Z. If only one particle is shown, the reason is that at the
other vertex of the triangle the flat-direction mass $m_2^2$ is
negative at scales between $10^5$ and $10^7$ GeV, and hence that
vertex would lead to improper electroweak symmetry breaking.

Returning for a moment to the universal scenario, we recall that if
$\lt^G = \lb^G$ then the universal assumption is incompatible with
even approximate PQ and R symmetries. Indeed, when the Yukawa
couplings are exactly unified the universal case requires a much
bigger tuning \cite{ref:hrsII}. The reason is that when the scalar
masses are universal the only sources of custodial breaking
available for splitting $m_U^2$ from $m_D^2$ are the {\it small}
effects of hypercharge and $\ltau$. Moreover, in order to obtain
$m_U^2<m_D^2$ the gauginos must be very heavy. This can be
represented by the low energy relation $m_D^2-m_U^2= \epsc
M_{1/2}^2$, where $\epsc$ is small positive coefficient representing
the custodial breaking induced by hypercharge. Using this relation,
we see that proper symmetry breaking, i.e.~$m_U^2 < 0 < m_D^2$,
requires tuning some parameter ($\lG$, $\mu^2$ or $M_{1/2}^2$) to a
precision of $\epsc$. Moreover, since $m_A^2=m_U^2+m_D^2<\epsc
M_{1/2}^2$, and since $\mu\sim{\cal O} (M_{1/2})$ to make $m_A^2 >
0$ when the gauginos are heavy, we must tune $B$ with a precision
$\epsb \sim (\mu \mwi/m_A^2) (1/\tan\beta) \sim \epsc/\tan\beta$.
Thus the overall tuning is at least $\sim \epsc^2/\tan\beta \sim
1/\tan^2\beta$, using the rough numerical approximation $\epsc^2
\sim 1/\tan\beta$. Some more tuning is required to achieve an
acceptably small rate for $b\to s\gamma$. And finally, with the
large $\db$ corrections that result from such a spectrum, the top
mass is quite light and is therefore in conflict with the recent
data on $m_t$ \cite{ref:cdf}.

\subsection{SU(5)-type GUT Masses}
\label{sec:sufive}
We can repeat the above analysis for SU(5)-type boundary
conditions, that is, when the soft scalar masses need only be
SU(5)-symmetric at the GUT scale. There are now four rather than
three independent initial masses, and we will choose them to be
$M_0^2$ [$ = \third \mtenh^2 + \twothr \msixth^2 = \sixth(\mfivh^2 +
\mfivbh^2 + \mfivth^2) + \half \mtenth^2$], $\mx^2$ [$\equiv
\fivefour(\mfivbh^2-\mfivh^2) + \fivefour(\mtenth^2-\mfivth^2)$], $2
\mtenh^2 - 3 \msixth^2$ [$=\mfivh^2 + \mfivbh^2 - \ninefour\mtenth^2
- \thrfour\mfivth^2$], and $\msufive^2$ [$ \equiv
\half(\mfivh^2-\mfivbh^2) + \half(\mtenth^2-\mfivth^2)$].  The
allowed region in the 3-dimensional projected space of initial
scalar mass parameters is now a volume bounded by planes, which in
many cases is a tetrahedron [corresponding to the allowed triangle
in SO(10)]. When the PQ and R symmetries hold, we find as for SO(10)
boundary conditions that the asymptotically focused case
$\lG\to\infty$ cannot be reached, since the four equations
$I_{1,2,3,4}=0$ have no nontrivial solutions. Hence the allowed
tetrahedron closes for a finite $\lG\simeq 1.7$, which can be seen
by considering the evolution of the combination $m_Z^2 +
{53\over34}(m_A^2 + \msb^2) + {23\over34}\mstau^2 = (-14 X_t + 39
X_b + 15 X_\tau)/34 - 6/34 M_0^2$ (neglecting the small hypercharge
D-term).

Table 4 summarizes the consequences of imposing SU(5)-type boundary
conditions on the soft mass parameters for various choices of $\lG$
and $\lt^G/\lb^G$, and for unbroken or broken PQ and R symmetries,
in analogy with Table 3. This time it is the allowed volume, rather
than the area, which is shown on the first line of each sub-box;
also shown are the sampled values of [$\mu/M_0$,$M_{1/2}/M_0$]. This
volume is also plotted, as a function of $\lG$ and $\lt^G/\lb^G$, in
Fig.~8(b). The
choice of coordinates in the 3-dimensional initial parameter space
was as usual a matter of taste, so there is no objective way of
comparing the allowed volumes in SU(5) with the allowed areas of
SO(10). Qualitatively, however, it seems clear from Table 4 and
Fig.~8 that SU(5)-type boundary conditions require less tuning,
mainly because the additional degree of freedom $\msufive^2$ makes
the Higgs splitting independent of the squark masses and hence more
easily allows $m_U^2 < m_D^2$ without lowering $m_{{\tilde
b},{\widetilde L}}^2$. The middle entry in each sub-box
describes the typical spectrum near the $m_Z^2 = 0$ face of the
allowed volume, except for cases of broken PQ {\it and} R in which
the spectrum is shown for a generic central point in the
tetrahedron. The bottom entry shows which masses are allowed to
vanish simultaneously with $m_Z$; in general there are corners of
the allowed region in which two of these may vanish along with
$m_Z$, but which pairs may do so varies from case to case. Finally,
notice that in some cases, such as $\lG = 1$ and $M_{1/2} = 0$, the
SU(5) entry shows that only $m_A$ and $\mstau$ may be light along
with $m_Z$, whereas the SO(10) entry indicates that only $m_A$ and
$\msb$ may do so. The reason is that SU(5) allows the larger value
$\mu/M_0 = 1$ indicated in that entry, for which indeed the sbottom
cannot be made light but the stau can; at the smaller value $\mu/M_0
= .5$ the SU(5) boundary conditions must and do allow the sbottom to
be light, since they contain the SO(10) boundary conditions as a
special case.

\begin{figure}[tb]
\centering
\leavevmode
\epsfysize=11.5cm \epsfbox[30 350 645 730]{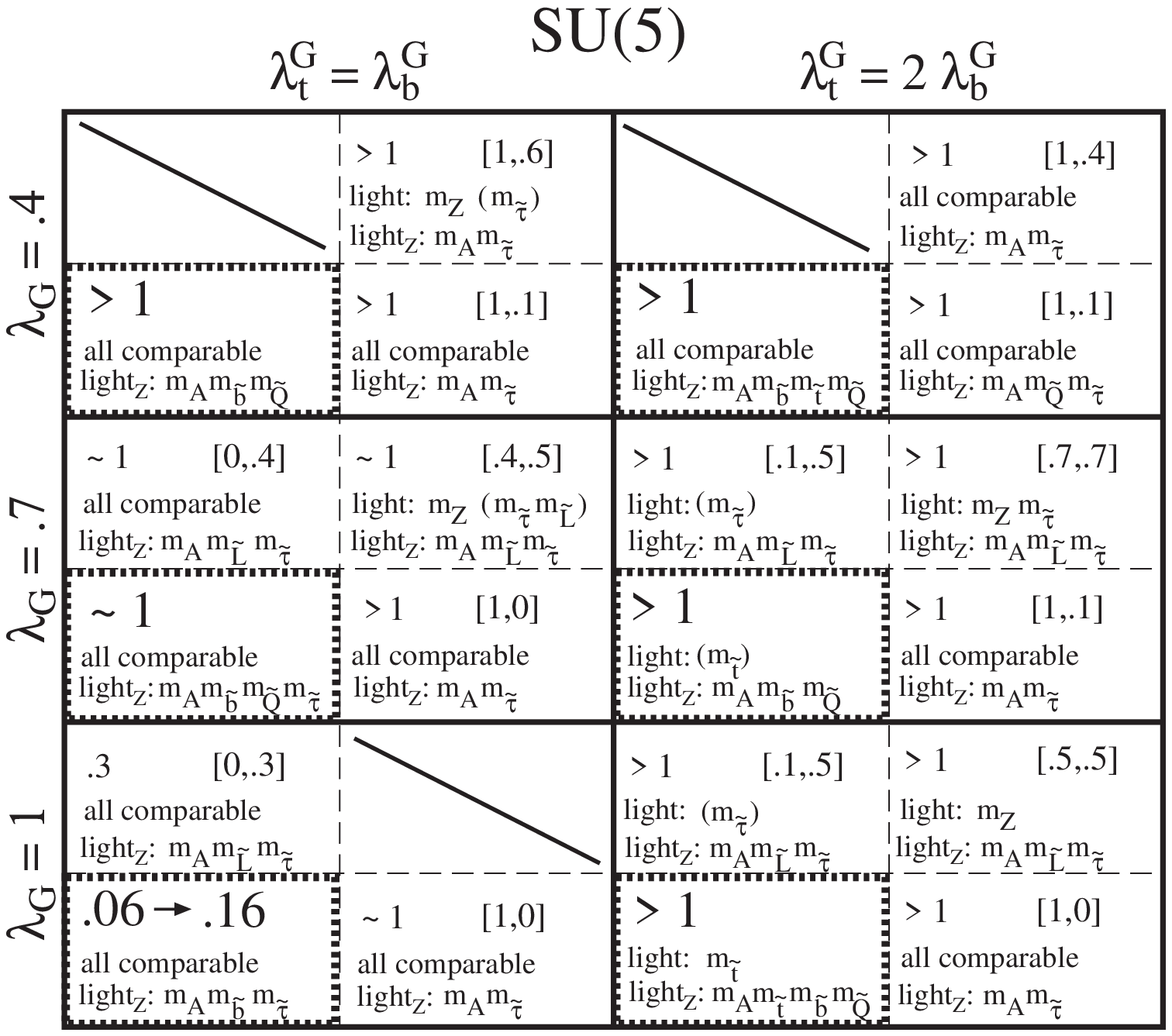}
\begin{quote}
{\small
Table 4. The characteristics of Yukawa unification with SU(5)-type
scalar mass boundary conditions, in analogy with Table 3.}
\end{quote}
\end{figure}

Recall that in the PQ- and R-symmetric SO(10) analysis, in which
$\msufive^2 = 0$, we were interested in the dependence of the
allowed area on $\lG$ and $\lt^G/\lb^G$, as depicted in Fig.~8(a).
The analogous contour plot of the allowed SU(5) volume is shown in
Fig.~8(b). One
could also ask for the volume allowed when the SO(10) scalar mass
boundary conditions are perturbed in the SU(5) direction to the same
extent that the SO(10) Yukawa coupling boundary conditions are
relaxed: namely, restrict $|\msufive^2/M_0^2| \roughly{<}
\lt^G/\lb^G - 1$ (and then normalize the volume by dividing it by
$\lt^G/\lb^G - 1$).
The answer is very simple: up to a normalization factor of order
unity (due to the arbitrary definition of unit area and unit
volume), the contour plot of this restricted allowed volume is
similar to that of the allowed area in pure SO(10). We learn that,
as we saw in the particular example of a light right-handed
neutrino, small scalar mass splittings and
small Yukawa coupling splittings affect the allowed area to a
similar extent.

Fig.~10 shows to what extent the PQ and R symmetries may be broken.
As in Fig.~8, we outline the ranges of $\mu/M_0$ and $M_{1/2}/M_0$
which lead to a nonvanishing allowed volume in the space of initial
scalar masses and to acceptable bottom-tau unification, for
different choices of $\lG$, $\lt^G/\lb^G$ and $\alpha_s(m_Z)$.
The permissible ranges extend to small $\mu$ and
$M_{1/2}$ even when a fairly large $\db$ is required, because within
the
SU(5)-allowed volume one can find corners where $\msq^2 \sim \msb^2
\sim m_Z^2 \sim \mu^2 \sim M_{1/2}^2 \ll M_0^2$ so $\db$ is quite
large.

\begin{figure}[tb]
\centering
\leavevmode
\epsfysize=12cm \epsfbox[0 270 635 730]{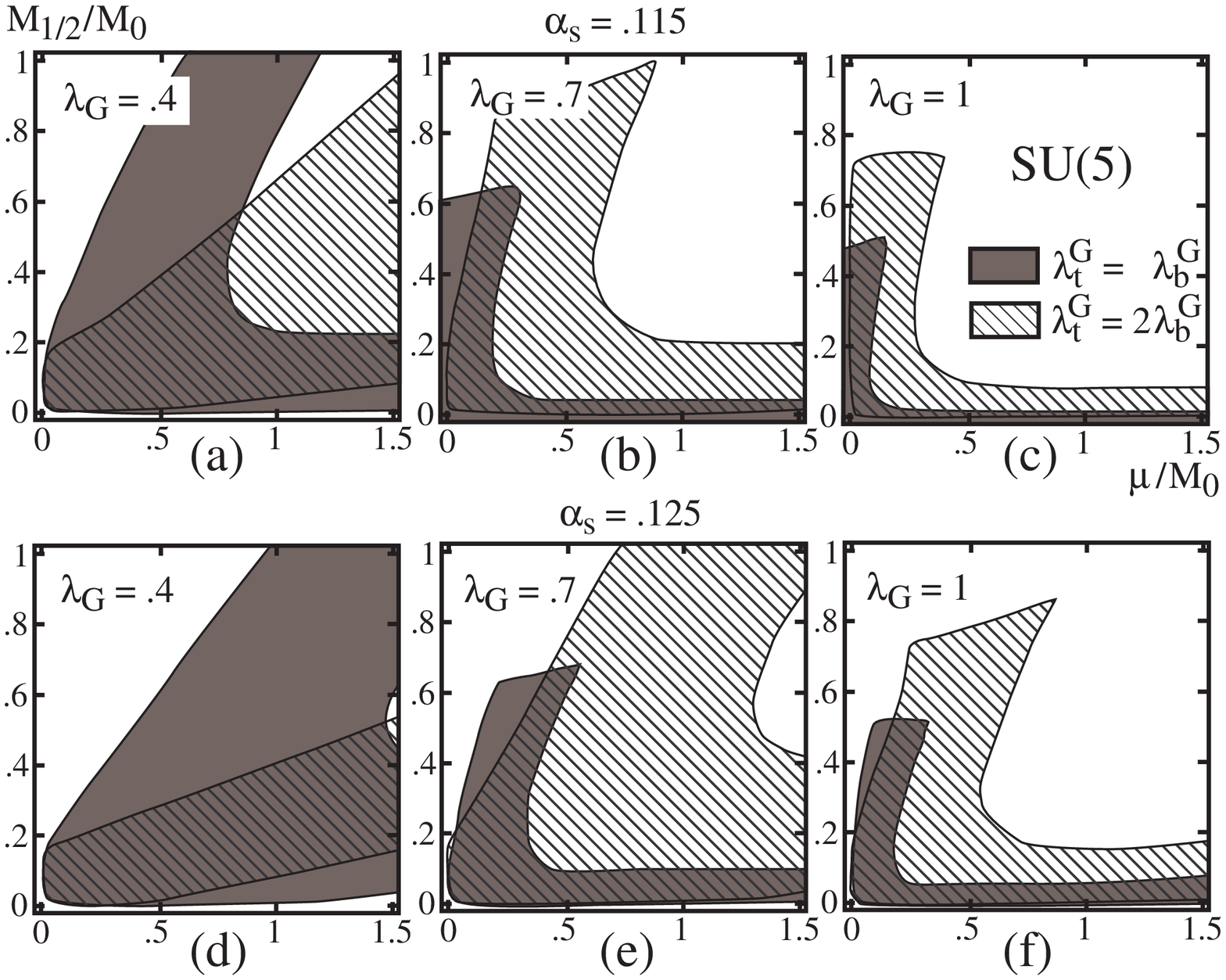}
\begin{quote}
{\small
Fig.~10. The allowed regions of PQ and R breaking at the GUT scale,
in analogy with Fig.~7.}
\end{quote}
\end{figure}

{Cosmological Bounds}
\label{sec:cosmo}
The stability of the lightest supersymmetric particle (LSP, which we
denote by $\chi$) can lead to serious cosmological bounds on the
parameters of the MSSM. In this section we discuss these bounds in
the large $\tan\beta$
scenario.

As a first constraint, the LSP has to be  both
electrically and color neutral \cite{ref:wolfram}, otherwise it
would have been found in searches for exotic isotopes.
This is typically not a problem for us.
As we have seen, in the most interesting large $\tan\beta$
scenarios, either $\mu$ or  $M_{1/2}$ or both are
considerably smaller than all the other SUSY parameters. In these
regions of parameter space the LSP can only be a neutralino or a
chargino. By a numerical study of the $4\times 4 $ neutralino mass
matrix, we find that for $\tan\beta\gg 1$ and with the LEP bounds
on $\mu$ and $M_{1/2}$ the LSP is always a neutralino.
In the limit $|\mu|+|M_{1/2}|\gg m_Z$, this property can be easily
checked by performing a perturbative diagonalization of the mass
matrix.

The second constraint arises from the LSP
relic mass density $\rho_\chi$, which must not exceed the critical
density of the universe today $\rho_c = (1.88\times10^{-29} {\rm g\,
cm^{-3}}) h^2$. We devote the rest of this section to this issue. We
base our discussion on Ref.~\cite{ref:relic}, where the lightest
neutralino relic density was studied but without an emphasis on the
parameter regions discussed in our paper. Recently the LSP abundance
in the large $\tan\beta$ scenario was also partially discussed in
Ref.
\cite{ref:pocolsp}, and where our analyses overlap there is
qualitative agreement.

The contribution of $\chi$ to the present $\Omega h^2$ (where
$\Omega \equiv
\rho/\rho_c$) is determined by how fast the LSP annihilates when it
is non-relativistic. In practice, $\Omega_\chi h^2$ is roughly
inversely proportional to the annihilation cross section
$\sigma_{\chi \chi}$ at a freeze-out temperature $T_F\sim m_\chi/20$
\cite{ref:lee}.
In our case the composition of $\chi$ and its annihilation
properties depend crucially on $\mu$ and $M_{1/2}$.
Thus the PQ and R symmetries provide once again the right
language for classifying the different situations.

Let us first consider the approximately PQ-symmetric and
R-asymmetric scenario
$m_Z \sim \mu \ll M_{1/2}$. In this case  the LSP is predominantly a
Higgsino. For $m_\chi>m_W$ the annihilation into W pairs proceeds
with full gauge strength via t-channel chargino exchange. The rate
is easily sufficient to ensure that $\Omega_\chi h^2 \ll 1$. For
$m_\chi<m_W$, one has to
rely on annihilation into standard-model fermion pairs  via
s-channel vector boson exchange.  Now the strength of the $\chi$
couplings to the Z plays a role, as does coannihilation both with
the
second-lightest neutralino $\chi'$ and with the lightest chargino
$\chi^+$ (in this case via s-channel W exchange). Define
$\tilde h_\pm^0=\tilde h_u^0\pm \tilde h_d^0$, where $\tilde
h_{u,d}^0$ are the neutral components of the Higgsino doublets. The
mass-eigenstate fields for the LSP $\chi$ and the second-lightest
neutralino $\chi'$ are given by
\be
\chi= \tilde h_+ + {\cal O}\left({m_Z^2\over \mu
M_{1/2}}\right)\tilde h_- \,,\quad \quad \chi' =\tilde h_-+{\cal
O}\left({m_Z^2\over \mu M_{1/2}}\right)\tilde
h_+\,.
\label{eq:pqcase}
\ee
(The plus and minus signs obviously depend on our conventions.) The
isospin quantum numbers of $\tilde h_u^0$ and $\tilde h_d^0$ are
such that the vertices $Z\tilde h_+\tilde h_+$ and $Z\tilde
h_-\tilde h_-$
vanish, while $Z\tilde h_+\tilde h_-$ has full gauge strength. For
large $M_{1/2}$, Eq.~(\ref{eq:pqcase}) implies that the
$\chi\chi Z$ vertex is suppressed by ${\cal O}\left(m_Z^2/\mu
M_{1/2}\right) \sim {\cal O}\left(m_Z/M_{1/2}\right)$ relative to
the $\chi\chi' Z$ vertex. A similar discussion applies to the
coupling
$\chi \chi^+ W$, which is not suppressed in this limit.
Furthermore, the splittings $ m_\chi-m_{\chi'}$ and $m_\chi
-m_{\chi^+}$ vanish like $m_Z^2/M_{1/2}$, so  $\chi$,
$\chi'$ and $\chi^+$ are all present just before the LSP freezes
out, and
coannihilation is important \cite{ref:griest}.
Thus for ultraheavy gauginos the self-annihilation rate
$\sigma({\chi\chi\to f\bar f})$ is negligible but coannihilations
 $\sigma({\chi\chi'\to f \bar f})$ and
 $\sigma({\chi\chi^+\to \nu \bar e, \dots})$
are important, since the mass splittings are very small.
The result\footnote{For quantitative estimates we used the
formulae of Ref.~\cite{ref:relic}, where only the effect of
coannihilation with $\chi'$ was included. We expect that accounting
also for coannihilation in the charged channel \cite{ref:mizuta}
will lower the final value of $\Omega_\chi$, and therefore
strengthens our conclusion that there the relic LSP abundance is
sufficiently small.} is that $\Omega_\chi h^2 \ll 1$
\cite{ref:relic}.  As $M_{1/2}$ is lowered below $\sim 400 \,\GeV$,
the $Z\chi\chi$ vertex becomes important and self-annihilation
becomes dominant, leading once again to $\Omega_\chi h^2 \ll 1$.
At intermediate values of the gaugino mass, $\Omega_\chi h^2$
reaches a maximum $\Omega_{\rm max} \sim 10\percent$.
So the PQ-symmetric case does not suffer from an
overdensity of LSP's.

As we lower $M_{1/2}$ down to $\sim m_Z$ we recover the maximally
symmetric case. Now the LSP contains sizeable portions of both
$\tilde h_+$ and $\tilde h_-$, and the $Z\chi\chi$ vertex has
essentially
full gauge strength. As a result $\Omega_\chi h^2$ is always well
below 1.

Finally, let us study the effects of raising $\mu$ to arrive at an
R-symmetric and PQ-asymmetric scenario $m_Z \sim M_{1/2} \ll \mu$.
Now the LSP is predominantly a bino (the hypercharge gaugino):
$\chi\sim\tilde B$. We will collectively denote the squark and
slepton masses and also $m_A$ by the single mass parameter $m_0$. In
the limit $\mu\sim m_0\to \infty$ the LSP is totally decoupled and
$\Omega_\chi h^2$ is extremely large: all cross sections for bino
annihilation vanish at least like $m_\chi^2/m_0^4$ or like
$1/(\mu^2\tan^2\beta)$. Clearly this poses a potential problem for
this scenario. For a quantitative study of the possible annihilation
rates, it is
useful to integrate out the heavy fields and obtain an
effective Lagrangian ${\cal L}_{\chi\chi}$ for the bino $\tilde B$
and the SU(2) gauginos $\wi_I$, $I=1,2,3$. We write ${\cal
L}_{\chi\chi} = \L{1}+\L{2}+\L{3}+\L{4}$ in two-component notation
as
\bea
{\cal L}_1 &=&{g'^2\over 2
\mu}\bi H_u^\dagger\left(1-{{\not \!\!D}\overline{\not \!\!D}\over\mu^2}+
\dots\right )\bi\times \\
&\phantom{x}& \qquad\left(1-{D^2\over m_A^2}+\dots\right) \left[
{1\over\tan\beta} H_u+ {1\over m_A^2} \left(\lb \overline b b +
\ltau\overline\tau\tau\right)\right] + {\rm h.c.}\\
{\cal L}_2 &=& {g'^2\over 2\mu^2}\bi H_u^\dagger i{\not \!\!D}\left (1-
{\overline{\not \!\!D}{\not \!\!D}\over \mu^2}+\dots\right)
\overline{\bi} H_u\\
{\cal L}_3 &=& {g'^2\over 2}\sum_f {Y_f^2\over m_{\tilde
f}^2}\overline\bi\overline f
\left (1-{D^2\over m_{\tilde f}^2}+\dots\right)\bi f\\
{\cal L}_4 &=& {gg'\over 2\mu^2}\bi H_u^\dagger i{\not \!\!D}\left (1-
{\overline{\not \!\!D}{\not \!\!D}\over \mu^2}+\dots\right)
\overline {\wi} H_u \,+\,{\rm h.c.}
\label{eq:leff}
\eea
where $D$ is the covariant derivative acting to the right, $Y_f$ and
$m_{\tilde f}$ are respectively the $\tilde f$ sfermion's
hypercharge and mass, $\wi=\wi_I\tau_I$ and $\tau_I$ are the Pauli
matrices. ${\cal L}_{1,2,4}$ are obtained by integrating out the
Higgsinos and the $H_d$ Higgs doublet, while ${\cal L}_3$ arises
from integrating out the sfermions. The dots represent higher
derivative terms. We have approximated $m_{H_d}^2\simeq m_A^2$.
Notice that $\L{4}$ introduces also an ${\cal O}(1/\mu^2)$ overlap
between the LSP and the $\wi_3$. The overall contribution of $\L{4}$
to the LSP annihilation amplitudes is ${\cal O}(1/\mu^4)$. In fact
the amplitude at ${\cal O}(1/\mu^4)$ gets contributions also from
virtual bino and wino diagrams.

Let us start from the PQ- and R-symmetric scenario and increase
$\mu$ above $\sim m_Z$. Then, for intermediate values of $\mu$,
the LSP annihilation cross section is determined by ${\cal L}_2$
and ${\cal L}_4$, since all other terms above are either suppressed
by a large scalar mass ($m_A^2$, $ m_{\tilde f}^2$) or by $1/\tan
\beta$. Focusing on ${\cal L}_2$ and ${\cal L}_4$, we notice that
the leading   ${\cal O}(1/\mu^2)$ contribution to the amplitude is only
in the p-wave. This contribution is determined by the first term in
${\cal L}_2$, and its p-wave character is easily seen  by using the
equations of motion in the limit of massless fermions (It is
accurate to neglect fermion masses since we
suppose $\chi$ to be below the top threshold in the R-symmetric
scenario.)
This fact is of some importance since the LSP's freeze out in the
non-relativistic regime in which p-wave
cross sections suffer a further suppression $\sim T_F/m_\chi$. For
$m_\chi<m_W$, the LSP's annihilate into $f\bar f$ pairs via
Z-boson exchange, at leading order from the first term in $\L{2}$.
We find that the relic density $\Omega_\chi h^2$ exceeds 1 already
for $\mu > 200-300\,\GeV$ (depending on how close $m_\chi$ is to
$\half m_Z$). The situation does not improve when $m_\chi > m_{W,Z}$
and all the bosonic channels WW, ZZ, Zh and hh are open and
dominant. There are two contributions to these channels: a
p-wave term $\sigma_p\sim 1/\mu^4$ from the first term in $\L{2}$,
and an s-wave term $\sigma_s\sim 1/\mu^8$ from the second term in
$\L{2}$ and from virtual gaugino diagrams involving the first terms
in $\L{2}$ and  $\L{4}$. (As already stated we use $\L{2}$ to
explicitly
display the suppression of s-wave processes, a fact also observed
in Ref.~\cite{ref:relic}. For quantitative estimates, we calculated
$\sigma_s$ by returning to the partial wave amplitudes given in
Ref.~\cite{ref:relic}.) In the limit $m_\chi\gg m_Z$, $\sigma_p$ is
dominated by final states with zero helicity (longitudinal vector
bosons and scalars).
The amplitude in this case is readily given by the annihilation into
Goldstone and Higgs bosons from the first term in $\L{2}$ via the
equivalence theorem.
For the purpose of our qualitative discussion we only kept this
leading term. (We expect the complete result not to be
drastically different in the region $m_\chi\sim m_Z$; in fact we
checked this explicitly for $\chi\chi\to WW$ by using
the formulae in Ref.~\cite{ref:relic}.) Using these estimates for
$\sigma_{s,p}$ we find, again, that $\Omega_\chi h^2>1$ for
$\mu\roughly{>} 250 \GeV$. We are thus led to the interesting
conclusion that we cannot
essentially move away from the PQ- and R-symmetric scenario towards
the PQ-asymmetric one, if all the other superpartners and the
pseudoscalar
Higgs are very heavy. Nonetheless the moderate $\mu\sim 200-300$ GeV
scenario is interesting, since in this case the LSP could account
for
all the dark matter and give $\Omega=1$.

When $\mu$ exceeds $\sim 300\,\GeV$ some other
particle (namely the pseudoscalar Higgs or some sfermion) has to be
lighter in order to avoid an overdensity of LSP's. A quick
inspection of the above effective Lagrangian shows that, by lowering
$m_A^2$ or some $m_{\tilde f}^2$, only the annihilation into $f\bar
f$ can be significantly affected [the amplitude into bosons from
$\L{1}$ is ${\cal O}(1/\tan\beta)$].
This process can be mediated by (I) t-channel $\tilde f$  exchange
via $\L{3}$, or (II) s-channel pseudoscalar Higgs exchange via
$\L{1}$. In case (I) the amplitude is p-wave and the effective
vertex is $\sim 1/m_{\tilde f}^2$. Supposing a sfermion $\tilde f$
is relatively light, we get the following estimate for the bino-like
LSP
relic density:
\be
\Omega_\chi h^2\simeq \left ({100\,\GeV\over m_\chi}\right )^2
\left ({m_{\tilde f}\over {r_f^{1/4}Y_f 100 \,\GeV}}\right )^4
\left ({x_F\over {25}}\right )^2
\label{eq:chitau}
\ee
where $Y_f$ is the fermion hypercharge (so $Y_Q = 1/3$) and $r_f$ is
the dimension of the $f$ multiplet (so $r_Q=6$). Thus, leaving all
the other fine tunings untouched, an acceptable $\Omega$ requires an
additional
tuning $m_{\tilde f}^2$ to at least an order of magnitude below its
natural scale. The amplitude for case (II) behaves
like $1/\mu m_A^2$, but is in the s-wave. In this case only the
final states $\bar b b $ and $\bar \tau \tau$ are relevant (the
lighter fermions are suppressed by the small Yukawa couplings).
We estimate:
\be
\Omega_\chi h^2\simeq \left
({{100\GeV}\over m_\chi}\right )^2
\left({\mu \over {350\,\GeV}}\right)^2
\left({m_A\over {350\,\GeV}}\right )^4\left ({x_F\over25}\right)\,.
\label{eq:chibbar}
\ee
(We are being a little sloppy in the above equations, by neglecting
$m_\chi^2$ terms in the sfermion and pseudoscalar Higgs propagators,
but the conclusions would not be changed much by a more careful
computation) Notice that even though the above $\Omega$ scales like
$m_0^6$,
the result is comparable to that in Eq.~(\ref{eq:chitau}). This is
partially due to the s-wave enhancement. We see once again that if
$\mu$ is increased from its minimally-allowed value, then some mass
parameter ($m_A^2$ in this case) must be made light to meet
cosmological bounds. Typically, such a requirement entails some
further tuning of some GUT-scale parameters.

Which annihilation channel is more likely? Assuming SO(10) boundary
conditions on the scalar masses, Table 3 shows that the only
particles that can be made very light (i.e. comparable to $m_Z$) are
the pseudoscalar Higgs and the sbottom, while for SU(5)-type
boundary conditions Table 4 adds the stau to this list. Hence
efficient fermion production through sfermion exchange [case (I)]
requires making $\msb\sim 100\,\GeV$ or $\mstau\sim 200\,\GeV$
(under
SU(5)-type conditions). Efficient fermion production through
pseudoscalar Higgs exchange [case (II)] need not require $m_A$ to be
quite as light (depending on $\mu$), but recall that lowering $m_A$
also increases the need to tune $B$ in order to generate large
$\tan\beta$. So both channel are roughly equally unlikely. There is,
of course, the possibility that the first- or second-generation
sfermions are light---their initial values need not be related to
those of the third generation, and their evolution is essentially
decoupled from the third generation and Higgs sectors---in which
case they could remedy the difficulties with the R-symmetric,
PQ-asymmetric scenario.

We conclude that whenever $\mu$ is small, whether the gauginos are
light or heavy, the cosmological density of the LSP is well below
critical. In the large $\mu$ but light gaugino case, LSP
annihilation is unacceptably suppressed if all the other
superpartner and pseudoscalar Higgs masses are large. The
annihilation rate can be sufficient if some of those masses are
lowered, either through fine-tuning of $m_A$, $\msb$ or $\mstau$, or
perhaps by appealing to the yet-unspecified first two generations of
squarks and sleptons. Needless to say, the course nature has chosen
will be definitively revealed by future measurements of the
superspectrum.

\section{Conclusions}
\label{sec:conc}
In this paper we have studied some of the consequences of large
third-generation Yukawa couplings in the minimal supersymmetric
extension of the standard model, subject to various grand
unification assumptions. We have focused our attention on the two
cases $\ltau^G = \lb^G = \lt^G$ and $\ltau^G = \lb^G \sim \lt^G$
imposed at the GUT scale $\sim 10^{16}\,\GeV$, but most of the
conclusions are qualitatively unchanged if these conditions are
instead enforced at the Planck or string scales, and a few general
features remain even if one only assumes that $\ltau\sim\lb\sim\lt$
at some very high scale. For example, the need to tune some
parameters to at least one part in fifty ($\sim m_t/m_b$) is a
generic consequence of LEP bounds and the structure of the MSSM
Lagrangian. This is also perhaps the most bothersome conclusion: as
long as the top and bottom Yukawa couplings start out comparable at
the GUT scale, there is no way to both explain the top-bottom mass
hierarchy naturally (in the technical sense)
and avoid tuning $m_Z^2 \ll \ms^2$ . But if this one bitter pill is
grudgingly swallowed, the remaining features of the large
$\tan\beta$ scenario are intriguing. Let us assume for the moment
that there is no theoretical bias about physics at the GUT scale
other than the existence of a GUT. Then, if we ask that an
$\CC$-invariant vacuum as well as the experimental rate of $b\to
s\gamma$ are to be typical rather than unlikely outcomes of the
GUT-scale parameters, and furthermore that the prediction for
$m_b/m_\tau$ agree with its experimental value as extracted using
QCD sum rules, then:
\begin{enumerate}
\item[(I)] the Lagrangian at the GUT scale should display
approximate PQ and R
symmetries;
\item[(II)] the value of the unified Yukawa coupling should either
be $\lG \simeq 0.6-0.7$ if the Yukawas are exactly unified ($\lb^G =
\lt^G \equiv \lG$), or be $\lG \roughly{>} 0.8$ if the
Yukawas are significantly split (for example $2 \lb^G = \lt^G \equiv
\lG$);
\item[(III)] if the Yukawas are exactly unified, the soft
SUSY-breaking scalar masses at the GUT scale should be SU(5)- but
not SO(10)-type, while if the Yukawas are split then they can be of
either type; and
\item[(IV)] threshold corrections to the $\rm SU(3)_c$ gauge
coupling at the
GUT scale must be significant and negative relative to the $\rm
SU(2)\times U(1)_Y$ couplings.
\end{enumerate}
All of these features have phenomenological, testable consequences.
We expect:
\begin{enumerate}
\item[(I)] light charginos and neutralinos, which may furnish
tantalizing signals at LEP II and would definitely be seen at the
LHC; and large masses for the squarks and sleptons (at least of the
third generation), the pseudoscalar Higgs and the charged Higgs
bosons [see Eq.~(\ref{eq:symspec})];
\item[(II)] either a top mass between 160 and 170 GeV if the Yukawas
are exactly unified, or $m_t \roughly{>} 175\,\GeV$ if they are
significantly split at the GUT scale; the Yukawa splitting at
$\Mgut$ is reflected in the superspectrum, for example in a light
stop, or better yet in the mass combination $(X_t-X_b)/(X_t+X_b)$
defined using Appendix A, which is very sensitive to both
$\lt^G/\lb^G$ splitting and to the departure from SO(10)
boundary conditions on the soft scalar masses (see the next point);
\item[(III)] a large value for the (almost) invariant combination
$I_4 = \mu_Z^2 + \mu_A^2 + {3\over2} \mst^2 - {3\over2} \msq^2 -
{1\over2} \mstau^2 + {1\over2} \msl^2$, at least if the Yukawas are
exactly unified, since SO(10)-type boundary conditions have
vanishing $\msufive^2 = - I_4$; and
\item[(IV)] a value of $\alpha_s(m_Z) \simeq .115$ somewhat below
the central gauge unification prediction.
\end{enumerate}
In particular, if feature (I) is actually borne out by future
discoveries, that is, if $m_Z^2 \ll \ms^2$ (where $\ms$ is a typical
soft-breaking scalar mass) is transformed from an unnatural
assumption to simply an experimental fact, then the large
$\tan\beta$ scenario is as natural as the small $\tan\beta$
conventional one. The two offer very different explanations of the
top-bottom mass hierarchy. But the large $\tan\beta$ scenario offers
a more robust test of the bottom-tau unification hypothesis, since
if $\lb$ and $\ltau$ are ${\cal O}(1)$ at the GUT scale they should
be much less subject to perturbations by other operators. In other
words, the conventional scenario suffers from uncertainties in
bottom-tau unification from physics at remote scales, while with
large $\tan\beta$ the uncertainties are at low energies and hence
are imminently accessible. As a result, only for large $\tan\beta$
is there a tight relationship between $m_t$, $m_b$, $m_\tau$ and the
superspectrum.

It is important to note that some of the above predictions strongly
depend on the allowed range for the bottom mass. The uncertainty in
this mass is dominated by our estimate of the theoretical error in
the QCD sum rule extraction. If, for example, we would know that
$m_b(m_b) < 4.15\,\GeV$, then the top mass would be at least 170 GeV
for exact Yukawa
unification or at least 180 GeV for split Yukawas, and the former
would be disfavored because of the tuning mandated by its large
$\lG$.  The upper bound on $\alpha_s(m_Z)$ would also be
strengthened, and would rely less on fine-tuning arguments. Of
course, the experimental uncertainty on $\alpha_s(m_Z)$ must be
reduced to convincingly test these predictions. But it is
conceivable that in the next decade we will know the superspectrum
well enough to calculate $\db$ and the logarithmic threshold
corrections; if we can also extract the bottom mass to ${\cal
O}(\alpha_s^2)$ and measure the top mass to within a GeV, then
precision tests of Yukawa unification at the GUT scale would be
within our reach.

After this work was essentially completed, the CDF and $\rm D\!\!{\not 0}$
collaborations at Fermilab announced the long-awaited discovery of
the top quark \cite{ref:cdf}. The PQ- and R-symmetric scenario we
have advocated predicts a top mass that agrees very well with the
values determined by these experiments: whereas we predict $160
\,\GeV < m_t \,\roughly{<}\, 190 \,\GeV$ (using the approximate
fixed-point value as an upper bound), CDF measures $m_t = 176 \pm8
\pm10\,\GeV$, while $\rm D\!\!{\not 0}$ makes the less precise
determination $m_t =
199^{\,+19}_{\,-21} \pm 22\,\GeV$. The argument can of course be
reversed: the measured value of the top mass lends further support
to a PQ- and R-symmetric Lagrangian. As the uncertainty in $m_t$ is
reduced, it will serve as an increasingly powerful test to
distinguish the various scenarios we have considered.

How did we arrive at the most likely set of parameters? As sketched
in Table 2, if the hierarchy $\vU \gg \vD$ and the suppression of
$\Gbsg$ are to be obtained most naturally, the PQ and R symmetries
should be approximately valid but without making the typical
superpartners too heavy nor the charginos and neutralinos too light:
the most desirable superspectrum hierarchy is $\mu/m_0 \sim \mwi/m_0
\sim 1/7$. (In highly-focused situations, such as SO(10) with $\lt^G
= \lb^G \roughly{>} 1$, there can be two mass hierarchies, but since
such situations are always more fine-tuned they are not presently
relevant.) From Fig.~3 and Table 1, we learn that the resulting
value of $\db$ ($\sim 5\percent$) is compatible with bottom-tau
unification only if: (1a) $\lG \roughly{>} 0.6$ and $m_t \roughly{>}
160\,\GeV$ for $\alpha_s \simeq 0.115$ or (1b) $\lG \roughly{>} 1$
and $m_t
\roughly{>} 180\,\GeV$ for $\alpha_s \simeq 0.125$, if the Yukawas
are
exactly unified; or (2) $\lG \roughly{>} 0.8$, $m_t \roughly{>}
175\,\GeV$ and $\alpha_s \sim 0.115$ if the Yukawas are
significantly split at the GUT scale. Then turning to Fig.~8, we
conclude that case (1a) can be saved from further tuning by allowing
SU(5) scalar mass boundary conditions and keeping $\lG$ below
roughly 0.7, but case (1b) would always require large tuning because
of its large $\lG$. The approximately-unified case (2), on the other
hand, can be naturally obtained by either SU(5)- or SO(10)-type
scalar mass boundary conditions. Scenarios (1a) and (2) are
therefore the two we have proposed as the most likely in the absence
of more specific model-building biases.

What GUT models would yield these preferred scenarios? Our original
motivation for studying unified Yukawa couplings was provided by the
minimal SO(10) scenario, in which both light Higgs doublets lie in
the same ${\bf 10}_H$ multiplet. This is the case, for example, in
the simplest implementation of the Dimopoulos-Wilczek missing VEV
mechanism \cite{ref:dimwil} for solving the doublet-triplet
splitting problem. In such models, the soft SUSY-breaking parameters
which remain after integrating out the heavy GUT sector can be
rather constrained. This is indeed the case when SUSY breaking is
communicated to the GUT sector only via gravitational interactions
with a hidden sector. Consequently, the structure of the soft terms
it tightly linked (see for example Ref.~\cite{ref:hito}) to the GUT
superpotential. It can then be shown that the only source of Higgs
splitting for minimal missing VEV models is
the D-terms, so the SO(10)-type scalar mass boundary conditions
hold. Therefore, to allow the freedom of SU(5)-type boundary
conditions favored in scenario (1a) above while preserving the
unified Yukawa relations $\lt^G = \lb^G = \ltau^G$, more general
soft terms would be required. With such terms, as may be produced
when there is moduli field  dependence of the GUT superpotential
couplings \cite{ref:louis}, it may be possible to induce additional,
F-type splittings between the MSSM particles. If scenario (1a)
were supported experimentally, it would thus shed some light on the
mechanism which breaks supersymmetry. On the other hand, when the
Yukawa couplings are split at the GUT scale, as in the second
favored scenario above, the tuning can always be made minimal by
using the PQ and R symmetries. Split Yukawas would be completely
expected in SU(5) models or as a consequence of string theory, but
they could also arise in SO(10) models when the light Higgs doublets
originate in several SO(10) multiplets. Note that even with
universal GUT scalar masses the fine-tuning can be minimal (that is,
$\sim 1/\tan\beta$) if $\lt^G/\lb^G$ is significantly greater than
unity.

If we do espouse some particular class of models, we may be willing
to accept a scenario which requires tuning to better than one part
in fifty. For example, the simplest SO(10) scenario with D-terms as
the only GUT source of Higgs doublet splittings requires roughly an
extra order of magnitude in tuning. But theoretically it is
appealing for its simplicity, and its tuning can be somewhat
mitigated by the phenomenologically-favored light right-handed
neutrino (which for large $\tan\beta$ does not impair bottom-tau
unification \cite{ref:hrsI,ref:vissani,ref:bmr}). Moreover, if a
model were sufficiently predictive to specify the GUT-scale boundary
conditions in terms of few unknown parameters or even none, then the
conditions we have specified for proper
electroweak symmetry breaking would either be fulfilled---in which
case the model would not be fine-tuned but rather remarkably
predictive---or not fulfilled, in which case it would be ruled out.
We do
not know of any such models at present; until a candidate is found,
we can only offer arguments of naturalness to point us in the right
direction.

If we are willing to sacrifice some naturalness, then also the PQ or
R
symmetries may be relaxed. Note that relaxing both would lead to a
large $\db$ and hence a light top, in contradiction with the recent
measurements of CDF and $\rm D\!\!{\not 0}$; therefore the typical
SUSY-breaking scalar mass {\it must} be significantly above $m_Z$.
Also, without these
symmetries other flavor-changing neutral current processes beyond
$b\to s\gamma$ may be problematic. In any case, the requirement of
proper electroweak breaking does not significantly favor one
symmetry over the other. Cosmological upper bounds on the relic LSP
density, however, favor a spectrum that is only PQ-symmetric over
one that is only R-symmetric. This is of benefit to models (see for
example Ref.~\cite{ref:ljh}) in which the $\mu$ problem is solved by
generating $\mu$ radiatively from gaugino masses or $A$ terms,
typically leading to $\mu \sim \alpha M_{1/2}$. On the other hand,
with light gauginos
and large $\mu$ the predominantly-bino LSP annihilates inefficiently
and ``overcloses'' the universe. To reconcile the predicted LSP
relic abundance with the measured age of the universe, one of the
superpartners or the pseudoscalar Higgs must be tuned light, or else
PQ symmetry must be partially restored by lowering $\mu$ down to
$200-300\,\GeV$. Of course, at the edge of the allowed range for
these parameters, the LSP is a prime candidate for the dark matter.

There are many aspects to the Yukawa-unified MSSM beyond the
question of naturalness. In Appendix A we give the exact,
semi-analytic solution to the complete 1-loop RG equations for the
third generation and the Higgs sector. Understanding their behavior
under various assumptions and boundary conditions was a prime topic
in our study. Secs.~\ref{sec:soten} and \ref{sec:sufive} addressed
the relationship between these boundary conditions and the
superspectrum, and presented the ranges in which $\mu$ and $M_{1/2}$
must fall to allow proper electroweak symmetry breaking. The effects
of a light (relative to $\Mgut$) right-handed neutrino threshold
were examined and found helpful to symmetry breaking, while it is
known that they have negligible impact on $b-\tau$ unification at
large
$\tan\beta$. The process
$b\to s\gamma$ was reexamined, and the possibility of cancellations
between various diagrams enhanced by large $\tan\beta$ were
identified for the first time. Finally, various issues regarding
proper electroweak breaking were raised and resolved: the two flat
directions which could destabilize the scalar potential, and the
scales at which they could pose a danger; the constraints on the
trilinear $A$ parameters even for third-generation sfermions due to
the hierarchy $m_Z^2 \ll \ms^2$; and similar constraints on the
trilinear $\mu$ couplings which are often neglected.

The implications of a hierarchy in the Higgs expectation values
rather than  in the third-generation Yukawa couplings are
surprisingly rich. Many aspects of the MSSM are qualitatively
changed by this assumption, and the phenomenological consequences of
these changes are clear and accessible to the next generation of
accelerator (and perhaps dark matter) experiments. Therefore the
large
$\tan\beta$ scenario offers a qualitatively different alternative to
the often-used small $\tan\beta$ ``standard'' supersymmetric model.
We have used criteria of naturalness to distinguish between the
various options for achieving large $\tan\beta$. Admittedly, these
criteria have also revealed that all large $\tan\beta$ models appear
to require some fine-tuning of the GUT-scale parameters which may
not be needed for small $\tan\beta$. In other respects, however,
such as bottom-tau unification, Yukawa unification has distinct
advantages over the conventional paradigm.  And in the near future,
most questions of naturalness will be replaced by solid experimental
data, which will be the final arbiter of all $\tan\beta$ scenarios,
large and small.

\section*{Acknowledgments}
We would like to thank L.J.~Hall, R.~Hempfling, C.~Kolda,
M.~Olechowski and F.~Zwirner for stimulating conversations on
various topics in this study, and the Aspen Center for Physics where
some of this work was done. U.S. would also like to thank the Theory
Group at CERN for its hospitality during the final stages of this
paper.  This work was supported in part by the
National Science Foundation under grants PHY-91-21039 (R.R.) and
PHY-8611280 (U.S.).
\newpage
\section*{Appendix A: Solving the RG Equations}

The 1-loop RG equations for the parameters of the MSSM are recounted
below. We use the notation $\dtau \equiv -8\pi^2 {d\over d\ln\mu}$
where $\mu$ is the mass scale, as well as $m_Z^2 = -2 m_U^2 =
\mu_Z^2 -2\mu^2$ and $m_A^2 = m_U^2 + m_D^2 = \mu_A^2 + 2\mu^2$
where $m_U^2
= \mu_U^2 + \mu^2$ and $m_D^2 = \mu_D^2 + \mu^2$ are the up- and
down-type Higgs mass parameters in the scalar potential. The soft
SUSY-breaking parameters run according to
\bea
\dtau \mu_U^2 &=&-3\lt^2\Xt^A  \phantom{-3\lb^2\Xb^A -
3\ltau^2\Xtau^A} -
\case{1}{2}\case{3}{5} g_1^2 S + \case{3}{5} g_1^2 M_1^2 + 3 g_2^2
M_2^2 \phantom{+\case{16}{3} g_3^2 M_3^2}\nonumber\\
\dtau \mu_D^2 &=&\phantom{-3\lt^2\Xt^A} -3\lb^2\Xb^A -
\spa\ltau^2\Xtau^A
+\case{1}{2}\case{3}{5} g_1^2 S + \case{3}{5} g_1^2 M_1^2 + 3 g_2^2
M_2^2 \phantom{+\case{16}{3} g_3^2 M_3^2}\label{eq:rgmass}\\
\dtau \mst^2 &=&-2\lt^2\Xt^A \phantom{ -\spa\lb^2\Xb^A  -
3\ltau^2\Xtau^A}
+\case{2}{3}\case{3}{5} g_1^2 S +\case{16}{15} g_1^2 M_1^2
\phantom{+
3 g_2^2 M_2^2} +
\case{16}{3}  g_3^2 M_3^2\nonumber\\
\dtau \msb^2 &=&\phantom{-2\lt^2\Xt^A} -2\lb^2\Xb^A   \phantom{ -
3\ltau^2\Xtau^A}
-\case{1}{3}\case{3}{5} g_1^2 S +\case{4}{15} g_1^2 M_1^2 \phantom{+
3 g_2^2 M_2^2} +
\case{16}{3}  g_3^2 M_3^2\nonumber\\
\dtau \msq^2 &=&-\spa\lt^2\Xt^A -\spa\lb^2\Xb^A   \phantom{ -
3\ltau^2\Xtau^A}
-\case{1}{6}\case{3}{5} g_1^2 S +\case{1}{15} g_1^2 M_1^2 + 3 g_2^2
M_2^2 +
\case{16}{3}  g_3^2 M_3^2\nonumber\\
\dtau \mstau^2 &=&\phantom{-2\lt^2\Xt^A -2\lb^2\Xb^A}    -
2\ltau^2\Xtau^A
-\case{3}{5} g_1^2 S +\case{12}{5} g_1^2 M_1^2  \phantom{+3 g_2^2
M_2^2+
\case{16}{3}  g_3^2 M_3^2}\nonumber\\
\dtau \msl^2 &=&\phantom{-2\lt^2\Xt^A -2\lb^2\Xb^A}    -
\spa\ltau^2\Xtau^A
+\case{1}{2}\case{3}{5} g_1^2 S +\case{3}{5} g_1^2 M_1^2 +3 g_2^2
M_2^2
 \phantom{+\case{16}{3}  g_3^2 M_3^2}\nonumber
\eea
and
\bea
\dtau \At &=&-6\lt^2\At -\spa\lb^2\Ab \phantom{. - 2\ltau^2\Atau}
+\case{13}{15} g_1^2 M_1 +3 g_2^2 M_2+
\case{16}{3}  g_3^2 M_3\nonumber\\
\dtau \Ab &=&-\spa\lt^2\At -6\lb^2\Ab- \spa\ltau^2\Atau
+\case{7}{15} g_1^2 M_1 +3 g_2^2 M_2+
\case{16}{3}  g_3^2 M_3\nonumber\\
\dtau\Atau &=&\phantom{-\spa\lt^2\At} -3\lb^2\Ab- 4\ltau^2\Atau
+\case{9}{5} g_1^2 M_1 +3 g_2^2 M_2
\label{eq:rgab}\\
\dtau B &=&-3\lt^2\At -3\lb^2\Ab- \spa\ltau^2\Atau
+\case{3}{5} g_1^2 M_1 +3 g_2^2 M_2
\nonumber
\eea
where
\bea
\Xt^A \equiv \msq^2 + \mst^2 + \mu_U^2 + \At^2
\equiv \Xt + \At^2\nonumber\\
\Xb^A \equiv \msq^2 + \msb^2 + \mu_D^2 + \Ab^2
\equiv \Xb + \Ab^2\label{eq:xdef}\\
\Xtau^A \equiv \msl^2 + \mstau^2 + \mu_D^2 + \Atau^2
\equiv \Xtau + \Atau^2\nonumber
\eea
and $S = S_1+S_2+S_3$ where $S_3 = -\mu_Z^2 - \mu_A^2 - 2 \mst^2 +
\msb^2 + \msq^2 + \mstau^2 - \msl^2$ and $S_{1,2} = -2 \muc^2 +
\mds^2 + \mqonetwo^2 + \memu^2 - \mlonetwo^2$. $S$ evolves according
to $\dtau S = b_1 g_1^2 S$, and therefore satisfies the useful
relation
\be
-b_1 \int_{\tau_G}^\tau g_1^2 S =
\left[1-{g_1^2(\tau)\over g_G^2}\right] S_G\,.
\label{eq:srel}
\ee
The gaugino masses are given by $M_i = M_{1/2} (g_i^2/g_G^2)$ where
$M_{1/2}$ and $g_G$ are the unified GUT-scale gaugino mass gauge
coupling, respectively. The $\mu$ parameter in the superpotential
runs according to
\be
\dtau \mu =\left(-3\lt^2 -3\lb^2- \ltau^2
+\case{3}{5} g_1^2 +3 g_2^2 \right)\case{1}{2}\mu\,.
\label{eq:rgmu}
\ee
Finally, the evolution of the gauge couplings is given by $\dtau
g_i^2 = b_i g_i^4$ where $b_1 = -33/5$, $b_2 = -1$ and $b_3 = 3$
[note that we always use $g_1$ normalized as an SU(5) coupling],
while the Yukawa couplings evolve according to
\bea
\dtau \lt^2 &=& \left(-6\lt^2\spa -\spa\lb^2\spa \phantom{. -
2\ltau^2\spa}
+\case{13}{15} g_1^2 \spa +3 g_2^2 \spa+
\case{16}{3}  g_3^2 \spa\right)\lt^2\nonumber\\
\dtau \lb^2 &=& \left(-\spa\lt^2\spa -6\lb^2\spa- \spa\ltau^2\spa
+\case{7}{15} g_1^2 \spa +3 g_2^2 \spa+
\case{16}{3}  g_3^2 \spa\right)\lb^2\label{eq:rgyuk}\\
\dtau\ltau^2 &=&\left(\phantom{-\spa\lt^2\spa} -3\lb^2\spa-
4\ltau^2\spa
+\case{9}{5} g_1^2 \spa +3 g_2^2 \spa \phantom{+ X
\case{16}{3}  g_3^2 \spa}\right)\ltau^2\,.\nonumber
\eea

We now present the solution to the RG equations for the dimensionful
parameters in terms of integrals over the dimensionless ones, namely
the gauge and Yukawa couplings. Notice that $\mu$ renormalizes
multiplicatively, in fact by a factor of order unity, and that is
does not enter into the RG equations of the other mass parameters.
For this reason, we may just as well treat $\mu$ at the electroweak
scale as the fundamental parameter, and thus we will not need to
refer to its GUT-scale value or its RG evolution.

The RG equations for the $A$ parameters take the form
\be
\dtau\vec{A} = H \vec{A} + M_{1/2}\vec{G}_A \,,\quad
H \equiv -\left(\begin{array}{rrr}
6\lt^2 & \lb^2 & 0 \\
\lt^2 & 6\lb^2 & \ltau^2\\
0 & 3\lb^2 & 4\ltau^2 \end{array}\right)
\label{eq:vrga}
\ee
where $\vec G_A = \left(\case{13}{15} g_1^4 +3 g_2^4+
\case{16}{3}  g_3^4, \case{7}{15} g_1^4 +3 g_2^4+
\case{16}{3}  g_3^4,\case{9}{5} g_1^4 +3 g_2^4\right)/g_G^2$. The
solution is given in terms of the ``time''-ordered exponential of
the integral of the matrix $H$,
\be
\HH \equiv T\left(\exp {\int^\tau \!H\,d\tau'}\right)
\label{eq:hhdef}
\ee
which satisfies $\dtau \HH = H\,\HH$. It may easily be computed
numerically or estimated analytically. For example, with a GUT scale
of $2.5\times 10^{16}\,\GeV$ and
$\alpha_G \simeq 1/24$, it is approximately given by
\be
\begin{array}{ccc}
\lb^G = \lt^G = 0.6: & \lb^G = \lt^G = 1.0: & 2\lb^G = \lt^G = 1.0:
\\
\HH = & \HH = & \HH = \\
\left(\begin{array}{ccc}
.239 &  -.056 &  .005 \\
-.061 &   .278  &   -.048 \\
.029 &    -.290 &   .610 \end{array}\right) &
\left(\begin{array}{ccc}
.115 &    -.042 &   .007 \\
-.048 &   .160 &     -.050 \\
.039 &    -.273 &    .430 \end{array}\right) &
\left(\begin{array}{ccc}
.103 &    -.041 &   .003 \\
-.073 &   .372 &     -.048 \\
.034 &    -.267 &    .670 \end{array}\right)
\,.\end{array}
\label{eq:numhh}
\ee
The trilinear couplings are then related to their GUT-scale values
and to the gaugino mass at the GUT scale via
\be
\vec A = \HH\vec A_G + M_{1/2} \HH\int_{\tau_G}^\tau \HH^{-1}\vec
G_A
d\tau'\,.
\label{eq:asoln}
\ee
The coefficient of $\vec A_G$ is typically an order of magnitude
smaller than that of $M_{1/2}$ in the solutions for $A_t$ and $A_b$,
so we will often assume that $A_{t,b}$ are essentially determined at
the electroweak scale by the gaugino masses. In the maximally
symmetric
case, both $M_{1/2}$ and $\vec A_G$ are negligible, while without
the
R symmetry the $M_{1/2}$ contribution is large and small effects due
to
$\vec A_G$ do not alter any conclusion substantially. (Of course one
could imagine a scenario with $\vec A_G$ much larger than the
gaugino mass and tuned to a particular value, for example to force
$A_t \to 0$ and thus suppress the
rate for $b\to s\gamma$, but we will not pursue this
further. Such a tuning is implicitly included in $\epsbsg$.) We
include approximate numerical expressions for $\vec A$ at the end of
this appendix.

The RG equations (\ref{eq:rgmass}) for the soft-breaking masses are
now
easily solved by noting that on their right-hand sides there are
only three homogeneous driving terms, the $X_i$, for seven mass
parameters. By taking linear combinations of the seven masses,
specifically
\be
\left(\begin{array}{l}
X_t \\ X_b \\ X_\tau \\ I_1 \\ I_2\\ I_3 \\ I_4
\end{array}\right) =
\left(\begin{array}{ccccccc}
 -1/2 & 0  &  1 &   0  &  1  &  0  &  0 \\
1/2 &  1  &   0  &  1 &  1 &  0  &  0 \\
1/2  &  1  &  0  &  0 &  0 & 1 &  1\\
-1 & 0 & -19/4 & -7/4 & 7/2 & 0 & 0 \\
0  &  0  &  -1/2 & -1/2 &  1 & 0  & 0 \\
-1 &  -1  & -4  & -1 & 5  &  0  & 1\\
1/2 & 1/2 & 3/2 & 0 & -3/2 & -1/2 & 1/2
\end{array}\right)
\left(\begin{array}{l}
\mu_Z^2 \\ \mu_A^2 \\ \mst^2 \\ \msb^2 \\ \msq^2 \\ \mstau^2 \\
\msl^2
\end{array}\right)
\label{eq:mdef}
\ee
or
\be
\left(\begin{array}{c} \vec X \\ \vec I \end{array}\right)
\equiv M\,\vec\mu
\label{eq:mdefvec}
\ee
we may write the RG equation of $\vec \mu$ in the form
\bea
\dtau\left(\begin{array}{c}
{\vec{X}} \\ {\vec{I}}
\end{array}\right)
&=&
\left(\begin{array}{cc}
H & 0 \\ 0 & 0
\end{array}\right)
\left(\begin{array}{c}
\vec{X} \\ \vec{I}
\end{array}\right)
-
\left(\begin{array}{c}
0 \\ \vec{v}_S
\end{array}\right)
b_1 g_1^2 S \nonumber\\
&\phantom{=}&
+
\left(\begin{array}{cc}
H & 0 \\ 0 & 0
\end{array}\right)
\left(\begin{array}{c}
\vec{A^2} \\ 0
\end{array}\right)
+
\left(\begin{array}{c}
\vec{G} \\ \vec{F}
\end{array}\right)
M_{1/2}^2
\label{eq:rgmassvec}
\eea
where $\vec v_S \equiv (-\case{25}{66},-\case{1}{33},
-\case{1}{3},\case{5}{22} )$ and $\vec G$ and $\vec F$ may be
expressed in terms of $g_i^6/g_G^4$ using Eqs.~(\ref{eq:rgmass}).
The
solution to these equations is straightforward [recalling also
Eq.~(\ref{eq:srel})]:
\bea
\vec{\mu} &=& M^{-1}\left[
\left(\begin{array}{r}
{\cal H}\vec{X}_G \\ \vec{I}_G
\end{array}\right) \right.
+ \left(\begin{array}{c}
0 \\ \vec{v}_S
\end{array}\right)
\left(1-{g_1^2\over g_G^2}\right) S_G  +
 \nonumber \\
&M_{1/2}^2&\!\!\!\!\!\left.
\left(\begin{array}{l}
\!\!\HH{\displaystyle \int^\tau}\!\HH^{-1}\left\{\vec{G} +
H \left[\HH \left({\vec A_G\over M_{1/2}} +
{\displaystyle \int^{\tau'}}\!\HH^{-1}\vec{G}_A\,d\tau''
\right)\right]^2
\right\}d\tau'\!\!\! \\
\phantom{\HH}{\displaystyle \int^\tau} \vec{F}\,d\tau'
\end{array}\right)
\right].
\label{eq:masssoln}
\eea
Assuming once again unification at the above values of the
GUT scale and $\alpha_G$, and ignoring the $A_G$ contribution,
yields
for the coefficient vector of $M_{1/2}^2 \simeq 1.6 \mwi^2$:
\be
\begin{array}{ccc}
\lb^G = \lt^G = 0.6: & \lb^G = \lt^G = 1.0: & 2\lb^G = \lt^G = 1.0:
\\
M^{-1}\left(\!\!\begin{array}{r}
7.32 \\ 7.64 \\ -1.83 \\ -18.17 \\ 0.46 \\ 2.73 \\ -0.46
\end{array}\!\!\right) =
\left(\!\!\begin{array}{r}
5.45 \\ -5.23 \\  4.77 \\  4.87 \\  5.28 \\  0.15 \\  0.53
\end{array}\!\!\right)
&
M^{-1}\left(\!\!\begin{array}{r}
6.84 \\ 7.22 \\ -1.96 \\ -18.17 \\  0.46 \\ 2.73 \\ -0.46
\end{array}\!\!\right) =
\left(\!\!\begin{array}{r}
5.87 \\ -5.61 \\  4.63 \\  4.75 \\  5.15 \\  0.17 \\ 0.54
\end{array}\!\!\right)
&
M^{-1}\left(\!\!\begin{array}{r}
6.71 \\ 8.23 \\ -1.51 \\ -18.17 \\  0.46 \\ 2.73 \\ -0.46
\end{array}\!\!\right) =
\left(\!\!\begin{array}{r}
6.19 \\ -5.25 \\  4.53 \\  5.11 \\  5.27 \\  0.13 \\ 0.52
\end{array}\!\!\right)\,.
\end{array}
\label{eq:mhcnum}
\ee

Finally, the RG equation for $B$,
\be
\dtau B = \vec H_B\cdot\vec A + G_B M_{1/2}
\label{eq:rgbvec}
\ee
where $\vec H_B = (-3\lt^2,-3\lb^2,-\ltau^2)$ and $G_B = (\case35
g_1^4 + 3g_2^4)/g_G^2$, is solved by simply integrating over the
gaugino
and $A$ contributions:
\be
B = B_G + \left({\displaystyle \int^\tau} \vec H_B\HH
\,d\tau'\right)
 \vec A_G + M_{1/2} {\displaystyle \int^\tau} \left(
\vec H_B\HH {\displaystyle \int^{\tau'}}\HH^{-1} \vec G_A \,d\tau''
+
G_B \right) d\tau' \,.
\label{eq:bsoln}
\ee
Under the same unification assumptions as before, we obtain for
$\lb^G = \lt^G = 0.6$: $B = B_G - (0.36,0.33,0.08) \vec A_G -
1.03\,M_{1/2}$; for $\lb^G = \lt^G = 1.0$: $B = B_G -
(0.41,0.36,0.11) \vec A_G -
1.25\,M_{1/2}$; and for $2\lb^G = \lt^G = 1.0$: $B = B_G -
(0.43,0.28,0.07) \vec A_G - 1.08\,M_{1/2}$.

The various integrals involving the gauge and Yukawa couplings may
be approximately evaluated analytically, since the evolution of the
$g_i$ is known and simple while the $\lambda_i$ may be approximated
in various ways, in particular near the fixed-point regime. However,
for our purposes the semi-analytic forms presented above are
sufficient. To get a feel for the results, we can evaluate the
integrals numerically. Using the same unification scale and gauge
couplings as above, inserting the initial conditions dictated by
SU(5) symmetry, and setting $A_{t,b,\tau} = A_G$ at the GUT scale
but neglecting (for ease of presentation) the small contributions of
$A_G$ to the soft-breaking masses, we obtain the following
(approximate) explicit solutions:
\bea
& & \lb^G = \lt^G = 0.6: \nonumber\\
{m_Z^2} &=&  \phantom{-}
 5.45\,{M_{1/2}^2} -
 1.29\,{\mtenh^2} +\,
 1.41\,{\msixth^2} +
 0.38\,{M_X^2} -
 0.11\,\msufive^2
 - 2 {\mu^2}   \nonumber\\
{m_A^2} &=&
 -5.23\,{M_{1/2}^2} +
 1.25\,{\mtenh^2} -\,
 1.50\,{\msixth^2}
 \phantom{.+038\,{M_X^2}} +
 0.05\,\msufive^2
 + 2 \mu^2 \nonumber\\
{\mst^2} &=& \phantom{-}
 4.77\,{M_{1/2}^2} -
 0.24\,{\mtenh^2} +
 0.53\,{\msixth^2} +
 0.09\,{M_X^2} +
 0.01\,\msufive^2\nonumber\\
{\msb^2} &=& \phantom{-}
 4.87\,{M_{1/2}^2} -
 0.21\,{\mtenh^2} +
 0.58\,{\msixth^2} -
 0.29\,{M_X^2} -
 0.51\,\msufive^2\nonumber\\
{\msq^2} &=& \phantom{-}
 5.28\,{M_{1/2}^2} -
 0.22\,{\mtenh^2} +
 0.56\,{\msixth^2} +
 0.10\,{M_X^2} +
 0.23\,\msufive^2\nonumber\\
{\mstau^2} &=& \phantom{-}
 0.15\,{M_{1/2}^2} -
 0.17\,{\mtenh^2} +
 0.66\,{\msixth^2} +
 0.12\,{M_X^2} +
 0.36\,\msufive^2 \label{eq:solone}\\
{\msl^2} &=& \phantom{-}
 0.53\,{M_{1/2}^2} -
 0.08\,{\mtenh^2} +
 0.83\,{\msixth^2} -
 0.31\,{M_X^2} -
 0.64\,\msufive^2\nonumber\\
{A_t} &=& \phantom{-}0.19\,{A_G} +
 2.2\phantom{0}\,{M_{1/2}} \nonumber\\
{A_b} &=& \phantom{-}0.17\,{A_G} +
 2.3\phantom{0}\,{M_{1/2}} \nonumber\\
{A_\tau} &=& \phantom{-}0.35\,{A_G} -
 0.13\,{M_{1/2}} \nonumber\\
B &=& \phantom{-}B_G
-0.76\,{A_G}
- 1.0\,{M_{1/2}} \nonumber\\
\mu  &=& \phantom{-}0.65\,{\mu^G}\,; \nonumber
\eea
\bea
& & \lb^G = \lt^G = 1.0: \nonumber\\
{m_Z^2} &=&  \phantom{-}
 5.87\,{M_{1/2}^2} -
 1.20\,{\mtenh^2} +\,
 1.61\,{\msixth^2} +
 0.38\,{M_X^2} +
 0.05\,\msufive^2
 - 2 {\mu^2}   \nonumber\\
{m_A^2} &=&
 -5.61\,{M_{1/2}^2} +
 1.14\,{\mtenh^2} -\,
 1.72\,{\msixth^2}
\phantom{.+ 038\,{M_X^2}}+
 0.06\,\msufive^2
 + 2 \mu^2 \nonumber\\
{\mst^2} &=& \phantom{-}
 4.63\,{M_{1/2}^2} -
 0.27\,{\mtenh^2} +
 0.45\,{\msixth^2} +
 0.09\,{M_X^2} -
 0.05\,\msufive^2\nonumber\\
{\msb^2} &=& \phantom{-}
 4.75\,{M_{1/2}^2} -
 0.23\,{\mtenh^2} +
 0.54\,{\msixth^2} -
 0.29\,{M_X^2} -
 0.47\,\msufive^2\nonumber\\
{\msq^2} &=& \phantom{-}
 5.14\,{M_{1/2}^2} -
 0.25\,{\mtenh^2} +
 0.50\,{\msixth^2} +
 0.10\,{M_X^2} +
 0.23\,\msufive^2\nonumber\\
{\mstau^2} &=& \phantom{-}
 0.17\,{M_{1/2}^2} -
 0.23\,{\mtenh^2} +
 0.54\,{\msixth^2} +
 0.12\,{M_X^2} +
0.41\,\msufive^2\label{eq:soltwo}\\
{\msl^2} &=& \phantom{-}
 0.54\,{M_{1/2}^2} -
 0.12\,{\mtenh^2} +
 0.77\,{\msixth^2} -
 0.31\,{M_X^2} -
 0.62\,\msufive^2\nonumber\\
{A_t} &=& \phantom{-}0.08\,{A_G} +
 2.0\phantom{0}\,{M_{1/2}} \nonumber\\
{A_b} &=& \phantom{-}0.06\,{A_G} +
 2.0\phantom{0}\,{M_{1/2}} \nonumber\\
{A_\tau} &=& \phantom{-}0.20\,{A_G} -
 0.22\,{M_{1/2}} \nonumber\\
B &=& \phantom{-}B_G
-0.88\,{A_G}
- 1.2\,{M_{1/2}} \nonumber\\
\mu  &=& \phantom{-}0.43\,{\mu^G}\,;\nonumber
\eea
\bea
& & 2\lb^G = \lt^G = 1.0: \nonumber\\
{m_Z^2} &=&  \phantom{-}
 6.19\,{M_{1/2}^2} -
 1.15\,{\mtenh^2} +\,
 1.69\,{\msixth^2} +
 0.38\,{M_X^2} +
 0.03\,\msufive^2
 - 2 {\mu^2}   \nonumber\\
{m_A^2} &=&
 -5.25\,{M_{1/2}^2} +
 1.24\,{\mtenh^2} -\,
 1.53\,{\msixth^2}
\phantom{.+ 038\,{M_X^2}} -
 0.07\,\msufive^2
 + 2 \mu^2 \nonumber\\
{\mst^2} &=& \phantom{-}
 4.53\,{M_{1/2}^2} -
 0.28\,{\mtenh^2} +
 0.44\,{\msixth^2} +
 0.09\,{M_X^2} -
 0.04\,\msufive^2\nonumber\\
{\msb^2} &=& \phantom{-}
 5.11\,{M_{1/2}^2} -
 0.18\,{\mtenh^2} +
 0.64\,{\msixth^2} -
 0.29\,{M_X^2} -
 0.54\,\msufive^2\nonumber\\
{\msq^2} &=& \phantom{-}
 5.27\,{M_{1/2}^2} -
 0.23\,{\mtenh^2} +
 0.54\,{\msixth^2} +
 0.10\,{M_X^2} +
 0.19\,\msufive^2\nonumber\\
{\mstau^2} &=& \phantom{-}
 0.13\,{M_{1/2}^2} -
 0.15\,{\mtenh^2} +
 0.70\,{\msixth^2} +
 0.12\,{M_X^2} +
 0.34\,\msufive^2 \label{eq:solthree}\\
{\msl^2} &=& \phantom{-}
 0.52\,{M_{1/2}^2} -
 0.07\,{\mtenh^2} +
 0.85\,{\msixth^2} -
 0.31\,{M_X^2} -
 0.65\,\msufive^2\nonumber\\
{A_t} &=& \phantom{-}0.07\,{A_G} +
 1.9\phantom{0}\,{M_{1/2}} \nonumber\\
{A_b} &=& \phantom{-}0.25\,{A_G} +
 2.4\phantom{0}\,{M_{1/2}} \nonumber\\
{A_\tau} &=& \phantom{-}0.44\,{A_G} -
 0.01\,{M_{1/2}} \nonumber\\
B &=& \phantom{-}B_G
-0.78\,{A_G}
- 1.1\,{M_{1/2}} \nonumber\\
\mu  &=& \phantom{-}0.57\,{\mu^G}\,.\nonumber
\eea

\newpage
\section*{Appendix B: The flat direction $\phi_2$}

In this appendix we discuss in more detail the constraints implied
by the flat direction $\phi_2$ of Sec.~\ref{sec:general}. As
discussed in that section, at high enough scales ($\Lambda >
\lhigh$) the direction may be stabilized by nonrenormalizable
operators regardless of the sign of $m_2^2$. At a lower scale, such
operators are ineffective, and a negative $m_2^2$ leads to an
instability unless the linear term $(\msl^2+\msq^2+\msb^2)
\left|{\mu\over \lambda_b}\phi_2\right| \equiv m_3^2 \left|{\mu\over
\lambda_b}\phi_2\right|$ in the scalar potential is significant at
such a scale. We need to estimate the scale $\llow$ down to which
this linear term may be ignored, and therefore above which
$m_2^2(\Lambda)>0$ must be enforced.

If $m_2^2(\Lambda) > 0$ for all $\Lambda$ between $\Mgut$ and $\ms$,
there is no instability. If $m_2^2(\Lambda) < 0$ for some $\Lambda >
\lhigh$, the dangerous minimum in the potential is only at field
values $\phi_2 \ll \lhigh$ (by construction of $\lhigh$), so to see
if it is a true minimum we must run to lower scales. If
$m_2^2(\Lambda) < 0$ when we reach $\Lambda = \lhigh$, then the true
minimum is at $\phi_2 \sim \lhigh$ and leads to unacceptable
symmetry breaking. If $m_2^2(\lhigh) > 0$, there is no dangerous
minimum at that scale, and we should continue running to lower
scales. If $m_2^2(\Lambda)$ gets to zero at a scale $\Lambda_C$
above the scale $\llow$ (to be determined below) so the linear term
in the potential may be ignored, we must minimize the full 1-loop
effective potential along the flat direction \cite{ref:CW}. At
1-loop order we parametrize the flat direction by a field $\phi_2$
with zero anomalous dimension, so $\vev{H_u}=\phi_2 z_u^{-1/2}$ and
$\vev{L}=\phi_2 z_L^{-1/2}$, where $z_{u,L}$ are wavefunction
renormalization coefficient functions satisfying the RG equations
$\partial \ln z_{u,L}/\partial\ln
(\phi_2/\Lambda)\equiv\gamma_{u,L}$. (Notice that $\vev{H_U} \not =
\vev{L}$ because the D-flatness condition which determines the VEVs
of $H_u$ and $L$ is corrected by 1-loop wavefunction
renormalizations.) Then in leading-$\ln (\phi_2/\Lambda)$
approximation the full 1-loop effective potential (neglecting the
linear term) is completely determined by the RG equation for $m_2^2$
calculated using Appendix A:
\bea
V_0+V_1&=&m_2^2 |\phi_2|^2 +
{1\over 8\pi^2}\left ( 3\lt^2 X_t+\ltau^2 X_\tau -
{6\over 5} g_1^2 M_1^2 -6 g_2^2 M_2^2 \right )|\phi_2|^2\ln \left
|{\phi_2\over \Lambda}\right| \nonumber\\
&\equiv& m_2^2 |\phi_2|^2 +
\Delta m_2^2 |\phi_2|^2\ln\left|
{\phi_2\over \Lambda}\right|\,.
\label{eq:oneloop}
\eea
At the scale $\Lambda_C$, where $m_2^2=0$, the above potential has
the well-known Coleman-Weinberg minimum at $\langle \phi_2\rangle
\sim \Lambda_C$. Moreover the vacuum energy at that minimum is
$\sim -\Delta m_2^2 \Lambda_C^2$, which is parametrically much below
the usual electroweak vacuum energy $-{\cal O}(m_Z^4/g^2)$. If
$\Lambda_C$ and therefore $\langle \phi_2\rangle$ get too low, the
linear term $m_3^2 \left|{\mu\phi_2\over \lambda_b}\right|$ in the
potential dominates, and the unwanted minimum disappears. To get a
rough estimate of when this happens, we just add the linear term to
Eq.~(\ref{eq:oneloop}) and again minimize at $\Lambda=\Lambda_C$. We
find that the dangerous minimum is eliminated because of the linear
term when $\Lambda_C \roughly{<} \half e^{3/2}  (m_3^2/\Delta m_2^2)
\mu/\lb$. Thus, $m_2^2(\Lambda)$ must be prevented from vanishing
only above scales of order
\be
\llow \sim {m_3^2 \mu \over \Delta m_2^2 \lambda_b}\sim
{2\pi\over\alpha} {\mu\over \lambda_b}\,,
\label{eq:crit}
\ee
where the quantities on the right hand side are evaluated at
$\Lambda_C$ and $\alpha$ is a combination of gauge and Yukawa
coupling strengths. The above is just an estimate. For instance,
2-loop RG
effects and finite parts in the 1-loop potential
modify the numerical prefactor on the right-hand side by ${\cal
O}(1)$. Moreover, notice that along $\phi_2$ there is a hierarchy
between the scale of SU(2) breaking ($\sim \vev{H_u}$) and
that of color breaking ($\sim \vev{Q}$). Therefore
we expect the 1-loop corrections to $m_3^2$, which we haven't
included
in our estimate, to be of order  $\alpha\ln(\vev{Q}/\Lambda_C)\sim
\alpha\ln (1/\alpha)$.

\section*{Appendix C: Approximations in the effective potential}

Throughout the paper our analysis has been based on the 1-loop-RG
improved tree
level scalar potential $\Vzero$. In this appendix we discuss the
possible
relevance of a more accurate treatment which would include the full
1-loop effective potential, 1-loop threshold corrections and 2-loop
running of the soft SUSY-breaking parameters. We will show that all
these refinements cannot change the basic conclusions of our study.

Let us first discuss the use of the 1-loop effective potential. This
becomes necessary when the tree-level potential
is almost flat (or even unbounded) along some direction in field
space. Then the quantum correction stabilizes the potential
at large field strength. Minimization of the potential then yields a
vacuum expectation value for that field of order the renormalization
scale
$\Lambda$ at which the potential becomes flat (dimensional
transmutation)\cite{ref:CW}. (Of course we are assuming that at
very high energy scales the tree potential is bounded from below).
In our study, as discussed in Sec.~\ref{sec:general}, we need only
worry about the two flat directions $\phi_1$ and $\phi_2$.
Hence the full 1-loop effective potential is only relevant for those
parameter ranges when $m_1^2$ or $m_2^2$ become very small, namely
very near the $m_1^2 = 0$ or $m_2^2 = 0$ planes in the space of GUT
parameters. Thus a more correct and involved calculation would only
change the margins of the allowed parameter space but could not
significantly alter any conclusions.

The remaining improvements are given by ({\it i}) GUT-scale
thresholds, ({\it ii}) SUSY-scale thresholds, and ({\it iii}) 2-loop
running.
The first type of effects are model-dependent and have essentially
been encompassed by our discussions of the various boundary
conditions. Let us then turn to the effects of the superpartner
thresholds on
the allowed (usually triangular) regions of SO(10) parameter
space---the extension to more general initial parameters will be
obvious. The first question is where to stop the running. In the
plots in Fig.~7 the running has been stopped at $\Lambda = m_Z$,
even though in the typical scenarios the superpartners decouple
closer to 1 TeV. A more exact analysis would then add 1-loop
threshold corrections to the various mass parameters.
(In the notation of Ref.~\cite{ref:grz}, such threshold effects
would appear as terms in the 1-loop effective potential.) These
corrections
are roughly proportional to $\alpha \ms \ln (\ms/m_Z)$, where $\ms$
collectively denotes the low energy values of the MSSM mass
parameters and $\alpha$ is the appropriate gauge or Yukawa coupling
strength. In the absence of strong premature focusing (for example,
$\lG$ well below 1.2 in the SO(10) scenario), when all the
superpartner masses are comparable, the effect on the triangle plots
is just to move the triangle boundaries by an amount which is
roughly ${\cal O }[\alpha\ln (\ms/m_Z)]$ times the size of the
triangle itself. The size of the allowed region in the GUT parameter
space is only slightly changed, and the same arguments we have made
can be applied to the slightly shifted GUT parameter ranges. If
there is strong premature focusing---that is, if $\lG$ is just below
its maximal value, the allowed region of parameter space is very
small, and some particles have masses $m_{0,L}^2$ well below the
rest of the superpartner masses $m_{0,H}^2$---then the threshold
corrections could be significant when  $\alpha m_{0,H}^2 \ln
(m_{0,H}^2/m_Z) \sim m_{0,L}^2$. But their only relevant effect
would be to slightly shift the maximal value of $\lG$, so once again
no conclusions are qualitatively altered.
Notice that it was crucial to establish that the
corrections are proportional to the {\it electroweak-scale}
values of the masses, which can be much smaller that the GUT values
when any focusing is relevant. The effects of 2-loop running,
however, are in general proportional to the value of the masses at
higher scales: we expect corrections $\sim \alpha M_0^2$ to the low
energy values of the masses (recall that $M_0$ is a typical soft
mass at $\Mgut$). When $\lG$ is large and $\lt^G = \lb^G$, the
low-energy masses in the allowed regions of parameter space are
focused to small values $m_0^2$. When the focusing is strong enough
that $m_0^2 \sim \alpha M_0^2$, some effects of 2-loop running are
large. For example, a value of $\lG$ which leads to a small but
finite allowed triangle with 1-loop running could lead to vanishing
allowed area using 2-loop running. So when there is large focusing
the more exact upper bound on $\lG$ could shift somewhat---but
because of the fixed-point evolution of $\lt$ at large $\lG$, the
corresponding values of $m_t$ will not change much. We therefore
expect that all our conclusions are robust.

\end{document}